%% file: thesis.tex
\documentclass{thesis}
\begin{document}
\pagenumbering{roman}

\include{cover}
\include{abstract}
\include{license}
\include{preface-ack}
\include{alt}

\include{chapter1}

\include{chapter2}

\include{chapter3}

\include{chapter4}

\include{chapter5}

\include{chapter6}

\include{chapter7}

\include{chapter8}

\cleardoublepage
\phantomsection
\addcontentsline{toc}{chapter}{Bibliography}
\bibliographystyle{plainnat}
\bibliography{references,rfc}{}
\pagebreak
\include{appendix}
\end{document}

%% file: cover.tex
\thispagestyle{empty}
\noindent
\hspace*{10mm}%
\parbox{0.75\textwidth-10mm}{\textbf{Adrián Yanes}}\par%
\vspace{6mm}%
\noindent

\hspace*{10mm}%
\parbox[t][120pt]{\textwidth-10mm}{\raggedright%
\usefont{T1}{phv}{b}{n}\fontsize{15}{15}\selectfont{OpenWeather: \vspace{5mm} \\ A Peer-to-Peer Weather Data Transmission Protocol{}}}\par%
\vspace{20mm}

\noindent%
\hspace*{10mm}%
\parbox{1\textwidth-10mm}{\raggedright\small%
{\usefont{T1}{phv}{b}{n}\fontsize{12}{11.1}%
\selectfont{\textcolor{black}{Faculty of Electronics, Communications and Automation{}}}}\\[1em]
}\par%
 
\noindent%
\hspace*{10mm}%
\parbox{1\textwidth-10mm}{\raggedright\small%
{\usefont{T1}{phv}{b}{n}\fontsize{10}{11.1}%
\selectfont{\textcolor{black}{Department of Communications and Networking (Comnet){}}}}\\[1em]
}\par%

\vspace{1ex}%
\noindent%
\hspace*{10mm}%
\parbox{1\textwidth}{ \small 
Thesis submitted for examination for the degree of Master of Science in Technology.\\
\\
Otaniemi, Espoo, 31.08.2011
}\par

\vspace{23mm}%
\noindent%
\hspace*{10mm}
\parbox{1\textwidth}{ \small 
\small \textbf{Thesis supervisor and instructor:} Prof. Jörg Ott
}\par
\noindent%

\vspace{5cm}%

\parbox [h]{0.75\textwidth-10mm}{
\hspace{10mm}
\includegraphics{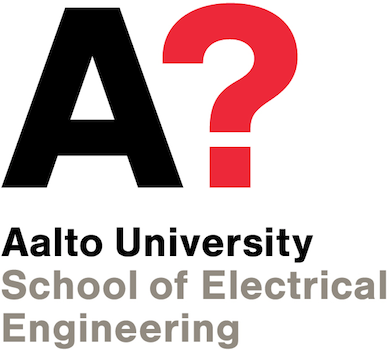}
}\par

\pagebreak

%% file: abstract.tex
\thispagestyle{empty}
\begin{tabular}{l r}
\begin{minipage}[c]{102mm} 
\textsc{Aalto University}\\ \textsc{School of Electrical Engineering}
\end{minipage}
 & \hfill
\begin{minipage}[c]{40mm} 
\begin{flushright}
\textsc{Abstract of the}\\ \textsc{Master's thesis}
\end{flushright}
\end{minipage}\\ 
\end{tabular}

\begin{tabular}{|l|}
\hline %
\begin{minipage}[c][26mm]{\textwidth} 
\vfill
\textbf{Author:} Adrián Yanes \vfill
\textbf{Title:} OpenWeather: a peer-to-peer weather data transmission protocol \vfill
\textbf{Date:} 31.08.2011 \hfill \textbf{Language:} English \hfill \textbf{Number of pages:} 115 + 23 \vfill
\vfill
\end{minipage}\\ 
\hline

\begin{minipage}[c][18mm]{\textwidth} 
\vfill
Faculty of Electronics, Communications and Automation \vfill
Department of Communication and Networking \vfill
\textbf{Professorship:} Networking Technology \hfill \textbf{Code:} S-38\vfill
\vfill
\end{minipage}\\
\hline

\begin{minipage}[c][12mm]{\textwidth} 
\vfill
\textbf{Supervisor:} Prof. Jörg Ott \vfill
\vfill
\end{minipage}\\
\hline
\begin{minipage}[t][133mm]{\textwidth} 
The study of the weather is performed using instruments termed \emph{weather stations}. These weather stations are distributed around the world, collecting the data from the different phenomena. Several weather organizations have been deploying thousands of these instruments, creating big networks to collect weather data.\\

These instruments are collecting the weather data and delivering it for later processing in the collections points. Nevertheless, all the methodologies used to transmit the weather data are based in protocols non adapted for this purpose. Thus, the weather stations are limited by the data formats and protocols used in them, not taking advantage of the real-time data available on them.\\

We research the weather instruments, their technology and their network capabilities, in order to provide a solution for the mentioned problem. OpenWeather is the protocol proposed to provide a more optimum and reliable way to transmit the weather data. We evaluate the environmental factors, such as location or bandwidth availability, in order to design a protocol adapted to the requirements established by the automatic weather stations.\\

A peer to peer architecture is proposed, providing a functional implementation of OpenWeather protocol. The evaluation of the protocol is executed in a real scenario, providing the hints to adapt the protocol to a common automatic weather station.
\vfill

\end{minipage}\\
\hline

\begin{minipage}[c][12mm]{\textwidth}
\vfill
\textbf{Keywords:} P2P, peer to peer, weather stations, real-time, protocol standardization, embedded system, IETF, RFC\vfill
\vfill
\end{minipage}\\
\hline
\end{tabular}

\pagebreak

%% file: license.tex
%

\thispagestyle{empty}
\begin{center}
\begin{minipage}{0.8 \textwidth}
\vspace{8cm}%
\emph{If you want to accomplish something in the world, idealism is not enough, you need to choose a method that works to achieve the goal. In other words, you need to be “pragmatic”. }\\
\\
Richard Matthew Stallman \\
\end{minipage}
\end{center}
\pagebreak
\thispagestyle{empty}

\begin{center}
\usefont{T1}{phv}{b}{n}\fontsize{20}{20}\selectfont{License}
\end{center}

\vspace{1cm}%

\begin{center}
\usefont{T1}{phv}{b}{n}\fontsize{20}{20}\selectfont{CC0 1.0 Universal (CC0 1.0)\\
Public Domain Dedication}
\end{center}

\begin{center}
\begin{minipage}{0.8 \textwidth}
\vspace{3cm}%
\textbf{No Copyright}\\

The person who associated a work with this deed has dedicated the work to the public domain by waiving all of his or her rights to the work worldwide under copyright law, including all related and neighboring rights, to the extent allowed by law.\\

You can copy, modify, distribute and perform the work, even for commercial purposes, all without asking permission. See Other Information below.\\

\textbf{Other information}\\

\begin{itemize}

\item In no way are the patent or trademark rights of any person affected by CC0, nor are the rights that other persons may have in the work or in how the work is used, such as publicity or privacy rights.
\item Unless expressly stated otherwise, the person who associated a work with this deed makes no warranties about the work, and disclaims liability for all uses of the work, to the fullest extent permitted by applicable law.

\item When using or citing the work, you should not imply endorsement by the author or the affirmer.
\end{itemize}
\vspace{1cm}
\begin{center}
This is a human-readable summary of the Legal Code: http://creativecommons.org/publicdomain/zero/1.0/legalcode
\end{center}
\end{minipage}
\end{center}
\vspace{0.2cm}
\textbf{Note: }this license does not apply for the following parts of this thesis: Figure 2.2, Figure 2.3, Figure 2.4, Figure 3.1,  Figure 3.5, Figure 3.7. The copyright of these figures belongs to their authors.

\pagebreak

%% file: preface-ack.tex
\sectionwp{Preface}

Before I started this thesis, my knowledge about weather stations and the technologies behind them was pretty limited. Nevertheless, in some way the weather data transmission got my attention. Probably my preference for open systems, libre software and my passion for network protocols, was the trigger to look for a topic that combines all of these areas.

During this thesis my main goal has been to show how a modern instrument as an automatic weather station, can be improved using concepts brought from open and standard technologies.

Furthermore, the impact that the weather has in our everyday deserves a deeper attention from the engineering point of view. Although the scientists are doing a great job finding new ways to understand the weather, they really need improvements in the technology field, to achieve even more better results.

OpenWeather looks for a digression. This research is looking for the attention of the scientists and the industry; for those vendors that are manufacturing instruments without a common standard, and for those scientists that are experimenting issues with the weather data acquisition. Both communities must find an agreement to standardize the methods and the technologies to transmit the weather data.

I truly think that if we start using protocols designed having in consideration the characteristics of the weather data, the result of their use will change completely the vision about what is the weather, what causes it and how it can be predicted.

\thispagestyle{plain}
\pagebreak

\sectionwp{Acknowledgements}

Too many people have been involved directly or indirectly in this thesis, however, nothing of this would have happened without the support of my parents, Luisa Pilar and José Emilio. Thanks for your support during all these years; from the beginning you have trusted me blindly, thanks for teaching me the value of the knowledge, I love you.

Thanks to the Aalto University, and especially to professor Jörg Ott, for supervising my thesis. 

Thank you to Antti Lauri from the SMEAR project \& University of Helsinki, for inviting me to spend a weekend in the premises of the SMEAR project in Tampere, having the possibility to discuss with some engineers and scientists, the issues and particularities of one of the biggest weather stations in the world. Also thanks to Pasi P. Aalto and Erkky Siivola, from the University of Helsinki. Your patience and experience with weather stations have been really useful to me.

I really need to say a big thanks to Vaisala corporation, especially to Pekka Korhonen and Jing Lin, thanks for providing me with all the necessary hardware -included an amazing last generation weather station- to perform my research and for the support offered.

A big thank you to Gonzalo Mariscal, for supporting me and my constant requests during two years.

Several friends have been involved in all the OpenWeather matters, thanks to the guys of the Polyteknikkojen Radiokerho (Radio Club), to provide me the materials and the place to install the weather station. Thanks to Jose Azeredo Lima, for his support with some of the figures. Thanks to those friends that probably read this thesis even more times than me: Borja Tarraso, Sergey Vetrogronov and David Fernández, thanks for your dedication helping me, your support has been really important for me.

Finally, I want to dedicate this thesis not to one, two or three persons, I want to do it to a community, the community of the libre and open source software. To all of those developers that are working so hard to make the world a better place, without your work this thesis would never exist: respect.

\vspace{1cm}
Otaniemi, Espoo, August 2011.

\vspace{0.5cm}
Adrián.
\thispagestyle{plain}
\pagebreak

\tableofcontents

\pagebreak

%% file: alt.tex
\newacronym{ASCII}{ASCII}{American Standard Code for Information Interchange}
\newacronym{AWS}{AWS}{Automatic Weather Station}
\newacronym{AWOS}{AWOS}{Automated Weather Observing System}
\newacronym{ASOS}{ASOS}{Automated Surface Observing System}
\newacronym{AWSS}{AWSS}{Automated Weather Sensor System}
\newacronym{CSV}{CSV}{Comma-Separated Values}
\newacronym{CPU}{CPU}{Central processing unit}
\newacronym{RS232}{RS-232}{Recommended Standard 232}
\newacronym{RS422}{RS-422}{Recommended Standard 422}
\newacronym{RS485}{RS-485}{Recommended Standard 485}
\newacronym{MBITS}{Mbits}{Megabits}
\newacronym{FTP}{FTP}{File Transfer Protocol}
\newacronym{TCP}{TCP}{Transmission Control Protocol}
\newacronym{GUI}{GUI}{Graphical User Interface}
\newacronym{IP}{IP}{Internet Protocol}
\newacronym{MB}{MB}{Megabyte}
\newacronym{KBITS}{kbit}{kilobits}
\newacronym{USB}{USB}{Universal Serial Bus}
\newacronym{HTTP}{HTTP}{Hyper Transfer Text Protocol}
\newacronym{RTP}{RTP}{Real-time Transport Protocol}
\newacronym{TSV}{TSV}{Tab-separated values}
\newacronym{SDI-12}{SDI-12}{Serial Data Interface at 1200 Baud}
\newacronym{SI}{SI}{Système international d'unités - International System of Units}
\newacronym{NMEA-0183}{NMEA-0183}{National Marine Electronics Association 0183}
\newacronym{MHz}{MHz}{Megahertz}
\newacronym{kB}{kB}{Kilobyte}
\newacronym{CWOP}{CWOP}{Citizen Weather Observer Program}
\newacronym{RAM}{RAM}{Random-access memory}
\newacronym{APRS}{APRS}{Automatic Position Reporting System}
\newacronym{IO}{IO}{in / out}
\newacronym{IETF}{IETF}{Internet Engineering Task Force}
\newacronym{RFC}{RFC}{Request for Comments}
\newacronym{SMB}{SMB}{Server Message Block}
\newacronym{GSM}{GSM}{Global System for Mobile Communications}
\newacronym{GPRS}{GPRS}{General Packet Radio Service}
\newacronym{UMTS}{UMTS}{Universal Mobile Telecommunications System}
\newacronym{AX.25}{AX.25}{Link Access Protocol for Amateur Packet Radio}
\newacronym{PROM}{PROM}{Programmable Read-Only Memory}
\newacronym{WMO}{WMO}{World Meteorological Organization}
\newacronym{FMI}{FMI}{Finnish Meteorological Institute}
\newacronym{NOAA}{NOAA}{National Oceanic and Atmospheric Administration}
\newacronym{API}{API}{Application Programming Interface}
\newacronym{ECMWF}{ECMWF}{European Centre for Medium-Range Weather Forecasts}
\newacronym{METAR}{METAR}{Meteorological Service For International Air Navigation}
\newacronym{ICAO}{ICAO}{International Civil Aviation Organization}
\newacronym{PTH}{PTH}{Pressure, Temperature, Humidity}
\newacronym{GOS}{GOS}{Global Observing System}
\newacronym{GTS}{GTS}{Global Telecommunication System and WMO Information System}
\newacronym{GDPFS}{GDPFS}{Global Data-processing and Forecasting System}
\newacronym{FEC}{FEC}{Forward error correction}
\newacronym{JSON}{JSON}{JavaScript Object Notation}
\newacronym{SOA}{SOA}{Service-oriented architecture}
\newacronym{OS}{OS}{Operating System}
\newacronym{P2P}{P2P}{Peer to peer}
\newacronym{NAT}{NAT}{Network address translation}
\newacronym{XML}{XML}{eXtensible Markup Language}
\newacronym{NTP}{NTP}{Network Time Protocol}
\newacronym{DTD}{DTD}{Document Type Definition}
\newacronym{SHA}{SHA}{Secure Hash Algorithm}
\newacronym{DNS}{DNS}{Dynamic Name Server}
\newacronym{UTM}{UTM}{Universal Transverse Mercator}
\newacronym{UTC}{UTC}{Coordinated Universal Time}
\newacronym{ISO}{ISO}{International Standard Organization}
\newacronym{PTU}{PTU}{Pressure, Temperature and Humidity}
\newacronym{UTF}{UTF}{Universal Character Set - Transformation Format}
\newacronym{URI}{URI}{Uniform Resource Identifier}
\newacronym{URL}{URL}{Uniform Resource Locator}
\newacronym{UML}{UML}{Unified Modeling Language}
\newacronym{BSON}{BSON}{Binary-JSON}
\newacronym{RTT}{RTT}{Round-trip time}
\newacronym{TLS}{TLS}{Transport Layer Security}
\newacronym{ACL}{ACL}{Access Control List}
\newacronym{DOS}{DoS}{Denial-of-service}
\newacronym{DDOS}{DDoS}{Distributed denial-of-service}
\newacronym{IANA}{IANA}{Internet Assigned Numbers Authority}
\makeglossaries
\printglossary[title=Acronyms, toctitle=ACRONYMS]
\pagebreak
\listoffigures 
\pagebreak
\listoftables
\pagebreak
\thispagestyle{plain}
\pagebreak
\thispagestyle{plain}
\pagebreak

%% file: chapter1.tex
\cleardoublepage
\setcounter{page}{1}
\pagenumbering{arabic}

\chapter{Introduction} 

From the beginning of the time, the weather has been an important factor in the
human life. Its impact of it in our everyday, gives as result that during centuries we
have been trying to understand and predict it as much as possible.

We all are familiar with some weather concepts, because it really has an impact on how we proceed in our life. For instance, it is really common to check the
forecast before we start some outdoor activity or even without any special
reason, only to know which kind of atmospherical conditions we are going to
experiment the following days; this is possible by the meteorology.

The science of meteorology takes the role of the scientific study of the atmosphere, this implies to know certain phenomena behave and which kind of predictions can
be made based on them, and of course the impact of them in our lives.
To achieve this goal, the science of meteorology has been developing different techniques and methods to measure and collect the necessary data to make these predictions.
The human history is full of inventions of different instruments designed to
make this possible. In the past, these instruments were based just in
mechanical principles with a high limited accuracy. Nowadays, we can find a huge
set of alternatives based in digital mechanisms which allow us to predict the
weather and understand the atmosphere phenomena with high precision and accuracy; giving us a better knowledge of our environment and at the end making our life easier.
Even if in the last years the transition from pure mechanical instruments to the
digital technology has been really fast, certain parts still have not been renovated
or are under development.

The purpose of this thesis is to study some possible
improvements of these parts, more specifically in the
protocols used to transmit the weather data collected in different
instruments to the places in which the data is processed for its broadcasting.

When I started researching some weather instruments their technology caught my attention, mainly in all the aspects of measuring a phenomenon with precision and feasibility, and at the same time I was confused about how the protocols used in them are full of legacy and low efficiency, in terms of data transmission and real time data availability.

Nowadays, we have functional and reliable weather data systems to study the different phenomena, however, the potential of the real time data gets blocked by the methods used on the weather data collection. Even if at the end, we have the capability to process and interpret the data, a huge amount of effort is needed to make this happen, due to the methods and technology used for the collection. This fact got my attention when I was trying to find some research area in which the protocols and the information theory could help to make this process more useful, faster and reliable.

After understanding and verifying how the weather instruments work, I found really important to ask some meteorological scientists what the state of the art is, concerning atmosphere data collection. I had the great opportunity to visit the SMEAR \cite{SMEAR} project for a weekend, study how the data is collected, transmitted, processed and stored. At the same time, the scientists that are using this data to study the atmosphere, could confirmed that some huge improvements can be made in order to improve the data transmission (this affirmation is mostly based on the technical issues that they are experimenting in their research).

This fact and the interest in peer to peer protocols and the real data transmission, were the final trigger to start this thesis and try to find a possible solution to improve the speed and reliability of the weather data transmission.

Applying the concept of "peer" to any group of sensors which are collecting weather data and assuming that also the scientist is a peer that fetches and exchanges data with other scientists (also considered peers), it was the foundation to research, looking for a protocol that allows the weather stations to exchange and route data with other weather stations and at the same time provides a infrastructure to access data collected in real time.

\section{Background}\label{1.1}

A weather instrument is an artifact which main task consists in the data collection from one to multiple atmosphere phenomena. These instruments are designed thinking in a specific use case: a particular natural phenomenon, and at the same time with a well defined goal: the collection of data that helps to study and predict  phenomenon.

Nowadays, we can find several solutions to achieve this goal. Science has found different ways to measure the same phenomenon in different ways and with different reliability. However, common techniques are used around world to measure the same phenomenon. Sometimes the reasons for using a certain technique can go from the complexity and reliability of it, to the cost of it. The standard way to measure a particular phenomenon is developing a specific instrument (also named \emph{sensor}) for it, this instrument is able to measure and understand it better.

Some popular concept to refer these sensors is \emph{"weather stations"}, nevertheless, this term is not correct at all due to the amount of instruments in a weather station can be barely different compare with other vendors' instruments.
Notwithstanding, this term is accepted as common to refer the group of sensors used to collect weather data (\textbf{we will use this concept from now on to refer to a group of sensors creating an identity named "weather station"}). 

The following list\footnote{These sensors are an example based on the market's offer, notwithstanding the amount of different sensors to measure the phenomena increases really fast, being difficult to track them all.} enumerates some common instruments in a weather station:

\begin{itemize}

\item Thermometer for measuring air and sea surface temperature
\item  Barometer for measuring atmospheric pressure
\item  Hygrometer for measuring humidity
\item  Anemometer for measuring wind speed
\item  Wind vane for measuring wind direction
\item Rain gauge for measuring precipitation
\item Disdrometer for measuring drop size distribution
\item Transmissometer for measuring visibility
\item Ceiling projector for measuring cloud ceiling

\end{itemize}

All of these instruments have a defined mechanism to measure a specific phenomenon and collect the data to be processed later. These instruments or sensors are applying some physic principle to get this data and converted it into digital information for future transmission.

After the data is collected in the instrument\footnote{A device named datalogger is involved in this process.},  it is transmitted to some organization, such as a meteorological institute, to interpret the data and get some conclusions concerning the current status of the weather and future predictions.

With information collected in different instruments around the world, we can know the status of the weather and how it will be in the future, all of these weather stations around the world are "weather data pickers", and the success of the final weather prediction resides in the efficiency and reliability in which this data is collected, transmitted and processed.

For a while, all of this process has been optimized in several ways, like creating better instruments, infrastructures and organizations focused only in this field. However, the standardization process only impacted on measure techniques and data units, putting in a secondary plane, other parts of the process such as communication protocols, digital interfaces used, etc.

\section{Problem statement}\label{1.2}

The nature of the data collected in the weather stations involves to place them in different locations around the world. It is creating a trickier scenario for the data collection.
Multiple weather stations are located in inaccessible places, but their location is mandatory to deploy feasible models for weather predictions. Commonly, these instruments are placed in different locations in which sometimes the environment is not friendly at all to be combined with digital technology; some examples of these are isolated places such as mountains, roads or forests. These environmental conditions bring issues as lower \textbf{bandwidth availability}, \textbf{difficulties to get enough energy 24x365} and the variable weather conditions in which some instruments are subdue with the implications of these in terms of lifetime.

The industry has been developing different instruments to achieve this objective and avoid the mentioned issues. However, the main effort has been to develop instruments with high accuracy, low power consumption, resistance, and small size; \textbf{resting importance to the methods used in the transmission efficiency of the data collected}. 

It is a fact that these instruments are getting more complex, reliable and tiny with the time. Nevertheless, there is non defined standard to transmit and process the data collected from the instruments to the locations in which this data is useful (meteorological organizations, computation centers, databases, etc). The common practice is that \textbf{the vendors choose their own data format / protocol for this purpose}, and depending on the manufacturer the instrument formats and transfers the information using some standard for peripheral devices such as \textbf{\gls{RS232}}, \textbf{\gls{RS422}}, \textbf{\gls{RS485}}, or \textbf{\gls{USB}}. At the same time one of the following serial communications protocol is commonly used to transmit the collected data:

\begin{itemize}
\item	RAW \gls{ASCII}\footnote{The concept of RAW refers to a serial communication in which is not used any special data format, just data formatted using ASCII as character-encoding scheme.}
\item \gls{SDI-12}
\item \gls{NMEA-0183}
\end{itemize}

These are the standards that the industry established to transmit the data from the instruments that they are manufacturing. However, \textbf{the mentioned standards are generic} for serial communications data transmission, \textbf{without any direct or indirect relation or adaptation to the weather data}. That means that the industry chooses only to take care of the data transmission for their own instruments, creating their own data formats, timings of transmission, data definition and so on. This common practice between vendors is causing the non-existence of an international standard and by default the incompatibility of these instruments with others brands, plus the possibility to combine the output data of different instruments from different brands.

The use of a non-adapted protocol for the data transmission decreases the efficiency and the possibility of a easy manipulation of the data. Even assuming that the industry chose this way to transmit the data based in the mainstream digital solutions, in terms of serial communications, \textbf{is possible and feasible to deploy a standard to format and transmit this information in a more optimized and reliable way}; this will imply the participation of different vendors to standardize this format. The process of standardization is a well-known practice in different fields of the industry due to the advantages that it brings in terms of compatibility, interoperability, safety, repeatability, or quality; at the same time standardization is supported in multiple cases (depending of the industry) for international laws.

Thus, the choice made by the industry makes the optimization of the data manipulation really painful, in addition, it is rare to have one point of weather collection with only one brand of instruments. It entails that at the end of the data transmission, the data collection scenario must be combined with different software from different vendors and different parameters. This makes the process of the weather data collection even more arduous, since the original format in which the data is transmitted is completely useless and must be converted to be combined with other data.

The absence of a standard is forcing to pre-process the weather data after its transmission, even if this is something needed in any network data transmission at some point; the format used in the process can save a lot of CPU cycles, memory and bandwidth. This absence forces the weather data collection centers to convert the data in a useful format for future computation, and this is happening through custom software developed by the vendor's instruments or in some cases, custom software developed by the organization itself. As an example of this, the SMEAR project\cite{SMEAR} has developed several parsers and scripts to manipulate this data before it can be processed, wasting time and resources that can be easily solve through a standardization.

We needed to highlight that most of the end users of this software are scientists that need the data to get some conclusions about the weather. It means that at the end of the data collection workflow, it is manipulated through software focused in mathematics computation like MathLab, which does not support any data format used by the weather instruments, forcing the scientists to have the data in dummy formats as \gls{CSV}, \gls{TSV}, or just plain text, to be able to use it.

The following figure shows an example of how the data is processed and where the conflictive points are:

\begin{figure}[H]
\centerline{\includegraphics[width=0.7\textwidth]{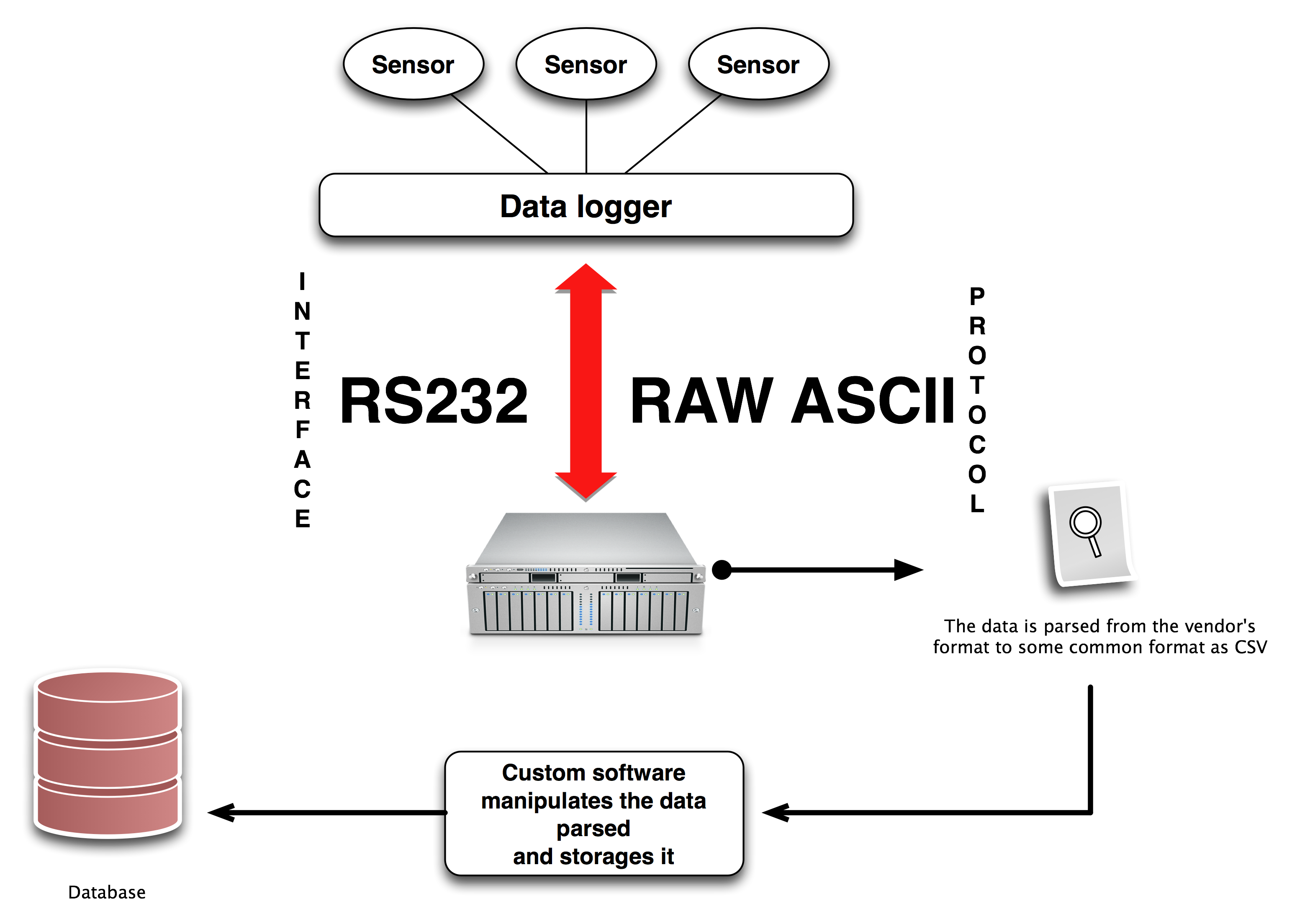}}
\caption{Common scenario to collect, transmit, manipulate and storage data in a weather station.}
\label{f1.1}
\end{figure}

As it is observable in the figure \ref{f1.1} the parsing and the implication of specific software in the process, is causing the implementation of unnecessary subprocess as parsing, packaging and data conversion. At the same time, the process described is decreasing the possibilities to have easily accessible information in real time.

Though meteorology needs big amounts of data collected in different places and the analysis of this data is made using different times frequencies, we can not ignore how useful the data of our environment can be if its accessible in real time. As example of this can be that industry has been focusing, in the last years, on developing technologies that allow the users to get information on demand and in real time, this is supported by the principle that with more detailed and updated information we can act with more precision and feasibility.

\textbf{The absence of a weather data transmission protocol} is impeding us to know how powerful can be the combination of multiple weather data sources in real time. It can provide the mechanisms to deploy different models and perform analysis of the data based in the real current situation of the weather, regardless the location or brand of the weather station. Even if nowadays we have enough precision understanding the atmosphere phenomena to predict future weather conditions, we still need to advance in the physics to deeper understand the impact of these phenomena and how they work, providing us a better knowledge of our environment, and at the end, improving our quality of life. 

Currently the weather data information is collected in real-time (because the sensors are taking samples of the current environment), notwithstanding the technology used in the process of the data transmission does not take this in consideration, non using a standardized and optimized process for this specific data.

This fact is, at some point, blocking the possibility to explore how useful this data could be for us, but it is not accessible at all because engineering issues. On the other hand, the absence of a common protocol even to exchange non real-time data, generates a big amount of issues in terms of data combination and comparison; causing several problems of incompatibility between the organizations focus in the weather  study, and forcing the use of extra resources in operations such as data normalization (something that can be fixed through a common data format). 

\section{Research objectives and scope}

The purpose of this thesis is to identify the points in the weather data transmission in which the process is not optimized according to the nature of the data. At the same time, a protocol is proposed as proof of concept, showing how the weather data transmission can be improved without too much effort from the vendor's side.

The foundation of this research is to find a path having in mind a real scenario as the SMEAR project\cite{SMEAR}, in which the process of the data transmission and manipulation can be improved offering new use cases for the data, in terms of real time acquisition, manipulation and storage.

The following points identify the approach of the research briefly:

\begin{itemize}
\item	Identify the blocker points in data transmission concerns
\item Study how the weather data transmission and manipulation can be improved 
\item	Develop a protocol prototype specification that provides an improvement in the current scenario
\end{itemize}

As final objective the author is looking forward to motivate the vendors to start a standardization process to improve the mentioned problems. Based on the opinions shared with atmospherical scientists, the absence of accessible real-time comes from the engineering side, and it is needed to develop some technology that ensures an easy a feasible method to access this data.

\section{Motivations}

After working with weather instruments, understanding how they work and how they transmit the data, I noticed the issues previously exposed. However, my vision was not enough to be sure about the key-problem treated (because it was only based in end-user weather instruments). When I had the opportunity to see how the biggest weather station in the world (concerning gas measurements) is fetching, transmitting and manipulating data, and at the same time, talk with some scientists about my suspicions were confirmed by them: this process can be optimized.

In addition, the absence of a standard in something so important as the weather data transmission, gave me enough reasons to perform this research, based in the idea that maybe some conclusions can be directly applied to the industry.

Finally, my vision about certain user rights is implied in this research as well. I am convinced that a society  informed has always more possibilities to have a better quality of life. In the last years several misunderstandings and confusions have been happening concerning the current situation of the climate in our planet. Unfortunately, the absence of accessible and understandable information generates confusion in our society. Although this is an issue in which the science has been leading from the beginning of the time, I support that the improvement of the methods used in the science, are always helping us to make the information more accessible, hence to have the possibility to spread the knowledge with less effort. In this case, I think this research can contribute to improve how we transmit and understand the weather data paradigm. It is a good moral reason to me to perform this study. 

History shows how the proper use of technologies adapted to specific scenarios, promotes the advance of linear sciences as Maths or Physics, and always these new findings are supported by new technologies. To find these new technologies, it is needed to analyze from the engineering point of view, which things can be improved and how; this philosophy turns this thesis in an exercise to find how a science as meteorology can benefit from communications technologies around it if they are optimized for its needs.

\section{Outline of the thesis}

This thesis is structured as follows: the second chapter gives a general overview about how the weather data collection is structured, and which organizations are interacting on this activity. The third chapter explains briefly how a weather instrument works and what kind of technologies are involved in the process, after that,  it is analyzed how the meteorological networks composed by these instruments work. The fourth exposes the technical deficiencies found by the author on the weather data transmission. In chapter five the OpenWeather protocol is presented, a prototype protocol developed by the author, adapted to the needs exposed in the previous chapters. Chapter six specifies from the technical point of view how the protocol works, its operations and architecture, accompanied with justification of the technical decisions taken on the thesis, concerning its implementation. Chapter seven evaluates the implementation of the protocol in a real scenario based in a specific weather instrument. Finally, chapter eight summarizes the conclusions of this thesis.
\pagebreak

%% file: chapter2.tex
\chapter{The impact of the weather data}

Even if it is obvious for all of us, weather is one of the most important factors of the environment, with a high impact in our life. At the same time most of us are not familiar with the repercussions of the weather, what is causing different phenomena and the implications of them. Finally, our needs concerning the weather are limited by the availability of the data that is given to us. The role of the weather forecast broadcasting resides in different organizations. However, the advantages of the technology are bringing us the capability to have a more frequent and reliable access. The following sections analyzes how the weather data is spread and in which points of its diffusion can be improved.

\section{Weather data collection and diffusion}\label{2.1}

Depending of the region of the world, we can find more or less geographical locations in which a weather station has been placed to collect information about different phenomena. It is important to clarify they are several categories of phenomena with different needs in terms of data collection requirements. In addition, we have different units and time frequencies to make this data useful.

Fortunately, nowadays, most of the known phenomena have a solid basement of understanding, meaning this that we can measure them and get some conclusions and to act in consequence.  The \gls{SI} is used as the recognized standard of units for these measurements\footnote{Some countries as Burma, Liberia, the United States or the United Kingdom, have other local standards coexisting with the \gls{SI}. This implies some adaptions concerning the weather data. Due to the local units it is necessary to include unit conversions in the data manipulation process.}.

The figure \ref{f2.1} shows the scenario abstracting the data to a generic input:

\begin{figure}[H]
\centerline{\includegraphics[width=1\textwidth]{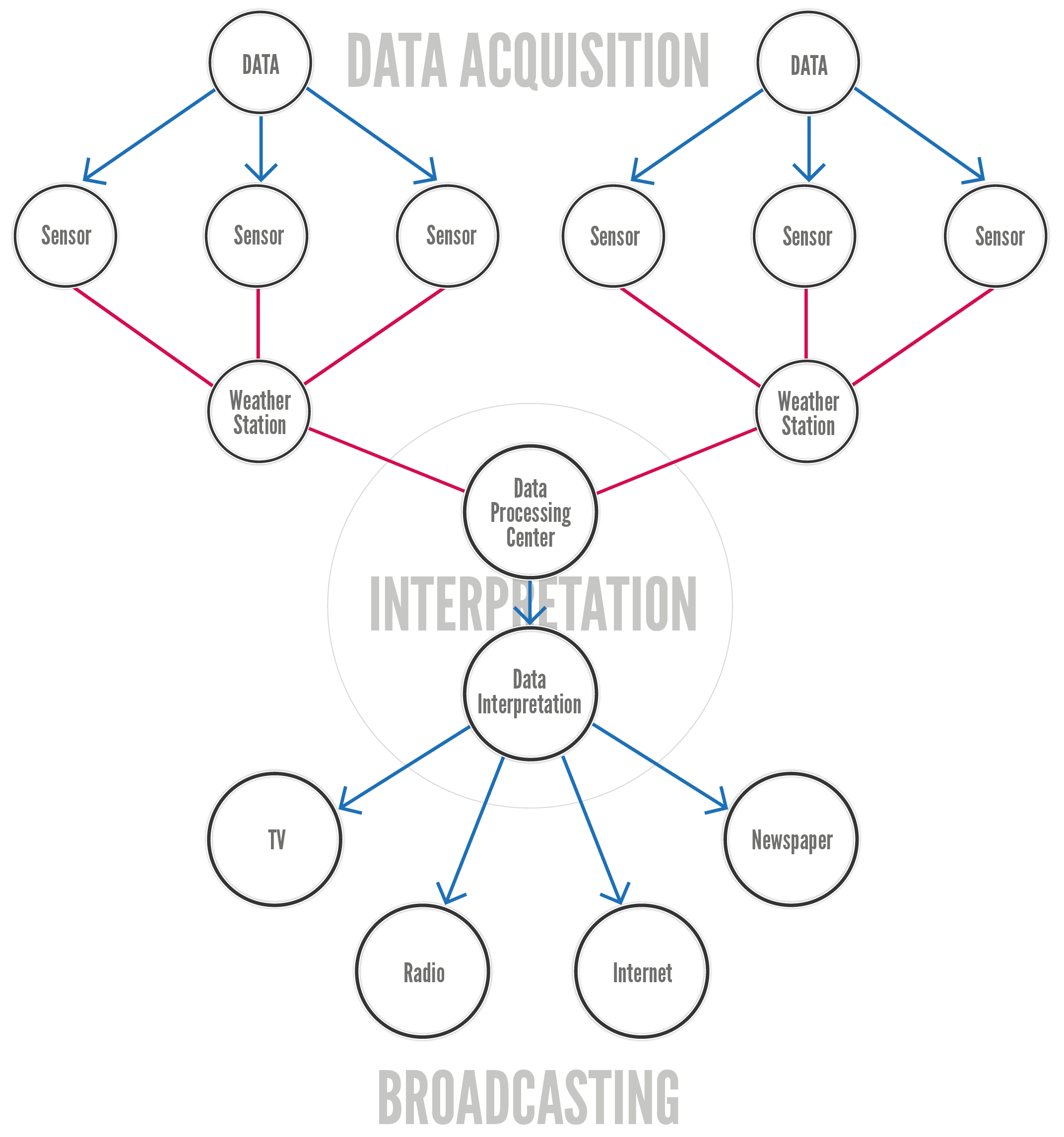}}
\caption{Layers abstracted in the weather collection data workflow.}
\label{f2.1}
\end{figure}

As we can see the scenario gives as an abstract input of data from the different environmental phenomena. After that, the data is sent to the data processing center (commonly a governmental \& scientist organizations). At the end, the data is interpreted and the conclusions are spread. The Physics are giving us the possibility to understand these phenomena based in the observation and correlation of them; for this it is needed to establish direct dependencies between the phenomena.

\begin{figure}[H]
\centerline{\includegraphics[width=1\textwidth]{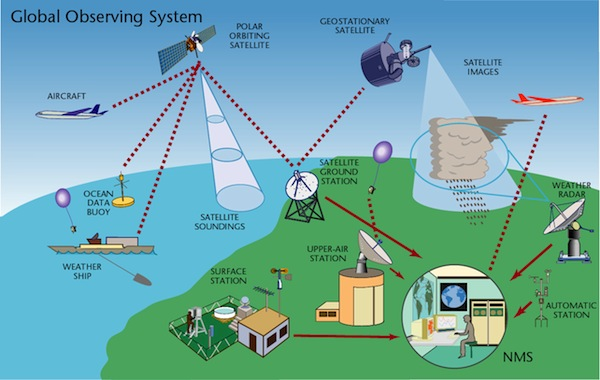}}
\caption{Weather data collection workflow. World Climate Data and Monitoring Programme.\protect\footnotemark}
\end{figure}

Commonly, we can find several governmental and scientist organizations around the world, focused in the weather data collection. As example of this,in Finland we have the Finnish Meteorological Institute(FMI) \cite{FMI}, or different example can be a worldwide organizations such as the \gls{WMO}\cite{WMO}, in charge of the coordination of the exchange and collection of weather data between organizations around the world. These organizations are the official source of information for weather data. Even so, they are not the only ones.

\footnotetext{The World Climate Data and Monitoring Programme (WCDMP) is a programme of the World Climate Programme that facilitates the effective collection and management of climate data and the monitoring of the global climate system, including the detection and assessment of climate variability and changes.\cite{WMO}}

Thousands of individuals are helping with the weather data collection as well. Those individuals in possession of some weather instruments can collaborate transmitting the data to some governmental organization, for instance the program \gls{CWOP} \cite{CWOP} has over \textbf{20,000 members in 149 countries}. This is possible using technologies like \gls{APRS} \cite{APRS} system, which is mentioned in CWOP website\cite{CWOP} as the following: 

\emph{"The Automatic Position Reporting System (APRS) is a part of ham radio that provides an ideal way for weather station operators to distribute their weather data much further than the regions within their transmitter range. APRS was originally intended for position information data but actually provides a means for automatic transmission of all sorts of digital data. This is especially true now that the original APRS packet radio concept has been enhanced to include the capabilities of the Internet. The reporting of citizen weather data is a particularly useful application of the APRS Internet Service (APRS-IS)."} 

\subsection{Governmental organizations}

Denominated as meteorological institutes or meteorological agencies, it is possible to find a big group of organizations around the world, which purpose is to study the weather. Almost all of these organizations are funded by the governments, moreover of these state and local organizations, other country-region organizations exist to coordinate the study of the weather in a bigger extension area. As an example, the \gls{FMI}\cite{FMI} is in charge of studying the weather in the region of Finland. At the same time the \gls{FMI} is member of the \gls{ECMWF}\cite{ECMWF}, organization in charge \emph{"to provide operational medium- and extended-range forecasts and a state-of-the-art super-computing facility for scientific research." }.The same scenario can be found in different continents as America with organizations as \gls{NOAA}\cite{NOAA}.

These worldwide organizations are creating the infrastructure to collect the weather data around the world. It is necessary to highlight that the study of the weather is an expensive activity, involving a big amount of resources such as high-tech instruments, installation of these instruments in different locations (with the extra cost that it implies) and use of computation centers to evaluate the data. Due to these facts, we can find that the amount of weather stations around the world and the effort or size of these organizations can vary significantly depending of the economy of the region. This means that the weather infrastructure in the occidental world is well designed, implemented and functional. However, in other areas like Africa, the amount of available weather stations decrease for economical reasons. In addition, and due to the nature of the weather, organizations like \gls{NOAA} and \gls{ECMWF} are installing weather collection points outside their official operation areas\footnote{Both organizations are restricted to America and Europe, nevertheless, these organizations have permission to place collection points out of their area to improve the quality of the studies and to encourage the international cooperation.}, thus getting better samples to evaluate the global weather conditions.

These state-region organizations have a huge cooperation between them. Scientists are pretty conscious about the need to get samples of weather data from different regions to evaluate it, thus, they are fomenting the cooperation of the weather data exchange. The \gls{WMO} defined the proceedings of measurement for meteorological variables\cite{GMIMO}, providing a common basement to perform the measurements related with the weather. Furthermore, the \gls{WMO} is conscious about the issue of data exchange, in chapter four the process of standardization that \gls{WMO} is supporting and the issues of it are analyzed deeply.

\subsection{Corporations}

As it was mentioned previously, weather has a big impact in our life. It implies that not only practical advantages can be extracted from the study of it, also the study of the weather is generating a big range of economical activities. 

Industries like construction or military, have even more interest in know which phenomena are occurring and the future predictions of them. This interest have fomented a whole parallel industry of services of weather data reports.

At the same time, some professional forecast services have appeared as an alternative for independent studies in particular regions of the world. Although this economical activity is mainly deployed by private corporations some governmental organizations are offering also private services.

\subsection{Individuals}\label{2.1.3}

The program \gls{CWOP} mentioned in the section \ref{2.1}, is a perfect example about how individuals can help to collect and to study the weather data. Furthermore, non official programmes have been appearing around the world; using the Internet as foundation, different communities of weather observers are contributing to create individual networks of data exchange, in which a user can access the data of different weather stations around the world. 
\begin{figure}[H]
\centerline{\includegraphics[width=0.7\textwidth]{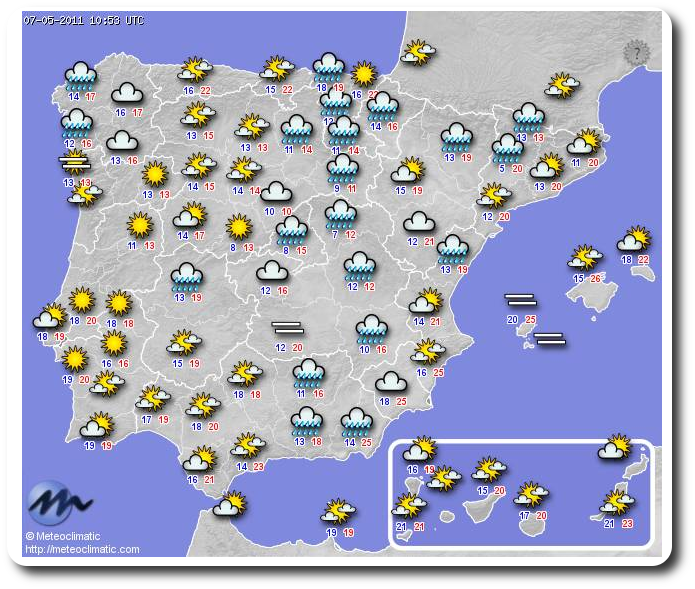}}
\caption{Meteoclimat screenshot showing weather forecasts.\protect\footnotemark}
\end{figure}
\footnotetext{This data is collected by individuals that have installed a specify software in their computers to send the data to Meteoclimatic servers.}

Meteoclimatic\cite{METEOCLIMATIC}, is a good example of this:" a big network of automatic non professional weather stations", in which hundreds of users share the data collected from their weather stations without any commercial purpose. 
Often, these communities share efforts with governmental organizations in programs as \gls{CWOP}, however, the turn up of theses communities are supported by the demand of the users to have a system in which their data is useful for other individuals, and at the same time give them some independency from governmental  organizations, in terms of data availability.

\subsection{Weather data publication}

The previous sections mention which organizations are involved on the process of data collection. However, the process does not end here; after the collection and evaluation of the data, the final step is to spread and make it useful. The implications of the broadcasting concerning the weather forecast are multiple and they are out of the scope of this thesis. Even so, the spreading of the data is limited for the protocols used in the acquisition of it. As mentioned in section \ref{2.1.3}, some communities of individuals appeared, taking the role of data availability disposal for the end user. Proving this the fact that the way in which the information is managed by the governmental and private organizations, sometimes does not fit with the end user's wishes.

In the past, the weather forecast was delivered through traditional methods as newspapers, radio and TV. Nevertheless, nowadays the Internet has taken this role in several aspects. Almost, all the governmental weather organizations mentioned in this chapter have a web site in which they publish -in different quantities and formats-, the information collected and extracted from their meteorological networks. Although traditional media still report the daily forecast, the tendency points to the Internet as the future mainstream channel of this information.

In addition, other commercial web sites offer this information partially free of charge. This practice caused the appearance of several sites offering \gls{API} services to fetch weather data, giving the possibility to the developers to get some storage data to perform some operations. Due to this \gls{API} availability, some organizations non related directly with the weather data collection workflow, have published some web sites that are exposing data fetched from different APIs and providing a different range of alternatives to the users.

\begin{figure}[H]
\centerline{\includegraphics[width=0.7\textwidth]{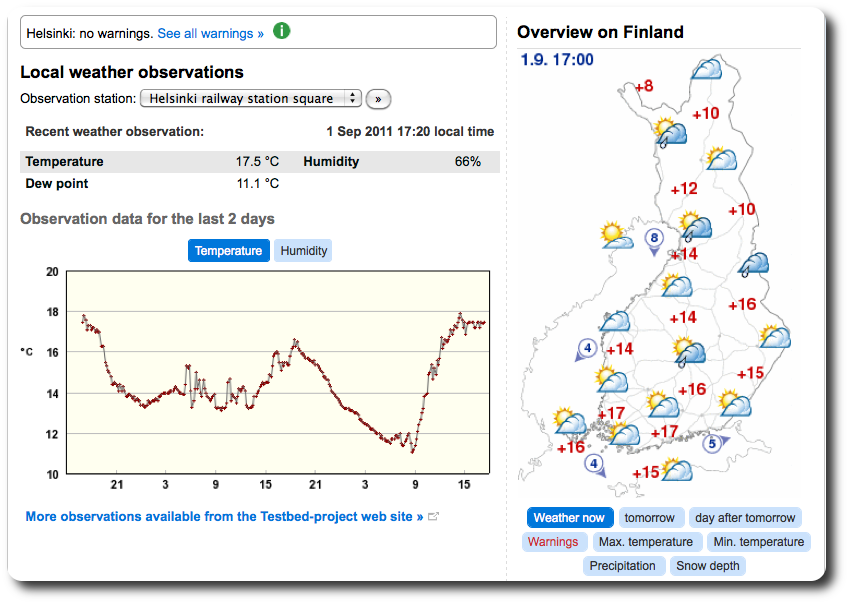}}
\caption{\protect \gls{FMI} website \protect \cite{FMI} spreading local weather observations.}
\end{figure}

The author could not find any \gls{API} offering the capability to connect directly to the weather instruments to fetch RAW data streams; all the APIs available are offering pre-processed data. 

\section{Summary}

In this chapter we have given general background information in order to make the scope of the thesis more familiar, in terms of which organizations are in charge of the weather collection and the structure and collaboration between them. Also, it has been analyzed how different organizations of the same field coexist.

We discussed how the same activity is performed in different layers, being involved in the process from official organizations to individuals. Some schemas have been presented, giving a global vision about how the weather data workflow works.

We know now that there is even a global organization named \gls{WMO}. This organization is only dictating some guidelines to perform the measurements. The next chapter introduces a general overview of a weather instrument, to understand how it works, its technologies and limitations.

In the next chapter some concepts and scenarios are explained to understand how a weather instrument works, the technologies that are conforming it, and giving us a global vision of the technologies to have in consideration when we are implementing a protocol for a weather instrument.

\pagebreak

%% file: chapter3.tex
\chapter{Infrastructure for the weather data}\label{3}

History is full of attempts to understand the weather. From the very beginning, humans have been focusing their attention in the weather, putting a lot of effort trying to understand and predict it. The first treatise concerning weather observations was \emph{Meteorologica}, written by Aristotle  (340 B.C.). Despite of this, \emph{"the birth of meteorology as a genuine natural science did not take place until the invention of weather instruments, such as the thermometer at the end of the sixteenth century, the barometer (for measuring air pressure) in 1643, and the hygrometer (for measuring humidity) in the late 1700s"}\cite{METO}. It was with the invention of the telegraph, in 1843, when the weather observations started to be useful owing to the capability to transmit the weather reports to different locations. Since this time elapsed, the industry has been developing and improving the weather instruments to achieve better measurements. Furthermore, the networks for weather data collection have been maturing. This chapter introduces the technology that is composing a modern weather instrument, its role in the weather's collection infrastructure and shows us some concepts to understand the conflicts of this setup exposed in chapter four.

\section{A meteorological instrument}

The purpose of a weather instrument is to measure a particular phenomenon under certain conditions, to collect some data that can be processed to obtain some conclusions (in terms of understanding and predictability). The success of the prediction and understanding comes supported by the accuracy that these instruments can provide. The industry has been creating new instruments based on new techniques discovered in Physics, to measure the phenomena; in addition, the advance of the digital technology, is providing to the  physicians a great scenario in which physical principles can be combined easily with digital technology, producing as result modern instruments with the ability to transform the result of these physical principles in digital data.

Despite their size and appearance, the weather instruments are complex artifacts. 
The materials used to build them are a combination between plastic and metal, this combination provides the necessary robustness to place the weather instruments at isolated places with all kind of degradation conditions. Furthermore, these instruments must have a low power consumption in order to fit the requirements of their locations. That forces the manufacturers to use more tiny and efficient technologies for measuring the phenomena without sacrificing energy and accuracy.

It is not possible to discuss all these instruments in this thesis. For this reason the following subsections of this chapter are focused on automatic weather stations(\gls{AWS}es). The \gls{WMO} defines an \gls{AWS} as: \emph{meteorological station at which observations are made and transmitted automatically}\cite{GMIMO}, at the same time this concept comes with other nuances as \gls{AWOS} and \gls {ASOS}: \emph{a combined system of instruments, interfaces and processing and transmission units is usually called an automated weather observing system \gls {AWOS} or automated surface observing system \gls {ASOS}. It has become common practice to refer to such a system as an \gls{AWS}}.

The focus on the \gls{AWS}es is supported by the popularity of these weather stations as main tools to measure the weather. \textbf{The author considers more useful to focus on this technology because a wide range of \gls{AWS}es is available for the end-non professional user}; meaning this that is possible to experiment with a new protocol using this scenario without affecting the current setups used for scientific purposes. In addition, later migration of the protocol to professional instruments should not be difficult because the manufacturers are using mostly the same technologies in the data transmission interfaces for both brands (professional and end-user).

\subsection{Industrial design}

Depending of the type of phenomenon to measure, the physical principle needed will require an instrument with certain sizes, materials and lifetime. It is rarely possible to measure multiple phenomena with the same instrument, this fact causes the creation of instruments focused only on one phenomenon \footnote{We refer here to high-tech and professional instruments for scientific purposes. It is possible to find several sensors giving an output for different phenomena in one instrument. However, this is not common in the instruments used for scientist observations; at the same time this configuration should be considered as a weather station not  as an "individual" weather instrument.} and even in only one specify and tiny part of it. 

The industrial design of an instrument is one of the keys for the success of the observations; the ability to put available the required technical conditions to perform the measurement through a digital interface, reside on it. To avoid conflicts in the study of the phenomenon, the materials should be chosen very carefully based on a complex equation between: robustness, durability, impact, impact assessment, etc. Furthermore, the shapes and sizes depend on the environment in which the instrument is going to be placed and the requirements needed for the physical principle used.

\begin{figure}[H]
\centerline{\includegraphics[width=0.5\textwidth]{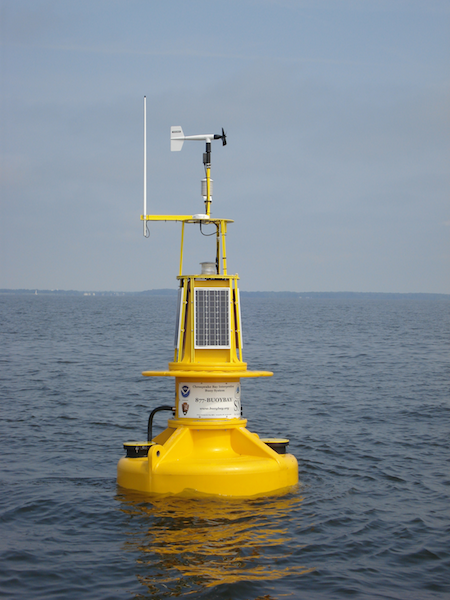}}
\caption{\protect \gls{NOAA} weather buoy \protect \cite{NOAA}, example of a complex an robustness \protect \gls{AWS}.}
\end{figure}

We can find in the market dozens of instruments for the same purpose, using in some cases the same principles to measure the phenomenon and even with some strong differences concerning the industrial design. Though, the instruments from different manufacturers have similar dimensions and they are build with similar materials, there is non available standard concerning all these characteristics, only some general guidelines are provided by the \gls{WMO}{\cite{WMO} suggesting dimensions and sizes for some instruments, an example of this recommendation is the following:

\emph{Wind-measuring systems can be designed in many different ways; [...]
The first system consists of an anemometer with a response length of 5 m, a pulse generator that generates pulses at a frequency proportional to the rotation rate of the anemometer (preferably several pulses per rotation), a counting device that counts the pulses at intervals of 0.25 s, and a microprocessor that computes averages and standard deviation over 10 min intervals on the basis of 0.25 s samples.}\cite{GMIMO}

\begin{figure}[H]
\centerline{\includegraphics[width=1\textwidth]{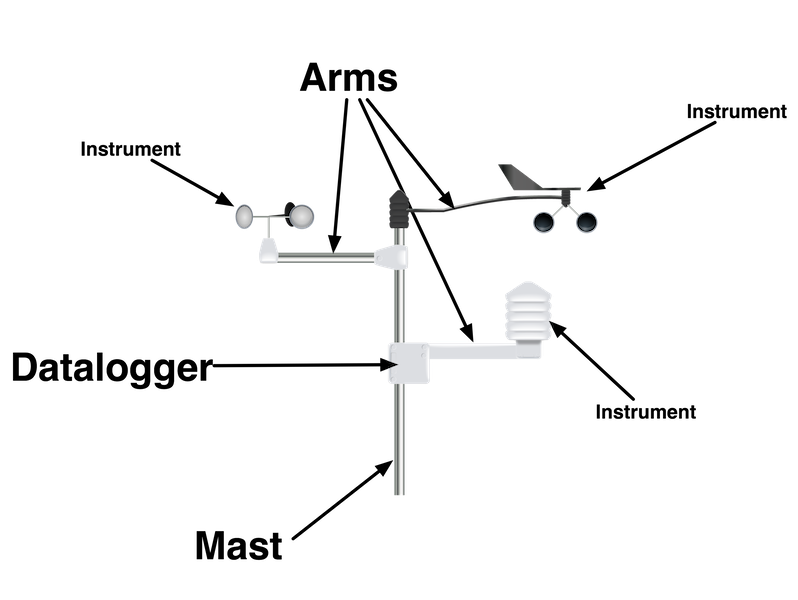}}
\caption{Generic \protect \gls{AWS} with different instruments and materials combination.}
\label{f3.2}
\end{figure}

The figure \ref{f3.2} shows a generic schema in which we can see different combinations of materials as plastic and metal, at the same time the instruments are placed in different heights due to technical requirements for the techniques used to perform the measurements. 

Most of the instruments available at the market are the result of the coordination between the requirements requested by the physicists and the possibilities that the technology developed by the industry. Notwithstanding, the instruments industry and their industrial design, is something really big and complex and it is out of the scope of the thesis. Furthermore, we need to be conscious about the industrial design of the instruments, because it is strong-linked to the electronics that they can house, conditioning this the digital interfaces for data transmission that we can install in them.

\subsection{Electronics and data handling}\label{3.1.2}

The electronics of a weather instrument are barely different irrespective of the phenomenon to measure. The industry is producing a wide range of instruments with a complete different set of sensors. Nevertheless, as embedded systems, all these instruments have a common need to conform these type of systems. The \gls{WMO} gives again some general guidelines with respect to electronics and weather instruments.  The following paragraphs summarize them.

 \subsubsection{CPU}\label{cpu}
 
As other electronic device in charge of process data, an \gls{AWS} has a \gls{CPU} running at clock frequency of a few \gls{MHz}. This CPU is microprocessor based with 8-bit wide.\footnote{Nowadays some manufacturers are introducing progressively new microprocessors using 32-bit wide.} Despite the low bit wide of these microprocessors, an \gls{AWS} does not needed more calculation power because the amount of data generated by the sensors will be rarely up of 1 \gls{kB}, meaning this that frequencies oscillating between 8-33 \gls{MHz} will fit perfectly in the requirements to process the data.

\subsubsection{Volatile Memory}

Often 32-64 \gls{kB} is the maximum amount of volatile memory available on an \gls{AWS}, it makes the instrument non capable to keep too much data on a \gls{RAM} at all. Forcing to the manufacturers to design the instruments with fast and reliable mass storages, ready to transfer the data from the volatile memory to the persistent storage.

\begin{figure}[htb]
\centerline{\includegraphics[width=0.7\textwidth]{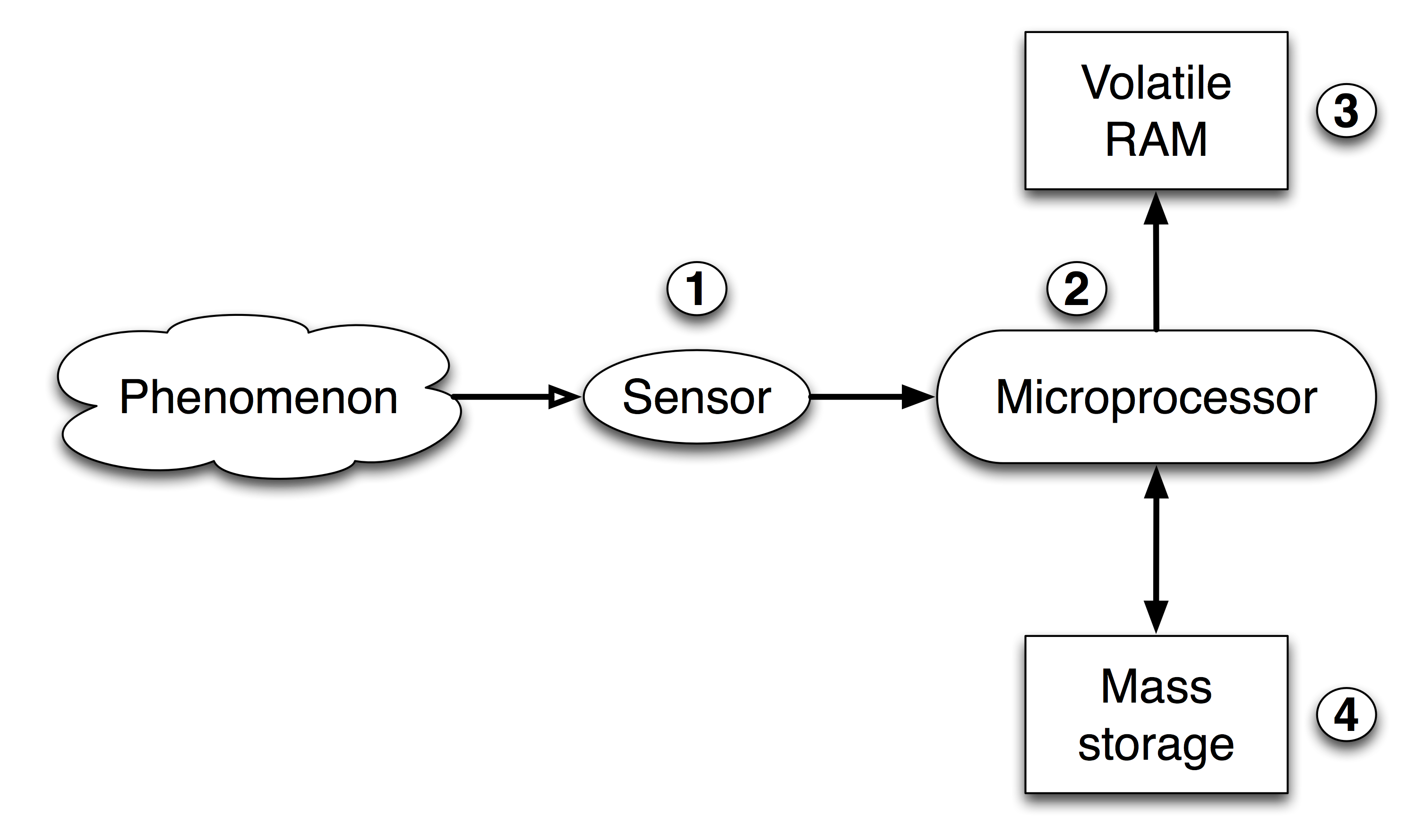}}
\caption{Abstracted electronic schema of an \protect \gls{AWS} reading data from one sensor.}
\label{f3.3}
\end{figure}

The figure \ref{3.3} shows the workflow data of an abstract sensor. In the first step the sensor generates the data from the phenomenon, based on the observation of some physical principle; the data acquired is processed by the microprocessor in the the second step, placing the data on the volatile memory. When the data is placed on \gls{RAM} the \gls{IO} operations start, transferring the data from the volatile memory to the mass storage (persistent memory). According to the Guide to Meteorological Instruments and Methods of Observation \cite{GMIMO} published by the \gls{WMO}, it is highly recommended to equip the \gls{AWS} with a battery backup dedicated to the volatile memory to avoid data loss due to some power fails. This non common feature in generic computers can be an advantage to have in mind when a protocol is implemented, because it enables the possibility to have some methodology in the protocol to recover the session after one power failure.

\subsubsection{Mass storage}

Typically, an \gls {AWS}, will have mass storage device to save the data collected from the sensors. The storage of data in the \gls{AWS} has been changing in the last years due to the continuously decreasing price of flash memories. It is common to find very different architectures in terms of data storage in the \gls{AWS}.

\begin{figure}[htb]
\centerline{\includegraphics[width=0.7\textwidth]{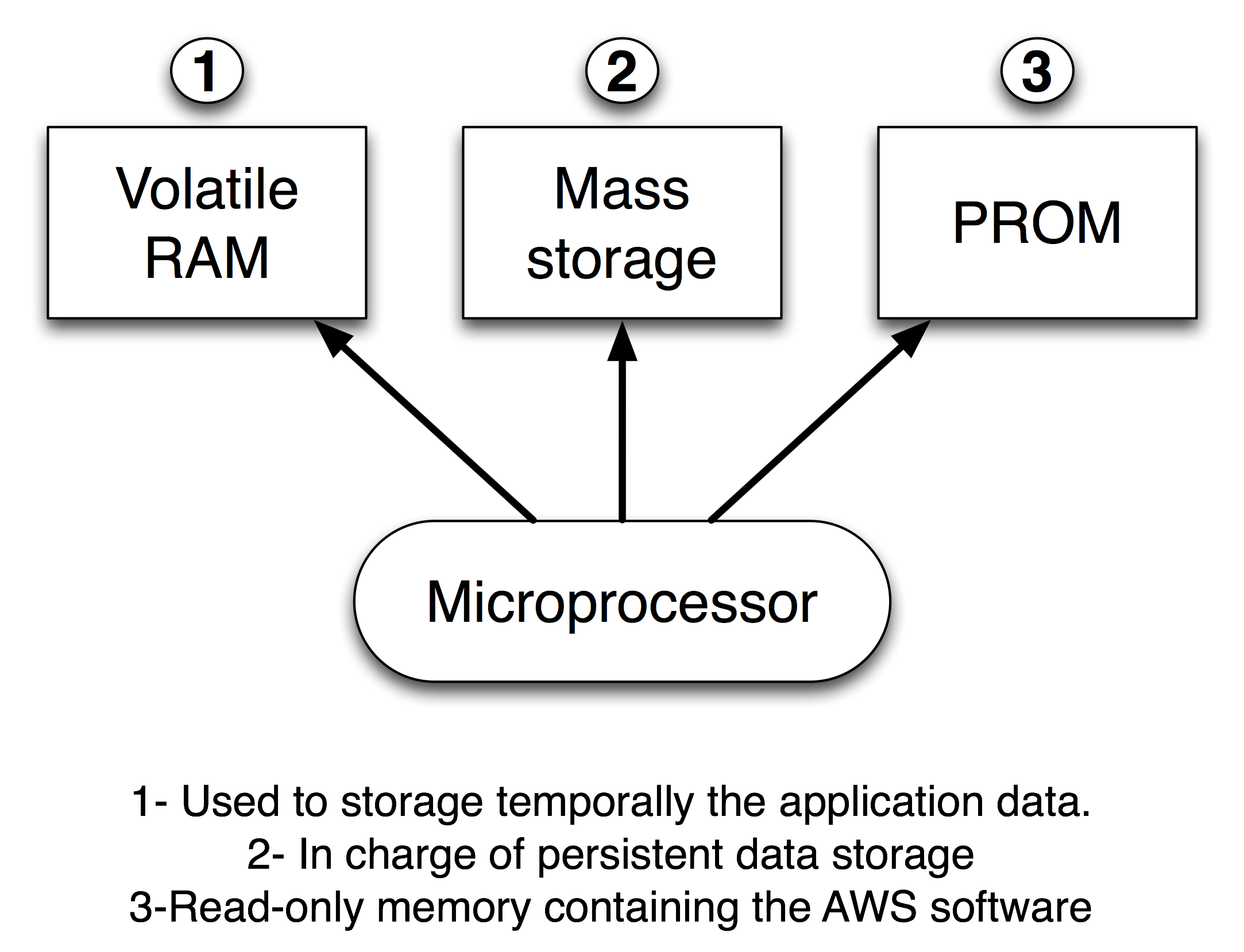}}
\caption{Types of storages available in an \protect \gls{AWS}.}
\end{figure}

The number of sensors and the frequency in which the information is transferred to the data centers, determines the size of available memory in an \gls{AWS}. Based on the market, the mainstream option in terms of memory size for mass storage is around 1 \gls{MB}, that space is more than enough to save thousands of samples in case that the \gls{AWS} has not send the data to the collecting point.

\subsubsection{Sensors}

The sensors are the digital interfaces that make an \gls{AWS} different from other embedded devices. As explained in section \ref{1.1}, a sensor is a digital interface using some physical principle to measure a particular phenomenon. Their principles, implementation and complexity are out of the scope of this thesis. Even so, we need to consider the sampling frequency of them because they are involved in the frequency in which the data is produced.

The sampling frequency of the sensor depends on the data required to understand the phenomenon. A big range of sampling frequencies are used to measure different phenomena. Nevertheless, the author is not assuming this frequencies as a need for the protocol. 

A correct behavior of the sensors requires a high-accurate calibration of them. The manufacturers have been developing several methodologies and mechanisms to calibrate the instruments and verify their correct behavior. These calibrations are not considered as part of the problem statement of this thesis because they are unrelated to the methods of the data transmission.

\subsubsection{Digital interfaces}

As mentioned in the section \ref{1.2}, an \gls{AWS} is equipped with at least one peripheral device to provide data interaction. These interfaces offer the possibility to configure the \gls{AWS} and transfer data from it. The type of device is a serial communication physical interface, and depending on the type and vendor of the instrument, it will be one the following\footnote{Other types of interfaces can be found in the instruments. However, the industry stablished —with non-written agreement— the use of the mentioned interfaces as mainstream.}:

\begin{itemize}
\item \gls{RS232}
\item \gls{RS422}
\item \gls{RS485}
\item \gls{USB}
\end{itemize}

These four types are well-known in the industry. They are available in almost all the modern computers, however the relation of them with this thesis is focus mainly in the bandwidth that they offer. 
The table \ref{Table3.1} shows a comparison between these physical digital interfaces and their bandwidth.
\begin{table}[h]
\centering
    \begin{tabular}{| l | l | r | r |}
    \hline
    \textbf{Standard} & \textbf{Bandwidth} & \textbf{Bytes/s}  & \textbf{kB} \\ \hline
    TIA/EIA-232-F\cite{RS232S} & 116 \gls{KBITS}/s & 14848 & 14.5 \gls{kB}\\ \hline
    TIA/EIA-422-F\cite{RS422S} & 200 \gls{KBITS}/s & 25600 & 25 \gls{kB}\\ \hline
    TIA/EIA-485-F\cite{RS485S} & 35  \gls{MBITS}/s & 4587520 & 4.375 \gls{MB}\\ \hline
    USB\cite{USBS}\protect \footnote{Referencing the USB in low power mode (specification 1.0)} & 1.5  \gls{MBITS}/s & 196680 & 192 \gls{kB}\\ \hline
\end{tabular}
  \caption{Comparison between standards and bandwidth offered.}
  \label{Table3.1}
\end{table}

Even so, as the table \ref{Table3.1} shows, the minimum bandwidth provided by theses interfaces (\gls{RS232}) should be enough. As described in the sensors section, the total amount of data generated by the sensors of one \gls{AWS} rarely exceeds 1\gls{kB}; fitting perfectly this in the bandwidth offered by the \gls{RS232}.

Due to the constant renovation in digital interfaces that the industry does, we do not consider other old interfaces in the analysis, assuming that the protocol will work with instruments manufactured in the last 10 years\footnote{Those should be equipped with the interfaces mentioned in the Table 3.1.}.

Although the interfaces are not conditioning our protocol implementation, it is necessary to highlight that most of the vendors offer the possibility to re-wire the \gls{AWS} to make it work with different physical interfaces.

\begin{figure}[H]
\centerline{\includegraphics[width=1\textwidth]{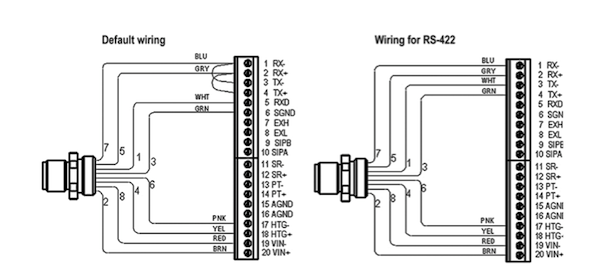}}
\caption{Wiring schema showing how to re-wire the \protect \gls{AWS} to use \protect \gls{RS422}.}
\end{figure}

\subsubsection{Datalogger}\label{dataloggersection}

The datalogger is one the most critical parts of an \gls{AWS}. It is in charge of the data logging produced by the sensors and deliver by the operating system. Its main task is to keep track of the data collected by the \gls{AWS}. This component plays an important role in the implementation of the protocol, because of the data of the protocol must be originated in this part.

Depending on the architecture of the \gls{AWS}, the datalogger can be an external embedded system with serial communication capabilities, able to send data through a network and with multi-station capability\footnote{Some dataloggers are able to track and to operate several \gls{AWS} at the same time.}. Small \gls{AWS}es can have datalogging capabilities, keeping the data in a persistent memory for a short period. To have the datalogger implemented internally implies increasing the complexity of the \gls{AWS}, converting it in a more complex embedded system with features as data delivery through a network, long-term data storage, etc.

Often, the architecture chosen for \gls{AWS}es is an external device connected through the physical interface. These devices are equipped with some kind of connectivity such as \gls{GSM}, \gls{GPRS} or \gls{UMTS} modems, using them to deliver the data to the collection point. 

\begin{figure}[H]
\centerline{\includegraphics[width=0.5\textwidth]{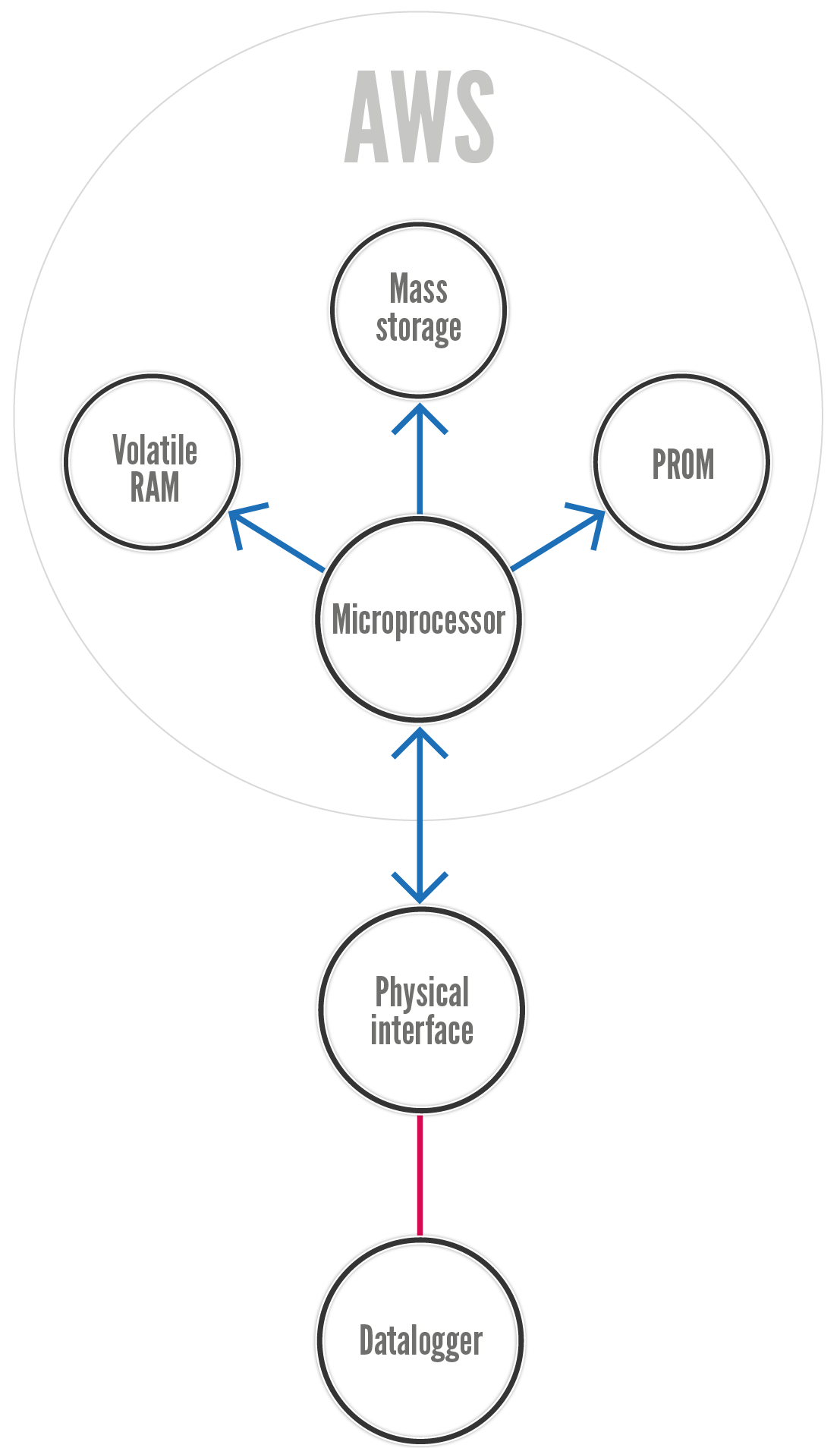}}
\caption{Location of the datalogger in an \protect \gls{AWS}.}
\end{figure}

\subsection{Software}

As it is common in the embedded systems, an \gls{AWS} has a tiny internal software. The programming languages used to develop this software have no relevance in this topic. We assume that the internal operating system of the \gls{AWS} will offer us the data collected from the sensors, moreover of some set of options to configure and calibrate the \gls{AWS}.  

We need to differentiate between the software embedded in the \gls{AWS} and the software at the end of the peripheral device. 

\subsubsection{AWS's Operating System}

The operating system installed in an \gls{AWS} resides in a \gls{PROM}. Its architecture is based in a real-time clock implemented on the mother board of the \gls{AWS}. The OS provides a limited set of options to interact with the \gls{AWS}, most of these options are focused in data acquisition, calibration and hardware configuration. This software is in charge of the formatted data of the \gls{AWS}, in other words, it gets the data from the sensors, applies the necessary formulas to extract a meaningful result and formats the data in one of the following serial communication protocols\footnote{We need to distinguish between the data format used to communicate with the interface (\gls{ASCII}, NMEA-0183, etc) and the format in which the data is formatted, this is explained the section \ref{4.2}.}:

\begin{itemize}
\item	RAW \gls{ASCII}
\item \gls{SDI-12}
\item \gls{NMEA-0183}
\end{itemize}

After the data is formatted, it is transmitted through the peripheral device to the the datalogger.

\subsubsection{External software used for datalogging / data distribution}

As explained in \ref{dataloggersection}, an \gls{AWS} needs a datalogger device to track the data collected from the sensors. Irrespective of the type of datalogger, at the end of it, we will find some computer in charge of the data manipulation and storage. The software installed on the computers can be really differently implemented and designed depending of the vendor, but its main task is to understand the data format chosen by the vendor to transmit information and take use of it.

The market offer concerning software for \gls{AWS}es is too big, even some companies not related with the manufacturing of the instruments, are releasing software for datalogging purposes. It is common that the \gls{AWS} is provided from the factory with its own set of software, nevertheless due to the serial communications protocols used by the \gls{AWS}, is simple to implement a software that interprets and takes advantage of the data format chosen by the vendor to implement new capabilities.

\begin{figure}[H]
\centerline{\includegraphics[width=1\textwidth]{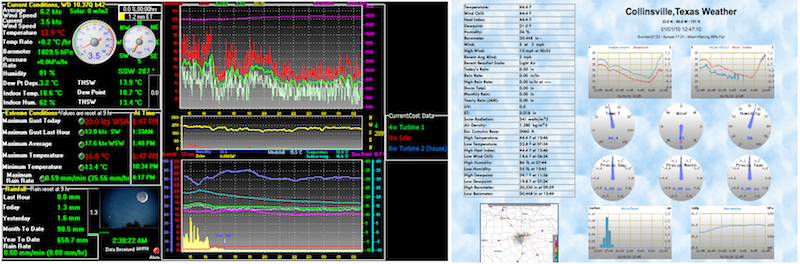}}
\caption{Screenshots of some popular desktop applications for \protect \gls{AWS}.}
\end{figure}

\subsection{Networking}\label{3.1.4}

As mentioned in the datalogger subsection, the connectivity capabilities in an \gls{AWS} resides on it. The industry offers multiple options to provide connectivity in an \gls{AWS}, nevertheless, most of these options are limited for bandwidth, energy and geographical limitations. It is possible to find \gls{AWS}es directly connected to a computer via \gls{USB}, providing this the connectivity, or we can find an isolated \gls{AWS} in the middle of a mountain connected through a radio-link to the closest place. 

The common technologies to provide connectivity to an \gls{AWS} are:

\begin{itemize}
\item \gls{GSM}
\item \gls{GPRS}
\item \gls{UMTS}
\end{itemize}

In places with better geographical location and energy availability, it is possible to find the following technologies offering connectivity:

\begin{itemize}
\item Ethernet
\item \gls{USB}
\item 802.11b/g
\end{itemize}

Whatever the connectivity on the \gls{AWS} is, the common pattern is that this connectivity is reliable but offers a rather low bandwidth. 

\section{Meteorological data networks}

The previous section gave a general overview of \gls{AWS}es, the relation between them and this thesis, is how they behave in terms of networking communication, which kind of topologies are used and in which points this communication can be improved.

To understand the workflow of the data in terms of weather data collection, we should see an \gls{AWS} as an individual node without interaction with other nodes, except the collection point.

The collection point is the place in which different data from different \gls{AWS} is received. It is not mandatory that this collection point is the end of the weather data workflow, for instance it is possible to find an intermediate collection point that has been stablished for geographical reasons to improve the connectivity\footnote{Some \gls{AWS} are located at inaccessible places, sometimes this implies to establish a collection point close to them to avoid issues such as lack of connectivity (\gls{GSM}, \gls{GPRS}, \gls{UMTS}).}. Even so, we consider the collection point, the place in which the data has been received and it is ready to be processed.

\begin{figure}[H]
\centerline{\includegraphics[width=1\textwidth]{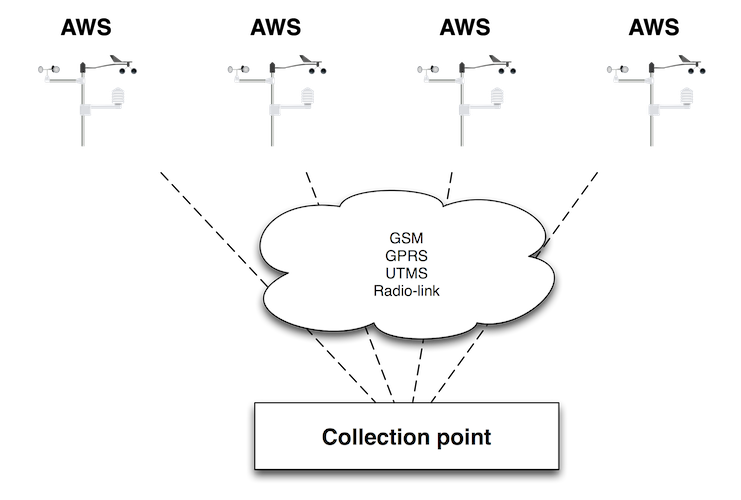}}
\caption{\protect \gls{AWS}es connectivity schema.}
\end{figure}

When the data arrives at the collection point, different mechanisms get activated to process it. As described in section \ref{1.2}, rarely, the data received comes from the same brand of instruments, meaning this that the data will be received in different formats and different time frequencies; this fact forces to implement these mechanisms to homogenize the data and make it understandable on the collection point. The collection point is the hop in which to have a standard protocol to communicate with the \gls{AWS} will have a bigger benefit, because it is in this hop in which the most effort is made, it in terms of data parsing, power calculation and data homogenization.

\subsection{Common architectures}

The definition of star topology fits in the methodology used to collect data from different \gls{AWS}es. The nodes have a strong dependency with the collection point, without it, an \gls{AWS} will have a high limited time to save data before it is fetched manually. Furthermore, the meteorological networks are not following the pure definition of star topology because different nodes are transmitting data with different connectivity technologies. Nevertheless, seems the nodes are not interacting between them, the network is not affected by bandwidth limitations. This topology is chosen by weather organizations based in the geographical limitations. However, the possibility to interconnect \gls{AWS}es between them has not been study deeply. The assumption for this is that the utility of the data is based on the availability of it, for this reason the data delivered with big delays is not considered at all in the weather data collection workflow. Interconnect the nodes of the meteorological networks it not feasible with the current technology at all for different factors such as bandwidth, geographical locations or absence of a common protocol.

\begin{figure}[h!]
\centerline{\includegraphics[width=1\textwidth]{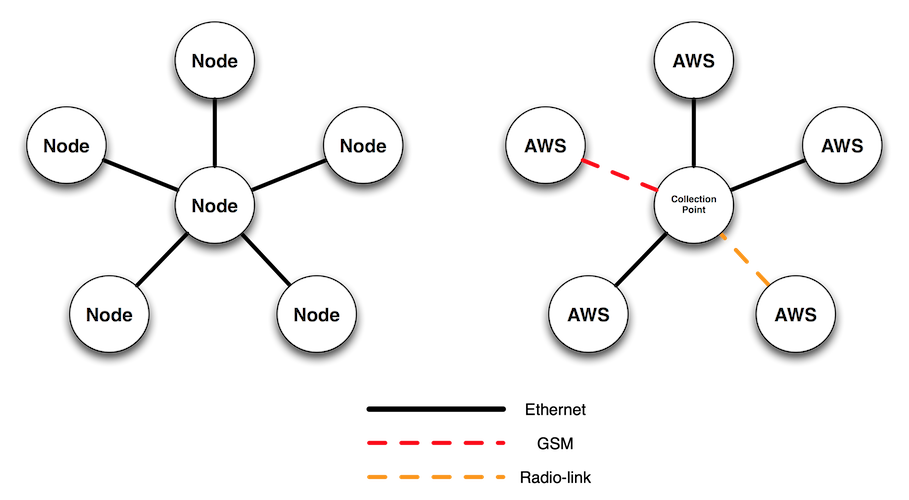}}
\caption{Comparison between pure star-topology against star-topology and the connectivity technologies used in \protect \gls{AWS}es.}
\end{figure}

Not only star-topology is used in the meteorological networks, the combination of different instruments can end in different topologies depending of the datalogger configuration. For instance, it is possible to have some local network of sensors connected to a datalogger that is part of a star topology, commonly, this topology will be a combination between bus-topology and star-topology. These combinations will not affect a common protocol in anyway, due to its implementation should happen on the datalogger's side, not mattering the combination of topologies behind it.

\subsubsection{APRS}

\gls{APRS} is using unnumbered \gls{AX.25} frames\cite{AX25}. \gls{AX.25} is a data link layer protocol without  too many capabilities in terms of bandwidth's offer, error correction and data integrity. Though it is used in some weather stations to spread the data, \textbf{it is not a good choice because it is not warranting a constant visibility and connection of the node.} 

The \gls{AWS} using the APRS technology are spreading the data based radio technologies. It is allowing to any node with a radio equipment to receive the information produced in the weather station. Furthermore this topology does not offer any warranty in data delivery because it does not use the collection point model.

\begin{figure}[H]
\centerline{\includegraphics[width=0.6\textwidth]{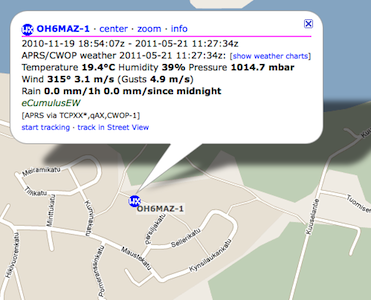}}
\caption{Example of an \protect \gls{AWS} using APRS at Helsinki area\protect \footnotemark.}
\end{figure}

\footnotetext{Source: http://aprs.fi}

\gls{APRS} has gained popularity inside the radio amateur community and programs as \gls{CWOP} due to the simplicity and technical requirements that it implies. The \emph{Weather Station Siting, Performance, and Data Quality Guide}\cite{CWOPGUIDE} explains how to setup an \gls{AWS} to get integrated in the \gls{CWOP} using \gls{APRS}.

\begin{figure}[H]
\centerline{\includegraphics[width=1\textwidth]{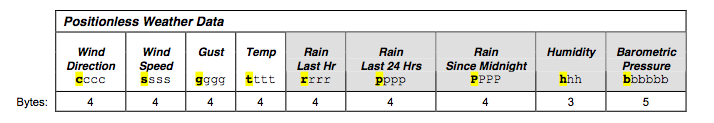}}
\caption{Weather data message using \protect \gls{APRS} \protect \cite{APRS}.}
\end{figure}

Nevertheless, \gls{APRS} is not used in scientific installations. Although it is not possible to re-implement \gls{APRS} to adapt it to OpenWeather, it will be possible to use the same data format as it used in OpenWeather under \gls{AX.25}. Thus, it will offer compatibility between applications using OpenWeather. To provide this capability, will involve modifying the way in which \gls{APRS} is used, one way to do it can be to send the same data beacon with different formats: standard \gls{APRS} messages for weather reports and after it a data message based in OpenWeather format.

Even with these incompatibilities the data provided by the \gls{APRS} data message can be transformed to OpenWeather's data format in a middle point having to modify the \gls{APRS} protocol. 

The author assumes that the \gls{AWS}es will behave as nodes with connectivity to a common point, being able to interact between them, through the collection point or point to point.

\subsection{Data distribution}\label{3.2.2}

Data distribution is the ultimate's reason for weather data collection. We can identify at least fours levels of different data in the process for weather data collection.

\begin{itemize}
\item \textbf{RAW data}, produced in the sensors'  instruments
\item \textbf{Network data}, used in the transmission from the instruments to the collection point
\item \textbf{Operational data}, result of the scientific's practices
\item \textbf{Informational data}, mainly focus in the general public (forecasts, climate reports, etc)
\end{itemize}

After the data is collected and processed, the conclusions made by the scientists must be spread to inform the society. It is necessary to highlight that only a few conclusions get to the general public, some of them are known as forecast or climate reports. Most of the data processed  is not useful for non scientists, because the complexity or amount of information on it. At the end of the work flow we have the data in two categories, the data that will be minimized to make it understandable to a general public, known as \textbf{informational data} \footnote{An example of this is the weather forecast shown every day in newspapers, TV, radio, etc.} and the data that must be shared between different international and local governmental organizations, known as \textbf{operational data}.

As part of the problem statement, the data distribution is one of the big efforts that these organizations need to do to make the data that they collect understandable. In 2002 the \gls{WMO} started a standardization process to create a metadata standard to fix part of this problem, however nowadays this standardization process is still on progress without any draft available\cite{WMOMETADA}. A standard protocol to communicate with the \gls{AWS} will help the development of a common data format between organizations because all of them will be fetching the data with the same methods and mechanisms.

\section{Summary}

We have now introduced the elements and process involved in measuring and collecting weather data and the technologies related with them. Some topics have been explained to provide a general understanding of how an \gls{AWS} works.

We have highlighted the limitations the \gls{AWS}es, concerning data storage and \gls{CPU} calculation; at the same time the maximum bandwidth available for the digital interfaces has been analyzed. The role of the datalogger has been exposed and its implications of it in the implementation of OpenWeather's format.

In addition, the connectivity technologies available in an \gls{AWS} have been enumerated, analyzing the bandwidth offered and concluding that only the interruption of the connection and not the bandwidth's offer can be an issue.

Finally, the topologies used in the meteorological networks haven analyzed briefly, clarifying that the \gls{AWS} are behaving as nodes without interaction between them, only sending data to a common point named "collection point" (the node that interacts with all the \gls{AWS}es). The \gls{APRS} protocol and its topology have been explained, taking in consideration the possibility to be compatible with the implementation of OpenWeather.

The next chapter describes the technical issues related with the data transmission in the \gls{AWS}es.

\pagebreak

%% file: chapter4.tex
\chapter{State of the art in the weather data transmission}\label{chapter4}

The previous chapters we have introduced a general overview of the basics needed to understand how weather data are collected and how a weather instrument is designed to undertake its function. Even though the purpose of this thesis is to analyze the issues found in the weather data transmission and to provide an alternative to fix these problems. Nowadays, the way in which a weather instrument is transmitting the data can be classified as generic, because the methodologies used in this task have not been optimized thinking in the data implied in the process. This practice limits the possibility to acquire data without the implementation of intermediary hops in which the data is parsed and converted to a useful data format. This results in an unnecessary investment  of CPU cycles, delays in the data delivery, incompatibility between difference brand of instruments, and at the end causing the investment of more resources and effort to exchange data between organizations. This chapter analyzes the technical points that are causing this issues in the weather data transmission.

\section{The evolution of the digital interfaces in a  \\ weather instrument}

As mentioned in chapter three, the meteorology did not advance until the invention of the telegraph. The value of the weather data resides in the ability to combine it with other sources to get some conclusions to make predictions. Nevertheless, this combination involves having the possibility to transmit this data fast and far enough. The telegraph brought this possibility, and with this new chance scientists had the opportunity to understand concepts as wind flow and storm movement\cite{METO} among others. During the 19th and 20th century the industry has been developing new improvements in the instruments manufactured; all of these improvements come supported for the new methods found by the physics to measure the phenomena, and the conversion of them to digital instruments. 

In 1969 the \gls{RS232}-C standard was published; this interface has been the mainstream technology used in the weather instruments for more than thirty years; only in the last decade some updates have been introduced in the industry, migrating to new standards as ANSI/EIA/TIA-232-F\cite{RS232S}, ANSI/EIA/TIA-422-F\cite{RS422S}, ANSI/EIA/TIA-485-F\cite{RS485S} or \gls{USB}. 

As far as we can judge this slow transition in as of the digital interfaces used in a weather instrument come supported for the fact of the wide use of \gls{RS232}-C in different fields of the industry, at the same time these interfaces fit perfectly in the needs of the weather data transmission: enough bandwidth, low cost and they are an international standard. If some updates have been introduced in the industry of the weather's instruments, they come supported by the need to adapt these interfaces to the hardware ports available at the moderns computers, seldom by the requirement of more bandwidth\footnote{In some big \gls{AWS}es in which have been placed many sensors and complex instruments, exists the possibility to need a bigger bandwidth, even so this is a specific case out of the mainstream setups.}.

It is an observable fact that the industry performs some updates in the technology to make it compatible with the moderns computers despite that the is not needed in terms of data  delivery. Moreover, the new standards are offering more capabilities a part of more bandwidth, for example, technologies as \gls{USB}, bring the opportunity to plug an \gls{AWS} to a computer and have it working without previous configurations as bit-rate, parity, etc.\footnote{Interfaces based in ANSI/EIA/TIA-232-F, ANSI/EIA/TIA-422-F, ANSI/EIA/TIA-485-F require to adapt the software to certain bit-rates, flow controls and other parameters.}

These interfaces provided by the industry are generic as in other technologies, not mattering the type of data transmitted through them; a well-known process of standardization has been performed to develop these interfaces. Though does not exist any standard specifying which type of interface should provide an \gls{AWS}, the \gls{WMO} recognizes the universality of the interfaces mentioned, and establishes them as requirement for the \gls{AWS}es performing official measurements for governmental organizations\cite{GMIMO}. Based on this we assume that a protocol implemented in an \gls{AWS} must work under these technologies; because these interfaces are generic, they have not any requirement for the data transmitted, giving complete freedom to us to implement any protocol over them.

As mentioned in section \ref{3.1.2}, the bandwidth offered for the different interfaces available in a weather instrument, are offering even more bandwidth than the amount of data that an \gls{AWS}'s CPU can process. Hence, a weather instrument has not limitations (concerning bandwidth) in the data interfaces that would prevent the possibility to implement a protocol to afford the needs of the data delivery.

Based on this retrospective we assume that the digital interfaces provided by the industry are well know and tested standards, providing mechanisms to achieve the goal of the data transmission. However, as it is explained in section \ref{4.2} no weather data transmission protocol has been defined for them. We identified this as \textbf{the first deficiency in the weather data transmission because of the potential offered by these digital interfaces is not used in the weather instruments}. 

\section{The absence of a protocol}\label{4.2}

The goal of the \gls{IETF}\cite{IETF} is to make the Internet work better. One of its multiple task implies to take care about the standardization process of the new Internet standards. A protocol is considered as standard when the IETF publishes a memorandum\footnote{This memorandums are named as \gls{RFC} for historical reasons.}, specifying all the aspects of the protocol and assigning a number in the STD series of it\cite{rfc2026}.

A research performed by the author in the \gls{RFC}s available at \gls{IETF}'s website \cite{IETF}\footnote{The searched has been performed over all the content of the RFC published: ftp://ftp.rfc-editor.org/in-notes/tar/RFC-all.tar.gz . Retrieved: 28-03-2011.}, looking for the following terms: "weather", "meteorology", "weather station", "atmosphere", "weather data", gave as result the following number of mentions. Only \textbf{9} \gls{RFC}s do direct or indirect mention to the weather data. 

The first \gls{RFC} mentioning a protocol related with the weather data is the RFC 765 \cite{rfc765} File Transfer Protocol (FTP): 

\emph{3.4.2.  BLOCK MODE
         The file is transmitted as a series of data blocks preceded by
         one or more header bytes.  The header bytes contain a count
         field, and descriptor code.  The count field indicates the
         total length of the data block in bytes, thus marking the
         beginning of the next data block (there are no filler bits).
         The descriptor code defines:  last block in the file (EOF) last
         block in the record (EOR), restart marker (see the Section on
         Error Recovery and Restart) or suspect data (i.e., the data
         being transferred is suspected of errors and is not reliable).
         This last code is NOT intended for error control within FTP.
         It is motivated by the desire of sites exchanging certain types
         of data (e.g., seismic or \textbf{weather data}) to send and receive all
         the data despite local errors (such as "magnetic tape read
         errors"), but to indicate in the transmission that certain
         portions are suspect).  Record structures are allowed in this
         mode, and any representation type may be used.}

Nevertheless, this reference of weather data is just an example (as the other references) that disappeared in later updates of the \gls{FTP}. 

The industry has focused its effort in improving the measure methodologies, the robustness of the instruments or other features as power consumption or life-time. Thus, the methodologies utilized to transmit weather data have been developed independently by the vendors, choosing their own data formats and techniques.

Nevertheless, the \gls{WMO} initialized different programs as \gls{GOS}, \gls{GTS} , \gls{GDPFS} \cite{WMO} among others, in which the weather data exchange is a key-component of the systems to archive the goals of these programs. In addition, as mentioned in the section \ref{3.2.2} the \gls{WMO} started a process of standardization 9 years ago.

Even assuming that the industry focused its attention on prioritizing measurements methods and product quality, the technologies related to the weather data transmission are outdated. The proof of this is that only a few governmental organizations have access to real-time information \footnote{All of these instruments are generating by default real-time data.} collected from the \gls{AWS}es\footnote{Note that these organizations can have this capability due to they invest a big effort in to develop custom systems for their weather instrument's setup.}, at the same time programs as \gls{CWOP} still depend of technologies such as \gls{FTP} or \gls{APRS}, that they do not contemplate scenarios in which scalability, data on demand or real-time data is needed.  Finally, as a real example, the SMEAR project\cite{SMEAR} is experimenting the issues of not having a standard protocol for the \gls{AWS}, producing as result the implementation of intermediary points to parse and normalize the data, incompatibility between different sources of data from the same phenomenon collected with different instruments and scalability of the system among others.

Based in these facts, we can say that during the last 40 years the industry unattended the communication's side of the \gls{AWS}, adapting the instruments to be capable to use protocols as \gls{FTP} to transmit the data from the \gls{AWS} to the collection point; focalizing the effort transmiting the data not mattering at all the technologies used or if they are or not optimized for that purpose. This practice gave as result multiple data formats implemented by the vendors without any common agreement, creating a huge incompatibility between the instruments and several bottlenecks in the data transmissions. 

The following subsections expose some data format used by the vendors to archive the data transmission and analyze why these data formats are causing bottlenecks.

\subsubsection{Data formats used by the vendors}

As mentioned in previous chapters, the format in which the data is produced by \gls{AWS} is formatted is up to the vendors. Nowadays the only standards used or involved in this process is \gls{ASCII} as character-encoding scheme or \gls{NMEA-0183}. Depending on the digital interface different control characters can be used, for instance is a common practice to generate one line of data follow by the carriage return (CR) or carriage return followed by line feed (CR+LF)\footnote{CR hexadecimal value: 0x0D. LF hexadecimal value: 0x0A. CR+LF: hexadecimal value 0x0D 0x0A.}.

\begin{table}[hc]
\centering

    \begin{tabular}{ | l | l | l | l |}
    \hline    
>"BARDATA"<LF>\\
<<LF><CR>"OK"<LF><CR>\\
<"BAR 29775"<LF><CR>\\
<"ELEVATION 27"<LF><CR>\\ 
<"DEW POINT 56"<LF><CR>\\ 
<"VIRTUAL TEMP 63"<LF><CR> \\
<"C 29"<LF><CR> <"R 1001"<LF><CR>\\
<"BARCAL 0"<LF><CR> <"GAIN 1533"<LF><CR> \\
<"OFFSET 18110"<LF><CR>\\
\hline
\end{tabular}
\caption{Example of data format used in a specific AWS to communicate the barometric pressure.}
\end{table}

Depending the \gls{AWS}'s brand the data's format is completely different from other brands and vendors. In most of the cases the data format is implemented based in the vendor's wishes. These wishes can be supported by technical reasons or not. Some vendors used acronyms to refer the data values returned by the sensors, others use the whole word to refer the phenomenon; not mattering the technique used in the data format, is a fact that they do not exist any compatibility of formats between vendors.

\begin{table}[hc]
\centering
\begin{tabular}{ | l | l | l | l |}
\hline    
0r2,Ta=10.6C,Tp=10.8C,Ua=74.6P,Pa=1006.0HKHK\\
\hline
\end{tabular}
\caption{Another example of data format used in a specific \protect \gls{AWS} to communicate different data as temperature or barometric pressure.}
\end{table}

A part of these big differences between the formats used in the digital interfaces, is needed to highlight that also the field's value used in \gls{CSV} or \gls{TSV} files producted by the \gls{AWS} are unique and incompatible between vendors. Thus, two levels of incompatibility exist, first the original data is delivered in a custom formatted untill the software's side. In the software's side this data is converted to a \gls{CSV} or \gls{TSV} format with the custom fields chosen by the vendors; this causes that even having the final data in a standard format as \gls{CSV} or \gls{TSV}, the order of the fields and their denomination will be different, forcing to the scientist to add an extra layer to the workflow to normalize this data and make it ready to be combined.

\subsubsection{Data formats used by governmental organizations}

Despite the fact that vendors used privative and non standard formats for the data, the \gls{WMO} has defined some specific data representation for certain users. An example of this is the \gls{METAR} format. Approved by the \gls{ICAO}, this format is the only one considered as official to communicate weather forecasting to the aviation and at the same it is widely use for other purposes as general weather forecasting.

\begin{table}[H]
\centering
\begin{tabular}{ | l | l | l | l |}
\hline
\textbf{Phenomenon} & \textbf{METAR's acronym} \\ \hline
cumulonimbus clouds & CB\\ \hline
thunderstorm & TS\\ \hline
moderate or severe turbulence & MOD TURB, SEV TURB\\ \hline
wind shear & WS\\ \hline
hail & GR \\ \hline
\end{tabular}
\caption{Some acronyms used in METAR format \cite{METAR}.}
\end{table}

However, this format has not relationship with the formats used by the vendors. Only a few \gls{AWS}es have the ability to product the \gls{METAR} format by default. The \gls{AWS} doing this are only focus in product data useful for the aviation, wasting the opportunity to provide the data in other formats for different use.

\begin{figure}[H]
\centerline{\includegraphics[width=1\textwidth]{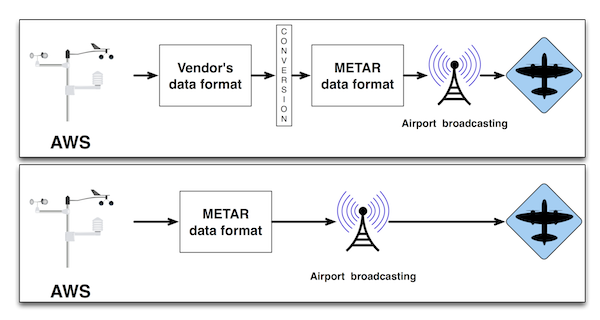}}
\caption{Weather data workflow, normal AWS VS METAR's AWS.}
\end{figure}

\gls{METAR} format is just an example of the multiple data formats invented for a specific purpose. The point to highlight is that often the weather data can be represented in a complete different format compare with the original format used for it. Nevertheless, the optimization of the data format until the point in which it is transformed marks a big difference in terms of data manipulation.

With the current technology the weather data arrives in different formats and with difference times frequencies, forcing to implement customized and particular mechanisms to transform this data to the format required. The complexity of this task resides in the requirement de facto requested by the \gls{AWS}es: they need intermediary points to convert the data because by default the data provided is useless for the required result.

In conclusion, it does not matter if the vendors provide a well known documented data format of their instruments. Because the observation of the weather is performed with different instruments, the data must be normalized to make it understandable.
Thus, at the end of the data workflow (when we take data from different sources and instruments), an intermediary layer to translate the vendor's data format to a common format is required.

\subsubsection{Mainstream architecture used for the weather data transmission}

To understand where are located the bottlenecks in the weather data transmissions is needed to understand the current architecture used by the vendors to archive this goal. As explained in section \ref{3.1.2}, an \gls{AWS} is an embedded system collecting information produced by the sensors attached to it. As embedded system, it has small capabilities to perform big CPU calculations, massive data storage or data delivery, however moderns systems are pretty balanced in terms of hardware and software to archive this goal. Although the \gls{AWS} have been optimized to collect and delivery the data, the protocols used for it are generic an non-specific. As explained in previous chapters the quality of the weather predictions reside in the ability to collect and process the atmosphere data with efficiency, reliability and fast delivery.

Despite of this, the methodologies for network communications are not optimized for this purpose. The following figure shows how the data is delivery.

\begin{figure}[h!]
\centerline{\includegraphics[width=1\textwidth]{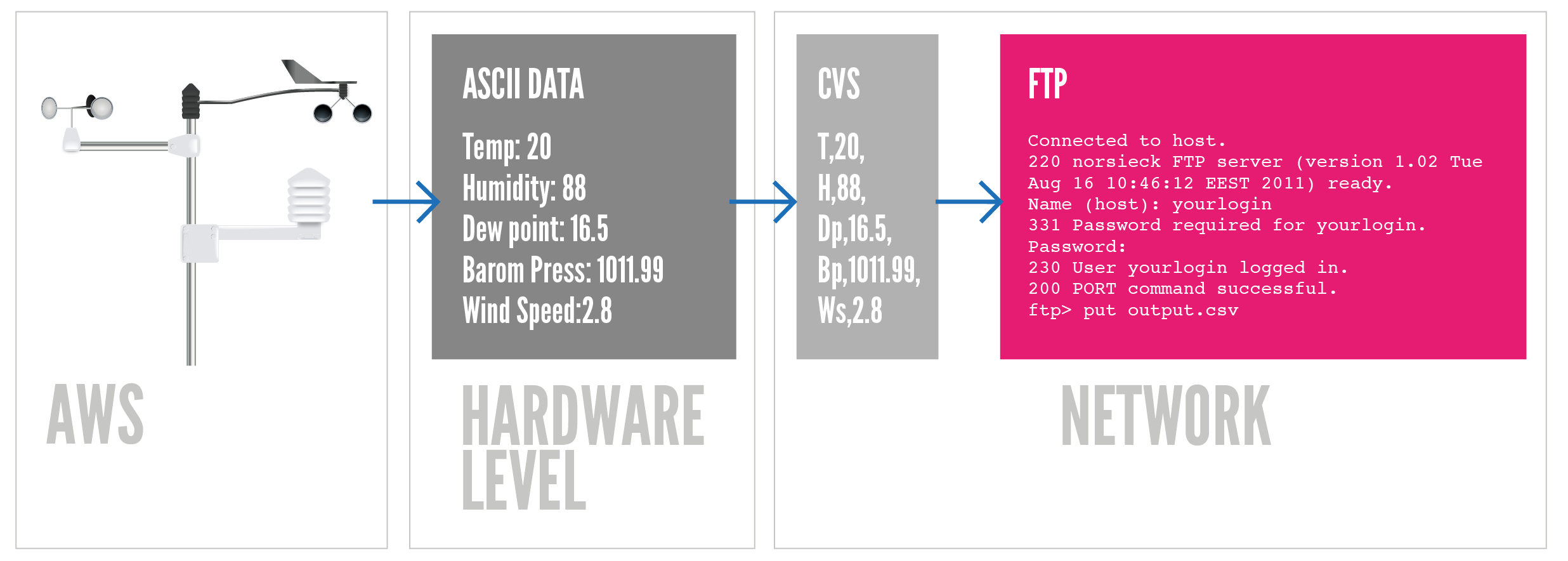}}
\caption{Example of an \protect \gls{AWS} transmitting weather data.}
\label{f4.2}
\end{figure}

In the figure \ref{f4.2} we can appreciate an example of the methodology used to transmit the weather data. In the hardware's level the data is delivered through a digital interface as explained in section \ref{3.1.2}, using some custom vendor's data format, commonly based in abbreviations as "Tmp (Temperature)", "Bp (Barometric Pressure )" "Ws (Wind Speed)", among others. These abbreviations are understood by the software. Depending of the AWS's setup this process can happen all together between the AWS and the datalogger:

\begin{figure}[h]
\centerline{\includegraphics[width=1\textwidth]{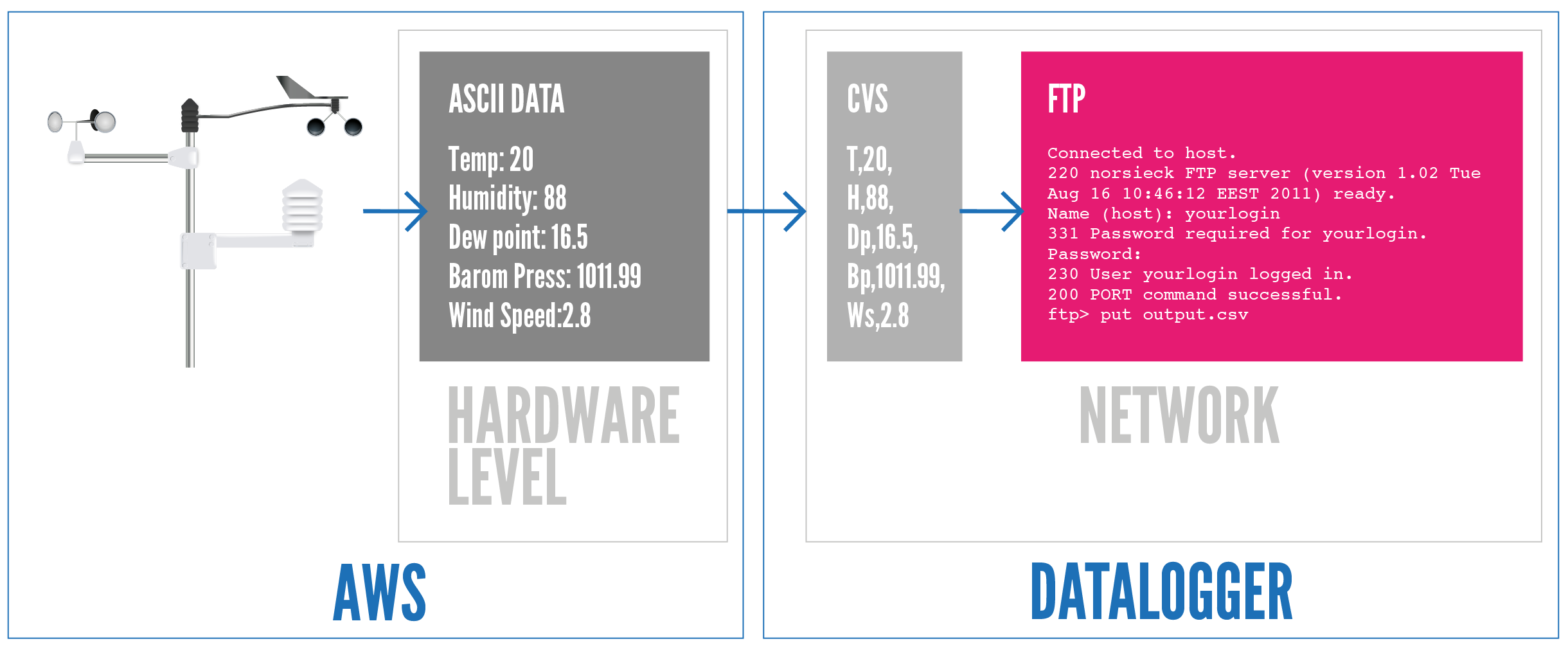}}
\caption{Example of a \protect \gls{AWS} and a datalogger transmitting weather data.}
\end{figure}

If the \gls{AWS} / datalogger has not network capabilities, a third entity can enter in the workflow. This entity is commonly a modern computer with the peripheral devices needed to interact with the \gls{AWS}. The computer takes the role of the weather data transmission, due to the possibilities that it offers, one computer can manage several \gls{AWS}es at the same time. Nevertheless, it does not introduce new protocols to send the data, it stills using protocols as \gls{FTP} or in some setups just shared folders using \gls{SMB}:

\begin{figure}[h!]
\centerline{\includegraphics[width=1\textwidth]{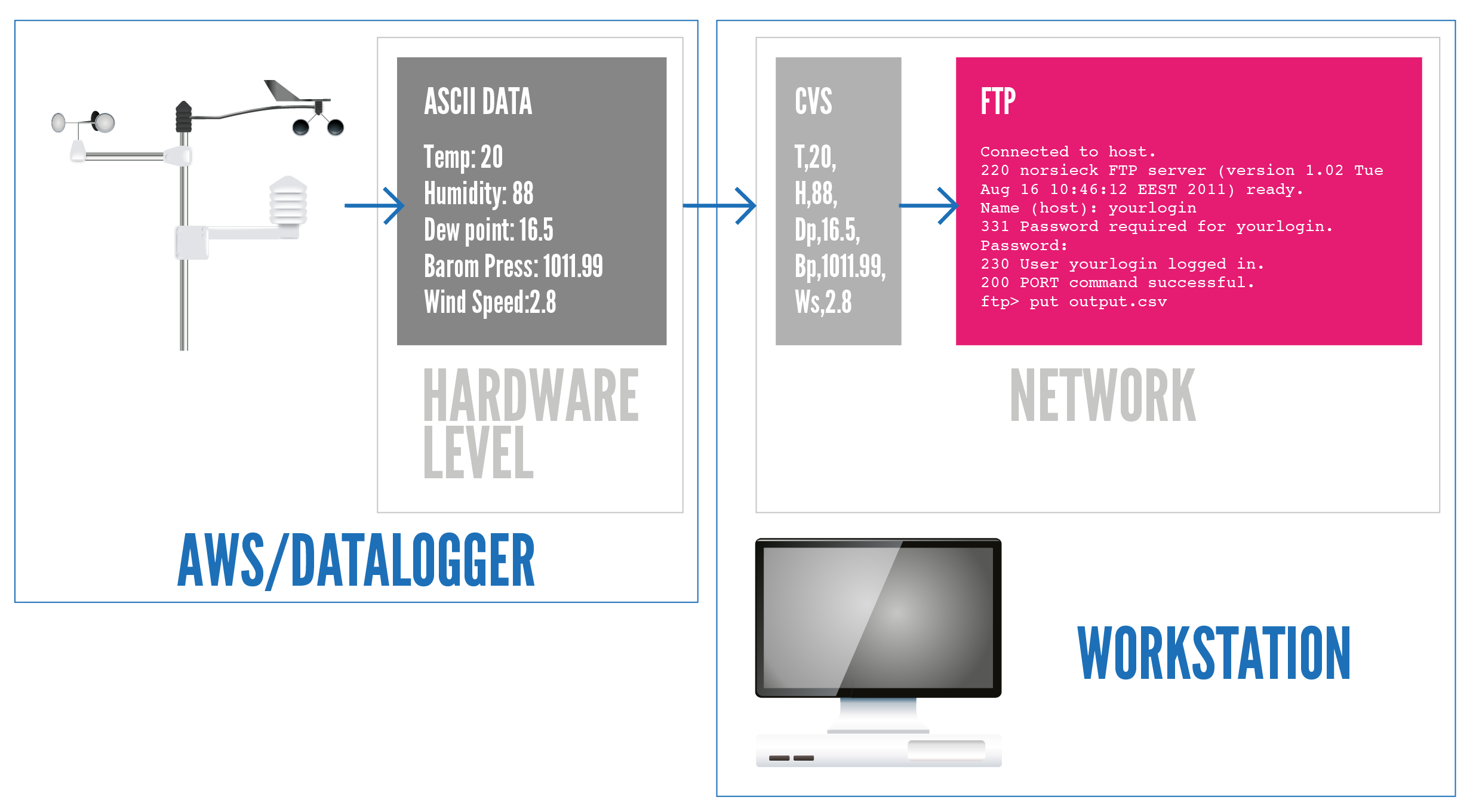}}
\caption{Workstation taking the role of the weather data transmission.}
\end{figure}

\subsubsection{FTP, the mainstream protocol in the weather data transmission}

Disregarding the setup used to send the data to the collection point, the protocol used will be generic and in most of the cases based in \gls{FTP}. Although \gls{FTP} has the capability to operate under stream mode \cite{rfc959}, the author could not find any vendor offering the capability to deliver the data through stream \gls{FTP} connections. Even being this possible, it will involve to use the image mode (commonly known as binary mode, thus, involving byte ordering choices) to transmit the data, however this choice will subject the data transmission to problems with the endianness\footnote{\emph{"Endianness describes how multi-byte data is represented by a computer system and is dictated by the CPU architecture of the system. Unfortunately not all computer systems are designed with the same Endian- architecture. The difference in Endian-architecture is an issue when software or data is shared between computer systems. An analysis of the computer system and its interfaces will determine the requirements of the Endian implementation of the software."\cite{endianness}}.}.

This setup can fill the requirements to delivery weather data collected over different time frequencies, however, it can not offer real-time capabilities, because the \gls{FTP} is not designed for this purpose. The author identifies \textbf{the use of FTP as a deficiency in the weather data transmission} \footnote{All the \gls{AWS} checked by the author are offering the data delivery based in \gls{ASCII} files using the \gls{FTP} \gls{ASCII} mode and sending the data using the \gls{FTP} block mode. Though is possible to find some \gls{AWS} using different methodologies as \gls{HTTP} get methods or email delivery, the FTP choice is mainstream overall the industry.}, the reasons for this are based in the fact that the protocol is designed to provide network capabilities to delivery data streams based in files. Notwithstanding, the \gls{AWS}es are producing data streams based in real-time data; \textbf{the use of FTP involves an intermediary step to convert these data streams to files, to continue after this sending theses these files to the collection point}. Even though to track this data in files is needed for storage and backup reasons, the data streams  generated in real-time by the \gls{AWS} are not used at all to send them directly to the collection point. In addition, the use of this methodology is forcing extra \gls{IO} operations required by the \gls{FTP}, that are not required in other protocols in which the data transfer does not involve the use of files.

Thus, it is not available any protocol taking advantage of all the capabilities offered by the \gls{AWS}es and its sensors, instead generic protocols as \gls{FTP}  or \gls{SMB} have been chosen to transmit data. These protocols are widely and accepted as the mainstream solutions for data transmission available on the weather instruments. 

\section{The missing standard}\label{missingstd}

One of the important factors of an implemented protocol, is to know how is going to be represented the data transmitted at the end of the transmission. This helps to design the best representation required by the data; for instance a protocol implementing real-time capabilities should be focus in fast data delivery and data integrity, among others.

In addition, to know the final representation of the data helps to implement a protocol optimized for the data that is transporting, this gives as result a better software for the protocol, besides it provides the capability to implement different protocols giving the same data result\footnote{A good example of this are the peer to peer networks, in which the protocol's designers know that at end of the process the data must be a file.}. 

Nevertheless, the weather data has certain particularities; the \gls{WMO} defines a set of methods to perform different measurements, notwithstanding theses methods are changing based in the advance of the physics, and these changes are causing an instability concerning what is the best way to measure a phenomenon, thus the data representation can get affected easily. Furthermore, the correlation between phenomena generates certain scenarios in which the data results can change completely if a new method is found to measure the phenomenon. This fact determines to which point we can have or not standards for these particular data. The \gls{WMO} defines which system of units must be used to represent the data for scientific purposes, in addition several guidelines are provided by the \gls{WMO} to perform the measurements under standard procedures. However, these guidelines are not enough to specify the final format of the data.

The \gls{WMO} started a process of standardization in \textbf{2002}, the goal is to create a data format to fit the requirements of the \gls{GOS}, in other words to provide a common basement to represent the data of the weather's observations. This is an arduous task, not only for the amount of data that is needed to manage, also for the big a mount of different phenomena in the atmosphere that are producing different data and their particularities. It is expected that in some point, the \gls{WMO} will publish a standard for weather's metadata representation, nevertheless, after 9 years this process still under development.

The absence of a standard for weather data representation is one of the key-issues of the current situation. Without knowing how must be formatted the data at the end of the collection workflow, is understandable that vendors ended implementing their own formats without compatibility. 

This is an open issue that unfortunately can not be treated in this thesis. The author recognizes that the implementation of a protocol to transmit the weather data without to know the final format of the data is a risky but an interesting feature. In chapter eight, an exposition of the solution chose (a software library to normalize the data) is explained.

\textbf{We identify the absence of a common format for data representation as one of the major technical deficiencies in the weather data transmission}. In addition, the absence of a common data format in the collection point as well, forces to convert the weather data multiple times to the final format. OpenWeather considers this issue and provides some mechanisms to implement smoothly and mostly transparent the conversion from OpenWeather's format to a future data standard.

\section{Data transmission and Automatic \\ Weather Stations}

As embedded systems the \gls{AWS} have more limitations that moderns computers, not having capabilities to perform complex \gls{CPU} operations or to manipulate a considerable amount of data. Most of the modern \gls{AWS}es offer the possibility to interact with them in a small scale. Commonly, this interact is focused in three tasks:

\begin{itemize}
\item AWS configuration
\item Sensor's calibration
\item Data retrieval
\end{itemize}

Even so in most of the cases the \gls{AWS}es behave as "broadcasters" of weather data. The tasks of configuration and sensor's calibration are performed only a few times in the instrument, happening this at the beginning of the \gls{AWS}'s installation and in some periodical calibrations during the life-time of the instrument; both operations are performed in most of the cases through command's line parameters or some \gls{GUI} developed for this purposed. As it was explained in section \ref{3.2.2}, the data transmission with an \gls{AWS} is performed through digital interfaces based in serial communications standards, it means that at the end all the data transmitted and received in an \gls{AWS} goes through some data format implemented by the vendor that provides a set of custom instructions. 

\begin{table}[hc]
\centering
\begin{tabular}{ | l | l | l | l |}
\hline    
>"BAUD 9600"<LF> \\
<<LF><CR>"OK"<LF><CR>\\
\hline
\end{tabular}
\caption{Example of command configuring the baud rate of the digital interface in an \protect \gls{AWS}.}
\end{table}

Even if this practice is something understandable\footnote{The author recognize that to have a proprietary set of instructions can be a method to keep some industrial's secret of the instruments, however this practice difficulties the implementation of standard methodologies to interact with multiple the instrument.}, an exception should be made in the data retrieval operation.

Most of the \gls{AWS}es offer the possibility to retrieve particular data if a specific command is sent to them. Again the method to obtain this data is up to the vendor, not being compatible these instructions between vendors, and even sometimes even not between the products of the same manufacturer.

The mechanisms to retrieve data from the \gls{AWS}es \textbf{are critical} in order to implement a protocol with real-time capabilities. We need to differentiate  two use cases on an \gls{AWS}. The first use case involves the data broadcasting that the \gls{AWS} is performing by default if it is configured as "automatic mode"\footnote{This is the default configuration used in almost all the scenarios.}. The \gls{AWS} just send the data through the digital interface in the time frequency configured, for this case is not required interaction with the \gls{AWS}; to read the data from the digital interface is enough to use it in the protocol. Nevertheless the second use case involves the retrieval of particular data. One example of this is a user interested in to know the average of temperature recorded by the \gls{AWS} in the last week. This data is not sent by default because it is not part of the information collected in real-time for the \gls{AWS}, to get the data the user must send a command asking for it to the \gls{AWS}:

\begin{table}[hc]
\centering
\begin{tabular}{ | l | l | l | l |}
\hline    
\textbf{Command}:  aR2<cr><lf> \\
\textbf{Response}: 0R2,Ta=23.6C,Ua=14.2P,Pa=1026.6H<cr><lf>\\
\hline
\end{tabular}
\caption{Example of command asking for \protect \gls{PTH} data.}
\end{table}

This second use case introduces much more complexity. If a particular data not send by default is needed, the interaction with the \gls{AWS} is mandatory, however, to interact with it implies to do it using the methodology specify by the vendor. To implement a protocol that takes this use case in consideration involves to implement a command-translator between the \gls{AWS} and the protocol implementation. \textbf{We identify this issue as another technical deficiency in the weather transmission in order to enable the capability to retrieve specific data on demand.}
\section{Summary}

In this chapter we described the state of art in the weather data transmission. We have been analyzing the different interfaces available in an \gls{AWS}, focusing on their bandwidth, and based in the bit-rate that they offer, concluding that the \gls{AWS} are not taking advantages of all the capabilities offered by the digital interfaces. This fact is enough reason to claim that the \gls{AWS}es are capable to manage more amount of data that the current quantities that they do.

Data formats used by the vendors and data format requested for the governmental organizations have been compared; finding that is not any relation between the original format used in the \gls{AWS}s and the final format in which the weather data is represented, being this one of the reasons that forces the implementation of intermediary points to translate the data to different data formats.

The absence of a protocol dedicated to the weather data transmission has been studied; the use of the \gls{FTP} has been explained and the limitations that it can involve to transmit data in real-time have been analyzed. We conclude that  \gls{FTP} is chose by the industry as non-optimal solution that fix partially the issue of the weather data transmission. In addition, the key issues of \gls{FTP} has been exposed in order to implement a system that use this protocol to delivery data in real-time.

We analyzed the implications of a missing standard to represent the weather data, concluding that without a consensus of the international community about how the weather data  should be represented, is really complex to implement a protocol to fit all the requirements needed.

Despite the absence of a protocol and the use of multiple protocols and data formats, the industry and weather organizations are using these methodologies to acquire weather data in their weather data networks.  Although projects such as \gls{GOS} or \gls{GDPFS},  are looking for technologies to optimize and standardize the weather data transmission, the current status of weather data acquisition is based on the methodologies that the industry provided without previous agreement. These methodologies have been accepted by the weather organizations as the standards for the weather data transmission, achieving until today their purpose.

At the end of the chapter we exposed how to retrieve particular data from the \gls{AWS} involves user interaction, adding complexity to the data workflow and requiring an intermediary step to translate the data requests to the native format used in the \gls{AWS} to retrieve the data. We identify this as an impediment in order to implement a protocol that provides data on demand.

The next chapter explains in which consists OpenWeather, its architecture and how it can fix the issues explained in this chapter.
\pagebreak

%% file: chapter5.tex
\chapter{Introduction to OpenWeather}

The previous chapter summarized the issues found by the author in the protocols used for weather data. It has been analyzed how the weather instruments use protocols as \gls{FTP} or \gls{SMB} to transmit data. Nevertheless, these protocols are not designed to be used in a scenario in which the data is generated based on real-time inputs. In addition, the current methodologies provided by the industry, are not efficient enough to interact with the \gls{AWS} without additional effort in performing data normalization or data delivery. This chapter gives a general overview of OpenWeather, the protocol developed by the author, in order to provide a solution to problems that weather instruments encounter during data transmission.

\section{Overview and goals}\label{5.1}

OpenWeather is an application layer protocol based on \gls{TCP}/\gls{IP}. It assumes a reliable transport layer (\gls{TCP}), in order to achieve a successful data delivery, based on such mechanisms as error detection, flow control, congestion control, etc.

The protocol is built assuming three principles:

\begin{itemize}
\item Every \gls{AWS} is considered to be a node
\item A node accepts incoming sessions from peering hosts and initiates outgoing sessions to peering hosts as well
\item	An \gls{AWS} must have the capability to provide and to request services from other nodes.
\end{itemize}

These principles are supported by assumptions that an \gls{AWS} is an embedded system with networking capabilities, able to interact via TCP/IP to deliver the data produced by its sensors. The sensors' output are considered to be services offered by the \gls{AWS} (node) to other nodes.
 
In addition, the star topology explained in section \ref{3.1.4}, disappears to give way to a decentralized topology based on a peer to peer architecture.

OpenWeather provides the capability to dispense a unique collection point. Instead, all nodes can be collection point and at the same time to be part of other collections points. In addition, the protocol offers a service oriented model (\gls{SOA}), to provide an easy way to interact with the nodes and retrieve or send data to them.

\begin{figure}[H]
\centerline{\includegraphics[width=1\textwidth]{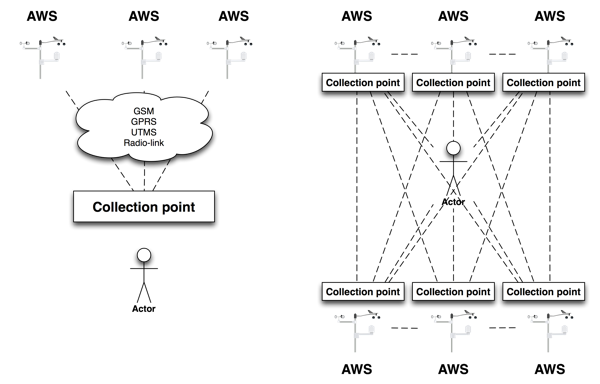}}
\caption{Comparison of the currently centralized architecture provided by the industry against OpenWeather architecture.}
\end{figure}

From the perspective of portability and data delivery, the protocol has been designed to avoid problems with the endianness and data normalization; to achieve this goal, \gls{JSON}\cite{rfc4627} has been chosen as data interchange format between nodes.

\gls{JSON} allows OpenWeather to use data streams based on parsable objects, facilitating the data manipulation and normalizing the data to one common format. Additionally, \gls{JSON} is well supported by several libraries\cite{JSONW}, bringing the possibility to easily create applications based on OpenWeather format.

\begin{figure}[H]
\centerline{\includegraphics[width=1\textwidth]{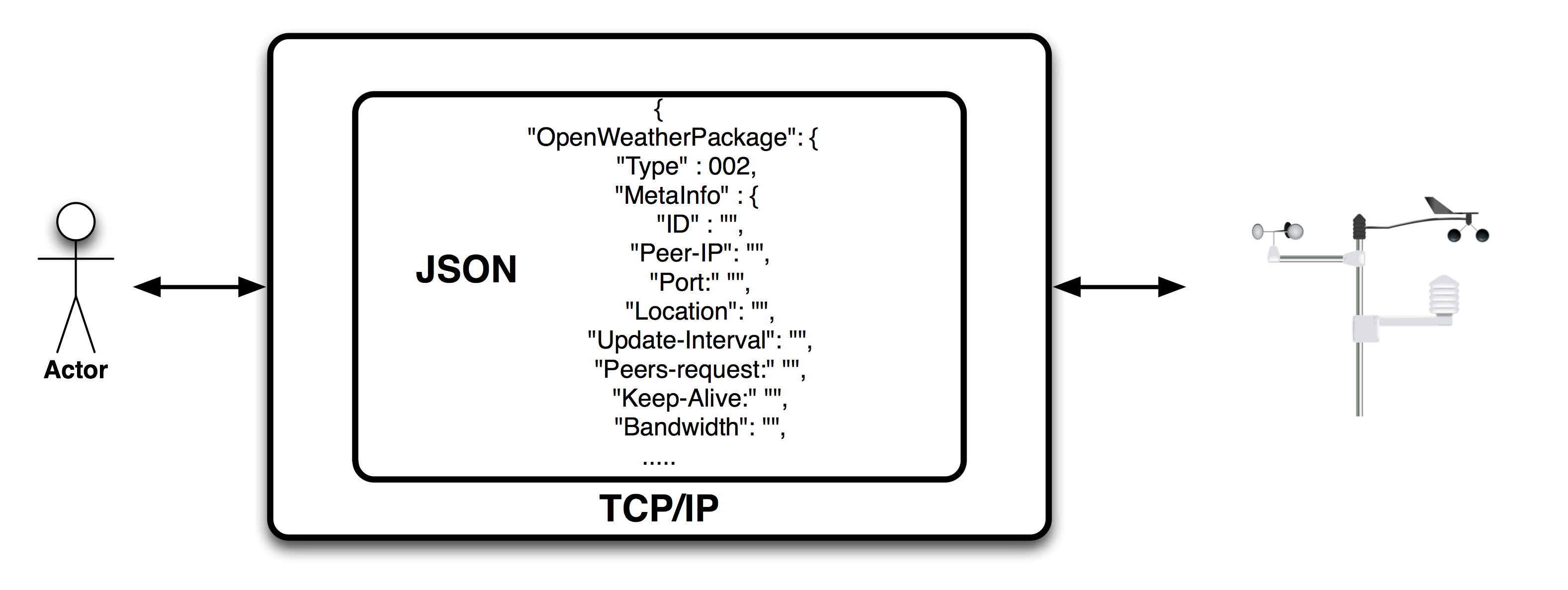}}
\caption{Example of a OpenWeather's JSON object inside of data message.}
\end{figure}

\subsection{Improvements in the current technology}

OpenWeather provides a new paradigm for weather data collection. Based on a \gls{P2P} architecture, it allows the users to interact between multiple nodes, retrieving and sending information inside of the network independently of the brand's instruments used. At the same time, it brings the possibility to combine the real-time data streams obtained from the nodes, providing a stack to build applications using multiple data sources without requiring extra resources on the data manipulation.

In addition, the protocol is designed to be extensible, adaptable to new types of data, while maintaining compatibility with future formats. Furthermore, the service oriented model (\gls{SOA}) of the nodes, allows the users to develop applications that only want to obtain some specific data from a particular service.

Finally, the protocol brings new opportunities to be operated under distributed models and to provide implementational basis for future standards of the weather data categorization. Because the data interchange format is text-based and human-readable, it provides the capability to combine the protocol with database applications without the need to develop extra \gls{API}s, facilitating even more possibilities to take advantage of the data.

\subsection{The role of OpenWeather and data spreading}\label{5.1.2}

OpenWeather is designed to fix deficiencies in weather data transmission, while helping with the tasks of spreading data to the end users. Though most of the phenomena require scientific analysis to make the data understandable, some phenomena as atmospheric temperature, pressure or wind speed, are simple enough and known to be spread across them directly to the end users without the need of additional processing. OpenWeather allows to connect to an \gls{AWS}\footnote{Through a intermediary layer implemented through software.}, to retrieve this type of data in real-time and —host to host— based, not needing more than a computer with software supporting the OpenWeather protocol and network connectivity.

In addition, the technologies used in OpenWeather can facilitate the creation of new \gls{API}s for web services oriented on weather's forecasts. Some websites offer the possibility for calling \gls{API}s to obtain weather data. However, these \gls{API} calls are completely different between websites, which leads with extra development time of web applications which utilizes different web resources for data extraction. This problem can be easily handled with OpenWeather, creating standard \gls{API} calls according to the protocol specification. This enables the use of such encapsulated protocols methods as \gls{HTTP} for creating for an intermediary bridge between the web application and the end nodes.  

\begin{figure}[H]
\centerline{\includegraphics[width=1\textwidth]{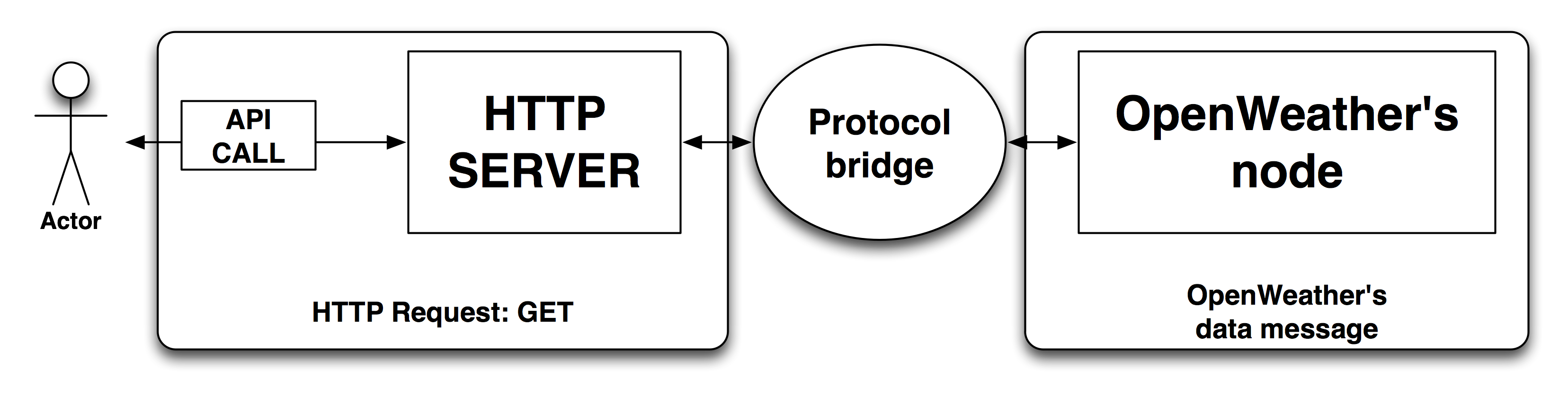}}
\caption{Example of an \protect \gls{API} call through \protect \gls{HTTP} and OpenWeather.}
\end{figure}

\subsection{Contribution to the current methodologies for weather data acquistion}

Even if OpenWeather is a proof of concept of an adapted protocol for \gls{AWS}, it proves how the problems exposed in chapter four can be resolved. The feasibility of migration of scientific installations for production, will be deemed feasible as the principles applied in OpenWeather, just adopting the \gls{P2P} architecture or the use of a human-readable lightweight format as \gls{JSON}, it will be enough to observe improvements in data delivery and acquisition. In chapter seven is analyzed the results of use OpenWeather.

As it was mentioned in chapter four, the \gls{WMO} has several worldwide projects, such as \gls{GOS}, in which different weather organizations around the world are involved in the process of creation of future basis for weather data processing. As described on \gls{WMO}'s website\cite{WMO}, one of the purposes of the project is: \emph{'The coordinated system of methods and facilities for making meteorological and other environmental observations on a global scale in support of all WMO Programmes''}. OpenWeather, as scalable and extensible protocol, can proven useful in certain areas of projects as \gls{GOS} or SMEAR\cite{SMEAR}, concerning data availability.

\subsection{Impact on weather instrument industry}

As it was analyzed in chapter four, the industry has not started the process of standardization for their instruments. Despite the issues that this practice causes, OpenWeather aims to be the first solution that tries to fix the absence of such protocol and at the same time provides a basis to be adapted for the future data standard format, providing better archiving mechanisms for a more efficient exchange of weather data.

Furthermore, the \gls{P2P} architecture brings such a new industry paradigm, allowing to develop new products in which real-time data retrieval will be put to use.

\section{Basic functionality of OpenWeather}\label{5.2}

Considering any \gls{AWS} a node, the implementation of OpenWeather should be done inside of the \gls{AWS}'s software itself. Nevertheless, the author \textbf{can not implement a fully functional prototype, because it is not available any open source / libre software version of \gls{AWS}'s \gls{OS}}. Instead, an intermediary layer has been created for the evaluation setup, to normalize data from vendor format into OpenWeather format.
\footnote{The removal of this layer depends on cooperation between vendors in order to implement a protocol inside of the \gls{AWS}'s \gls{OS}.}.

\begin{figure}[H]
\centerline{\includegraphics[width=1\textwidth]{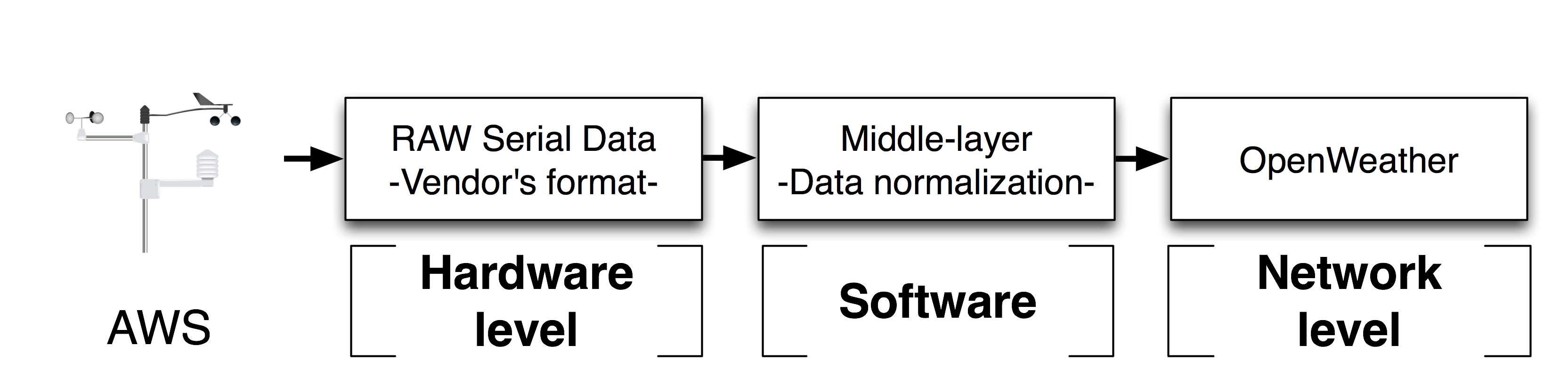}}
\caption{Middle-layer for data normalization.}
\end{figure}

This layer provides the conversion from native vendor format explained in section \ref{4.2}, to an operational format in which OpenWeather can work. When the data is pulled through a digital interface, the middle-layer recognizes the vendor format and converts it according with OpenWeather requirements.

This middle-layer is located between the hardware and the network level, giving as a result formatted data ready to be used in the protocol. With the introduction of this layer, \textbf{the steps mentioned in previous chapters\footnote{Concerning data parsing.} disappear}. The data normalization occurs only once at a time, instead of multiple times along the data workflow.

\begin{table}[H]
\centering
\begin{tabular}{|l|l|l|p{8cm}|}
\hline    
\textbf{Original sender \gls{AWS} data}:0r2,Ta=10.6C,Tp=10.8C,Ua=74.6P,Pa=1006.0HKHK\\
\hline
\textbf{OpenWeather's format}:  \\
\begin{minipage}[t]{\linewidth}
	\begin{verbatim}
"Data" : { 
            "PTU" : {
                "Air-Temperature" : "23.6", 
                "Relative-Humidity" : "14.2", 
                "Air-Pressure":  "1026.6" 
            }
           \end{verbatim}
\end{minipage} \\
\hline
\end{tabular}
\caption{Comparison of one vendor format against OpenWeather \protect \gls{JSON} format.}
\end{table}

When data is normalized by this intermediary layer, the \gls{AWS} is ready to operate inside of OpenWeather network. This intermediary layer will not be needed if the vendors establish a process of standardization.

\subsection{Peer to Peer Architecture}\label{5.2.1}

As mentioned in section \ref{5.1}, OpenWeather is designed based on a \gls{P2P} architecture. The \gls{RFC} 5694 (Peer-to-Peer (P2P) Architecture: Definition, Taxonomies, Examples, and Applicability)\cite{rfc5694}, defines a \gls{P2P} system as the following:

\emph{[...] We consider a system to be P2P if the elements that form the system
   share their resources in order to provide the service the system has
   been designed to provide.  The elements in the system both provide
   services to other elements and request services from other elements. [...]}

OpenWeather is according with the definition established by the \gls{RFC} 5694 \cite{rfc5694}. The protocol is thought to share the resources available in an \gls{AWS} and at the same time request services from others. In order to function properly the OpenWeather network requires a minimum activity that must be performed by the nodes (as peers's list exchange).

Note that user itself is considered to be a node. \textbf{It is not necessary to have an \gls{AWS} in order to be considered a node}. A node is part of OpenWeather network, interacting with other nodes, sending and retrieving data, while time offering services to them\footnote{Thus, a user without an \gls{AWS} can interact with other nodes offering for example peer list exchange.}. 

An OpenWeather node possesses the following properties:

\begin{itemize}
\item A node has a \textbf{unique ID} within OpenWeather's network
\item The geographical location of a node \textbf{is essential to its connection in order} to OpenWeather's network
\item A node of the OpenWeather network can require the use of \gls{NAT}\cite{rfc1631} \footnote{As described in \gls{RFC} 5128 (State of Peer-to-Peer (P2P) Communication across Network Address Translators (NATs)\cite{rfc5128}, will be recommendable to implement the TCP/UDP Hole Punching technique in OpenWeather's software, in order to avoid peer connectivity issues.}
\end{itemize}

Opposed to other \gls{P2P} networks, OpenWeather does not use the \gls{P2P} architecture to archive a better performance transmitting big amounts of data\footnote{In fact, as explained in section \ref{3.1.2}, the amount of data generated by a node is insignificantly small.}; the justification of use of \gls{P2P} architecture in OpenWeather is based on the distribution of the nodes and for better interaction with them. The centralized model, fails to utilize its ability to use weather data from different collections points without a pre-normalizing data. In addition, the \gls{P2P} architecture enables scaling of the network as well, as giving the advantage of not being restricted by the limitations of a central node.

\subsection{Service Oriented Architecture in nodes}\label{5.2.2}

As explained in section \ref{3.1.2}, an \gls{AWS} produces real-time data collected by its sensors. At the same time some \gls{AWS} are able to store specific data in persistent memory such as averages figures, daily reports, etc. These features provide two data use cases for OpenWeather:

\begin{itemize}
\item Data becomes available in real-time
\item Data can be retrieved on demand without the need to be real-time specific
\end{itemize}

OpenWeather handles these use cases providing an extra layer based on \gls{SOA}. In order to achieve this, OpenWeather provides a mechanism to discover which services being available in a particular node, being possible after the initialization of the session, to interact with these services.

\begin{figure}[H]
\centerline{\includegraphics[width=0.5\textwidth]{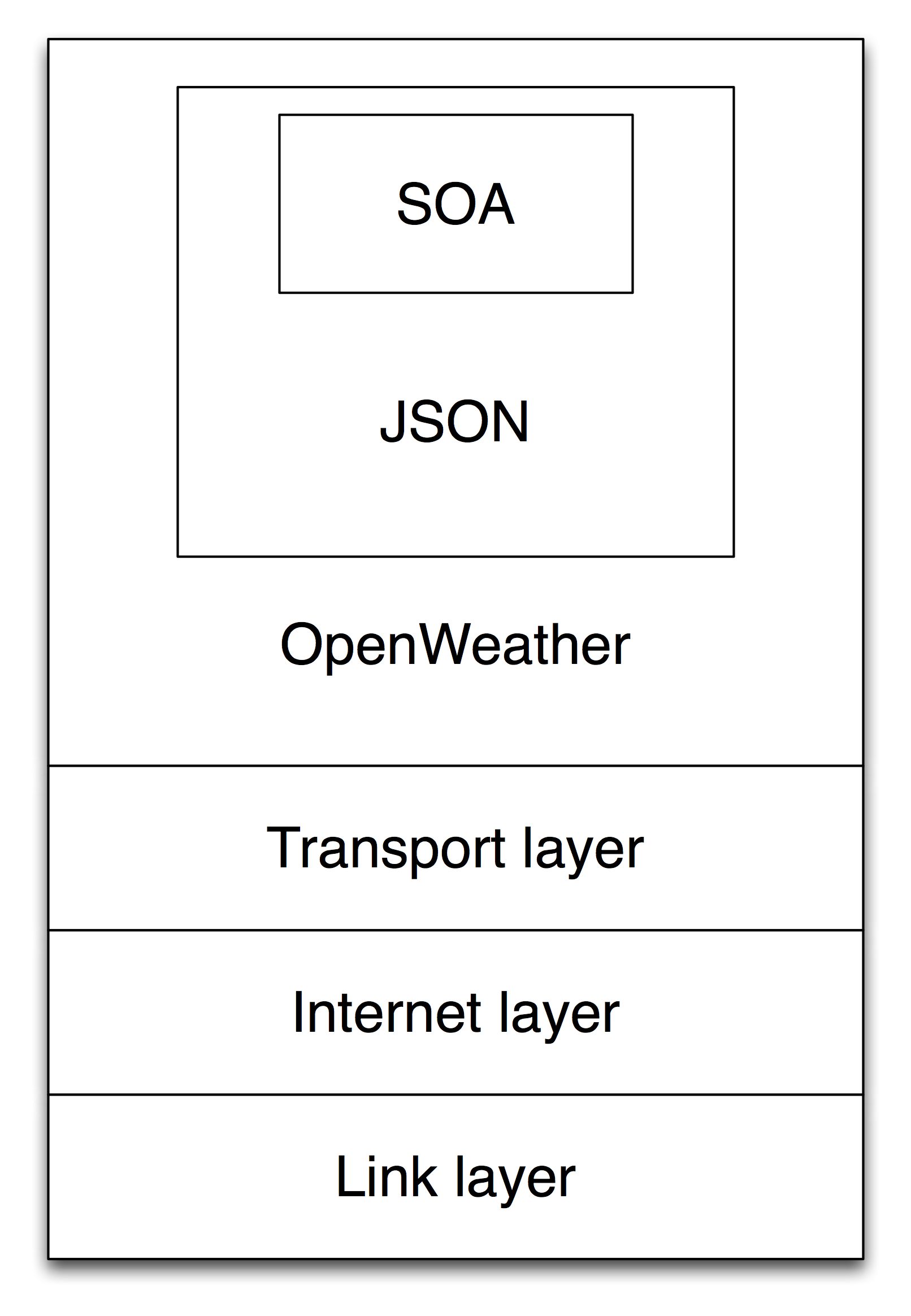}}
\caption{OpenWeather stack over TCP/IP.}
\end{figure}

The fundamental reasons of choice of \gls{SOA} for OpenWeather, is to facilitate the accessibility of the data. A user can be both interested in receiving only real-time data or in to retrieving a particular chunk of data. To provide this capability, the protocol must be \gls{SOA} oriented, in order to alleviate data access through these services.

\begin{figure}[H]
\centerline{\includegraphics[width=1\textwidth]{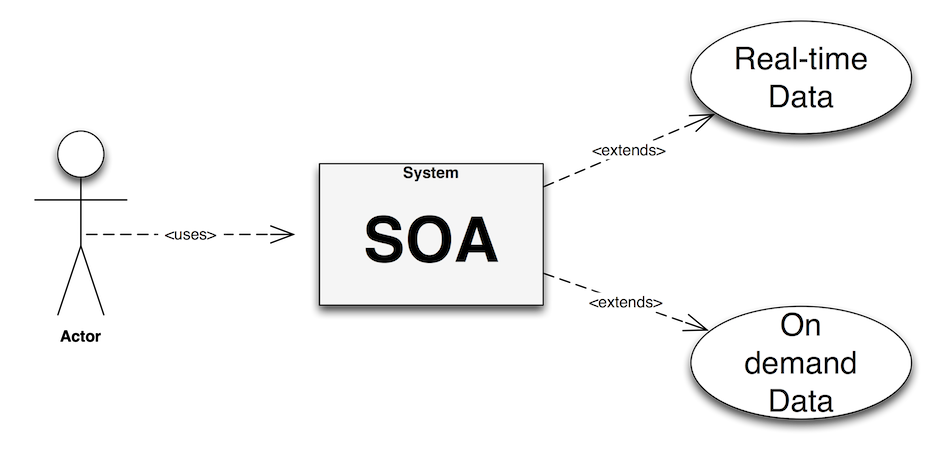}}
\caption{Uses cases available in OpenWeather via \protect \gls{SOA}.}
\end{figure}

Real-time data messages flow is considered to be as a continuos service offered by the \gls{AWS} via OpenWeather. Additionally, the possibility to retrieve saved data in the \gls{AWS} exits. Both real-time data and data on demand, is sent and retrieved through OpenWeather data message system, using \gls{JSON}. Thus, OpenWeather offers the same possibilities as the common methodologies currently used by the vendors explained in chapter four, moreover the chance to get real-time data through a reliable and efficient way.

\section{Summary}

In this chapter we gave an introduction of OpenWeather, highlighting the general guidelines applied in its design. We exposed some of the principles used in OpenWeather nodes we considered some possible examples of future applications using OpenWeather. 

We introduced the areas in which OpenWeather can have a contribution or impact. Projects as \gls{GOS} or SMEAR\cite{SMEAR} seeking for new technologies for data acquisition, could get a positive use of OpenWeather concepts.

In addition, the basic functionality of the protocol, such as its architectural principles or software model implementation have been introduced as well.

\pagebreak

%% file: chapter6.tex
\chapter{Protocol specification}

In this chapter, the OpenWeather protocol specifications are explained. 

\section{Definitions}

The following subsections summarized the role of the elements involved in the protocol. Some of the definitions are widely used in other protocols.

\subsubsection{OpenWeather network}

The nodes used in OpenWeather protocol conform to OpenWeather network standards. Inside of this network a node is able to interact with other nodes, requesting and delivering services to other nodes. These services are oriented to provide weather data. Because OpenWeather is based in a \gls{P2P} architecture, its topology is decentralized. This topology makes the nodes independent of central nodes in order to interact between them.

\subsubsection{OpenWeather node}

A node is an active or passive element connected to OpenWeather network. One node can offer none to multiple services. An element is considered a node when it has a working implementation of OpenWeather protocol and is connected to the network.

\subsubsection{Peer}

Every node is considered a peer of OpenWeather network. All nodes in OpenWeather network are able to be clients and servers at the same time. This is establishing the basis of the \gls{P2P} architecture used in OpenWeather. A peer must be able to offer services to others peers, however it is not mandatory to offer a service\footnote{Doing reference here to high-level services related with the data delivery, de facto, a peer is always offering a minimum amount of services integrated within the protocol, needed to interact in the network.} in order to be connected to OpenWeather's network.

\subsubsection{Weather data}

The purpose of OpenWeather is to create a network in which the data exchange comes from the weather data sources. To obtain this data the nodes can be connected to an \gls{AWS} or other system of weather data collection. OpenWeather does not differentiate between the original source of the instrument's brand, because data normalization\footnote{As it was explained in chapter 3 and chapter 4, this step is required because it is not possible to modify \gls{AWS}'s \gls{OS} without the vendor's collaboration.} is required in order to make the data network available.

\section{Architecture}

As it is mentioned in section \ref{5.2.1}, the architecture used in OpenWeather matches the requirements mentioned in the \gls{RFC} 5694 \cite{rfc5694}, with OpenWeather containing nodes offering and requesting services between them.

The technical reasons why a \gls{P2P} architecture is a better network solution for a topology as define by default by the \gls{AWS}, are supported in the following points:

\begin{itemize}
\item	An \gls{AWS} is an individual entity being part of a bigger network that does not need a centralized model except for data processing.
\item	The process executed over the weather data in order to extract meaningful  conclusions does not posses a technical requirement to be linked to the network layer.
\item	The collection point model forces the node to depend exclusively on one node in the network, adding unnecessary risks to the data flow.
\item	The common architecture used in the weather data flow, is forced by the legacy of the protocols used within it.
\end{itemize}

The \gls{P2P} architecture  is chosen by OpenWeather because it brings autonomy and robustness to the nodes. In addition, it provides the network the capability to scale and to share resources without single dependencies. Moreover, the geographical situation of the nodes, is suitable for developing models in which the nodes can collaborate to distribute the data. Finally, the \gls{P2P} architecture provides the capability to retrieve data directly from the node, without going through a common point that can be collapsed or not available.

\subsection{Standards used for data units}

OpenWeather does not provide the  weather date measurement units. The protocol is designed to deliver weather data formatted according to the data units specified in \gls{ISO} 80000 \cite{ISO80000} family and the \emph{Guide to Meteorological Instruments and Methods of Observation} \cite{GMIMO}.

Table \ref{t6.4} provides the data units used in the prototype:

\begin{table}[H]
\centering
    \begin{tabular}{ | l | l | l | l |}
    \hline
    \textbf{Data field} & \textbf{Data unit} & \textbf{Acronym} \\ \hline
    Air-Temperature & Celsius & C\\ \hline
    Relative-Humidity & Percentage& \% RH\\ \hline
    Air-Pressure & Hectopascals & hPa\\ \hline
    Wind direction & Degress & degrees \\ \hline
    Wind speed & Meters per second	& $m \over s$ \\ \hline
    Rain accumulation & Millimeters & mm\\ \hline
    Rain duration & Seconds & s \\ \hline
    Rain intensity & Millimeters per hour & $mm \over h$\\ \hline
    Rain peak &  Millimeters per hour & $mm \over h$ \\ \hline
    Hail accumulation & Hits per square centimeter & $Hits \over cm^2$\\ \hline
    Hail duration & Seconds & s \\ \hline
    Hail intensity & Hits per square centimeter per hour & $Hits \over cm^2 h$ \\ \hline
    Hail peak &  Hits per square centimeter per hour & $Hits \over cm^2 h$ \\ \hline
    \end{tabular}
  \caption{Data units implicit on the data fields.}
  \label{t6.4}
\end{table}

Since the data units have a known standard, the author considers that it is not necessary to increase data messages sizes and data fields, but only to provide the data units. Instead, it is more pragmatic and efficient to assume that weather data will be supplied with appropriate data units. It is necessary to highlight that despite the absence of network protocol for weather data, the vendors maintain a strict control of data units used in the \gls{AWS}, facilitating this the implementation of OpenWeather across vendors.

\subsection{Nodes}

A node connected to a OpenWeather network behaves as a deterministic finite automaton, not executing without a clear definition operations or a definite result. All the operations performed by the nodes are identified by codes placed in the MetaInfo data field. Any data message delivered in OpenWeather protocol contains all information\footnote{Through the protocol code.} required to identify the type of operation to be performed by the software when the data message is received / delivered.

Any data requested or delivered by a node using OpenWeather is based on a request and a confirmation of it. With this mechanism the nodes are notified of status of the operations of execution in the application layer are successful or not. This same mechanism is implemented in protocols as \gls{HTTP}\cite{rfc2616} in order to control the status of retrieval and delivery operations.

A node is able to interact with multiple nodes, being only limited by the bandwidth and system resources availability. OpenWeather does not define a minimum or maximum of connections needed, however a node requires a \textbf{>=1} number of peers on its internal list in order to interact with OpenWeather network.

\subsubsection{Automatic Weather Stations as individual nodes}

The section \ref{3.1.2} explains how the \gls{AWS} are categorized as embedded systems. By the definition, an embedded system has certain limitations in data processing and data delivery. Nevertheless the \gls{AWS} are still able to do some networking operations and data processing when the size of them is small. OpenWeather has been designed to work around these limitations.

Taking this as a basis, OpenWeather transforms the centralized model currently used by the industry, to a decentralized model taking advantage of a \gls{P2P} architecture. OpenWeather considers every \gls{AWS} as a node using \gls{SOA}. Because the \gls{AWS} are under constant connection and deliver data to collection points, the only modification needed in the equipment is to change the network protocols used to deliver this data\footnote{An adaption of the \gls{AWS}'s \gls{OS} will be required in order to integrate the OpenWeather's stack inside of the \gls{AWS}.}.

Instead of using an architecture in which the \gls{AWS} plainly sends the data over the network without any further interaction, OpenWeather provides the mechanisms to convert the \gls{AWS} to an entity able to respond to the data requests made by the user in real-time. Although all of this process can be handled through the centralized model, the independence of nodes from the collection point  is mandatory in order to achieve scalability and data accessibility. For instance, a user located outside of a specific network of \gls{AWS}, can not access the data produced by them without the need to go through the collection point\footnote{If the \gls{AWS} work but the collection point is down, the data will not be accessible.}, this use case avoids any possibility to combine data in real-time from different \gls{AWS} in different geographical locations, restricting any possibility to interact directly with the \gls{AWS}.

Enabling the \gls{AWS} to behave as a nodes, the protocol provides the basis to take advantage of the real-time data and at the same time fix the issues exposed in chapter four. Though this thesis sets an ambitious goal: the transition from a centralized model to a decentralized model, it has to be noted that the industry has been using the same technologies for decades, not taking advantage of the improvements made in the networking technologies, concerning data delivery and acquisition. The decentralized models have a proven successful track, offering scalability and robustness.

As any other network protocol, OpenWeather has a defined set of operations. These operations provide the core principles to deliver and retrieve data from nodes. However, these principles do not need to contain the whole data flow.

\subsubsection{Super-nodes}
 
OpenWeather refers to super-nodes to those nodes with static \gls{IP} / hostname, which are always available to exchange peer lists. Unlike other \gls{P2P} applications, an OpenWeather super-node does not have any other extra property, except its bandwidth availability \footnote{It must be higher than average so that it may process higher network traffic.} and an updated list of peers, to deliver to the other nodes. The role of a super-node is to be always available and to provide updated peer lists to those nodes without one. This is enough to guarantee that the nodes will be able to connect to it if they can not find other nodes available.

\subsubsection{Peer list calculation algorithm}

One of the biggest challenges of the \gls{P2P} architecture is to identify which peers are superior to others. This issue is mostly found in those architectures in which the purpose of the network is to transfer data based on user reputation\footnote{Meaning the amount of data shared and uploaded to other peers.}. Since all nodes are consider peers containing unique data, OpenWeather does not make distinction amount them.

Even so, for practical reasons, it is necessary to develop an algorithm to calculate which peers are better than others in terms of connectivity and bandwidth availability, to provide a list to the nodes to guarantee the connection to OpenWeather network. 

The author considers that due to the nature of the data and the main factor of its importance is availability. Thus, the algorithm shall be a node bandwidth, network latency and geographical location.

Bandwidth and latency are two obvious and common used factors in other \gls{P2P} architectures. However in this case is important to note that most of these nodes are going to have better network visibility with nodes are located in proximity. The geographical location of the node, available in the MetaInfo data field through the "Location" data field, can be used to calculate the closest peers.

The algorithm to calculate the best peers to keep on the internal list, is too a vast and complex topic to be analyzed in this thesis. In the prototype created, the author used random peers in order to verify the protocol specifications. It is necessary to highlight that the peer list calculation must be analyzed deeply in order to implement OpenWeather in production scenario.

\subsubsection{Node identification}

In section \ref{5.2.1} is mentioned that a node \textbf{has a unique ID}. This ID is used to identify the node and at the same time by the user/software to recognize which node is currently active. The value of this ID is based on \gls{SHA}-256\cite{SHA}. Nevertheless, the length of it and its alphanumeric composition make it really difficult to remember the node ID, even when using some mnemonic techniques. However, it can be easily fixed with a proper algorithm, based on a standardized \gls{AWS} system for identification and use of the \gls{RFC} 3986 \emph{\gls{URI}: Generic Syntax} \cite{rfc3986}. 

As example, the \gls{CWOP} uses different parameters\cite{CWOPID} to identify the \gls{AWS}; some of them are:

\begin{itemize}

\item Block number 2 digits representing the WMO-assigned block

\item Station number 3 digits representing the WMO-assigned station

\item Place name: common name of station location

\item Country name: country name is \gls{ISO} short English form

\end{itemize}

The block number refers to the geographical region\footnote{Extracted from station index numbers database, \gls{CWOP} Meteorological Station Location Information \cite{CWOPID}.} of the \gls{AWS}, and the station number is assigned base on \emph{the nearest 10 degree meridian which is numerically lower than the station longitude}\cite{CWOPID}. The place name and country name are values assigned based on the geographical location of the \gls{AWS}. Although \gls{CWOP} also provides the latitude and the longitude, their introduction in \gls{URL} generation, will cause greater complexity.

\begin{table}[H]
\centering
02;974;EFHK;Helsinki-Vantaa;;Finland;6;60-19N;024-58E;60-19N;024-58E;51;56;P
\caption{Example of \protect \gls{CWOP}'s \protect \gls{AWS} identification.}
\label{CWOPNOTA}
\end{table}

The table \ref{CWOPNOTA} shows all data used by \gls{CWOP} to identify an \gls{AWS}, the  the following syntax is used to generate the \gls{URL}:

\begin{table}[H]
\centering
\begin{tabular}{|l|l|l|p{10cm}|}
\hline
\textbf{owp}://Country Name/Place Name/Block number + Station number \\
\hline
\end{tabular}
\caption{ID partially based on \protect \gls{CWOP} notation.}
\label{t6.1}
\end{table}

Based on the data used in the table \ref{CWOPNOTA} the output will be:

\begin{table}[H]
\centering
\begin{tabular}{|l|l|l|p{10cm}|}
\hline
\textbf{owp}://finland/helsinki-vantaa/02974\\
\hline
\end{tabular}
\caption{ID's partially based in \protect \gls{CWOP}'s identification system.}
\end{table}

The scheme is denominated as \textbf{owp} (OpenWeather Protocol), the authority field is used for country name, the absolute path is based on the place name and the station number assigned by the \gls{WMO}. This combination is enough to guarantee the uniqueness of the node accessed through the \gls{URL}.

The value of the ID used in the OpenWeather data message will be the resulting hash of the data "02;974;Helsinki-Vantaa;;Finland" generated with \gls{SHA}-256.

\begin{table}[H]
\centering
\begin{tabular}{| p{15.6cm} |}\hline
\begin{minipage}[t]{\linewidth}
	\begin{verbatim}
{
    "OpenWeatherMessage": {
            ...
"ID" :"a88a9b6b4c0381e0509ce36cadb5fd06e5446ab23881020b9f212db24b16ee75",
            ...
},
      \end{verbatim}
\end{minipage} \\
\hline
\end{tabular}
\caption{IDs based in the \protect \gls{SHA}-256 result of the \protect \gls{CWOP} notation.}
\end{table}

\section{Protocol operations}

The protocol allows the following operations:

\begin{itemize}
\item Session establishment
\item Service discovery
\item Real-time data retrieval
\item Data on demand
\end{itemize}

Note that all of these operations have an implicit internal functional workflow, based on the requests and retrievals and their results. The following sections analyze the functioning of these operations.

\subsection{Session establishment - Peer handshake}\label{7.3.1}

The first operation needed for OpenWeather is session establishment. The elements involved in this operation can go from \textbf{2..n} nodes. Thus, a node can execute the operation to establish session with multiple nodes at the same time, nevertheless, the session establishment is always an isolated process between two nodes.

These nodes must offer the basic services integrated in the protocol, as peer exchange information or peers-list exchange. The session establishment between nodes is denominated \textbf{peer handshake}. At this point the nodes exchange their information in order to identify each other, sending a data message with the parameters mentioned in section 6.2.4. This operation is categorized as an internal protocol requirement, using the code \textbf{100} as type of data message.

\begin{figure}[H]
\centerline{\includegraphics[width=1\textwidth]{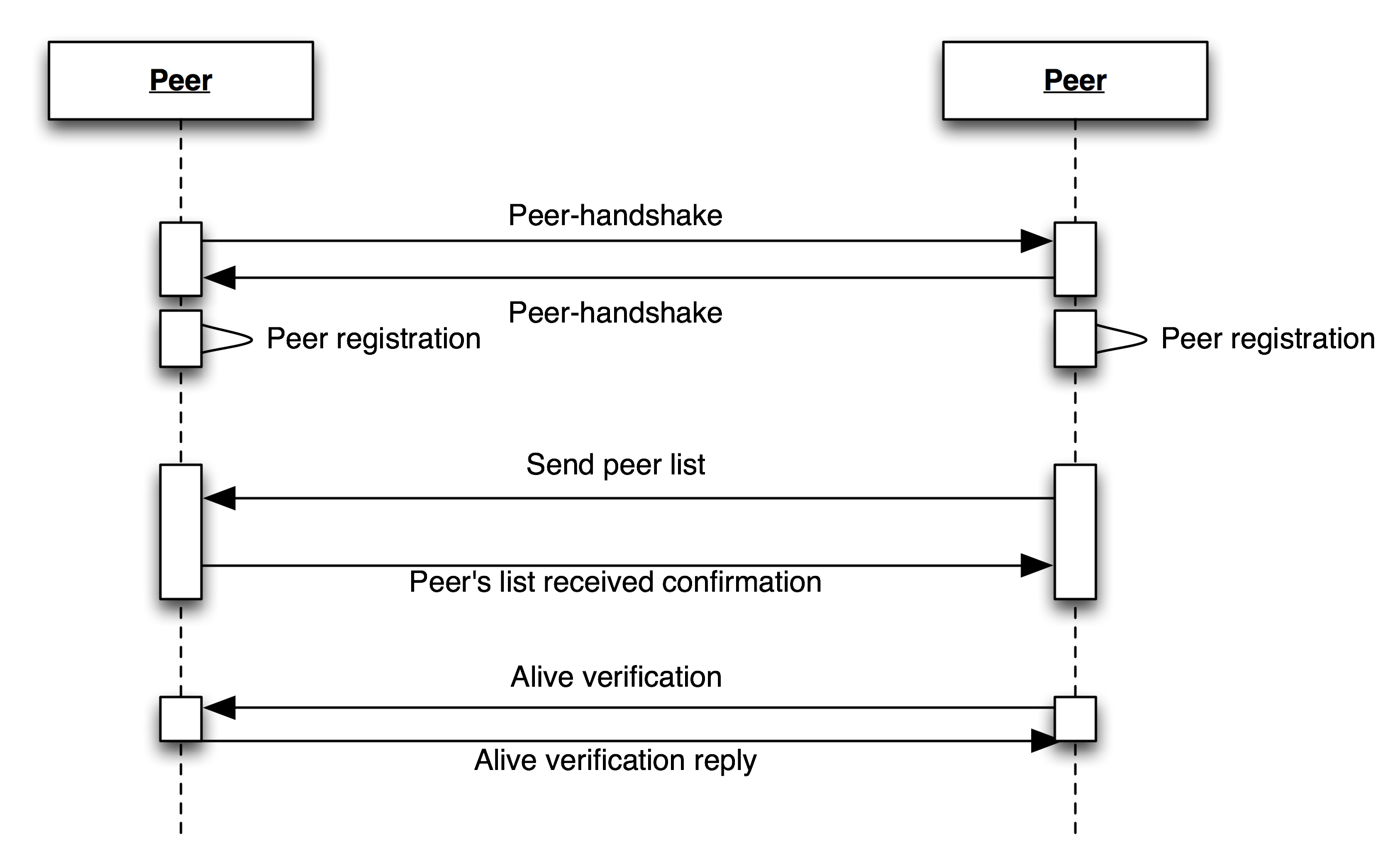}}
\caption{Session establishment sequence diagram.}
\end{figure}

When the nodes establish a session, two operations are performed

\begin{itemize}
\item	Peers-list exchange.
\item Alive verification.
\end{itemize}

The first operation —peers-list exchange— is performed in order to verify if the nodes can update their internal list of peers available.

The second operation performed is alive verification. The peers send a data message after the exchange of the peer list, in order to verify that the nodes are ready to request weather data\footnote{Note that this check is realized to ensure the availability of the node twice.}. If the alive verification is not successful, the node executing it will close the \gls{TCP} connection with the node that is not responding to it.

\subsection{Service discovery}\label{6.3.2}

OpenWeather assumes that when two nodes establish a session, the purpose of it is to exchange certain data, even if it is just for protocol requirements. As it is explained in section \ref{5.2.2}, OpenWeather is designed according to a service oriented architecture \gls{SOA}. All data sent  or receive by a node goes through services provided in the OpenWeather software implementation.

\begin{figure}[H]
\centerline{\includegraphics[width=1\textwidth]{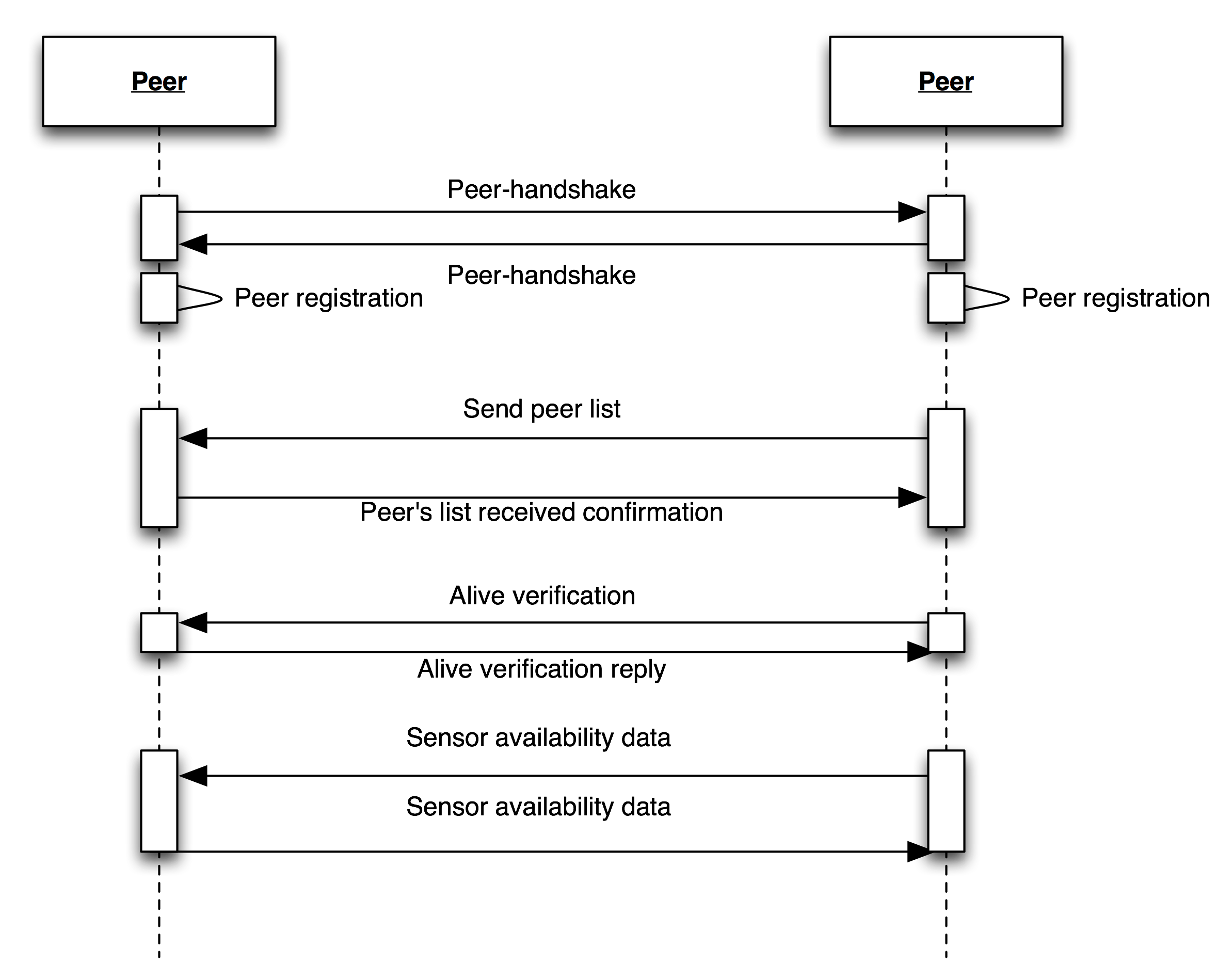}}
\caption{Service discovery sequence diagram.}
\end{figure}

The nodes involved in the session must exchange the type of data messages, in order to be aware of services available to the nodes. Note that this operation informs the nodes which sensors are available to other nodes and which kind of weather data can be retrieved from them. After the nodes communicate through the services available, other operations as real-time data retrieval or data on demand, can be performed.

\subsection{Real-time data retrieval}\label{7.3.3}

When the nodes establish the session  and service discovery operations is performed successfully, they are consider to be ready to send and receive weather data between them. As it is explained in section \ref{5.2.2}, the data can be real-time data or data on demand. In case of real time data, the node requesting it, must send a type of data message with code \textbf{200}, immediately after, the other node involved in the session, must start to delivery real-time data messages.

\begin{figure}[H]
\centerline{\includegraphics[width=1\textwidth]{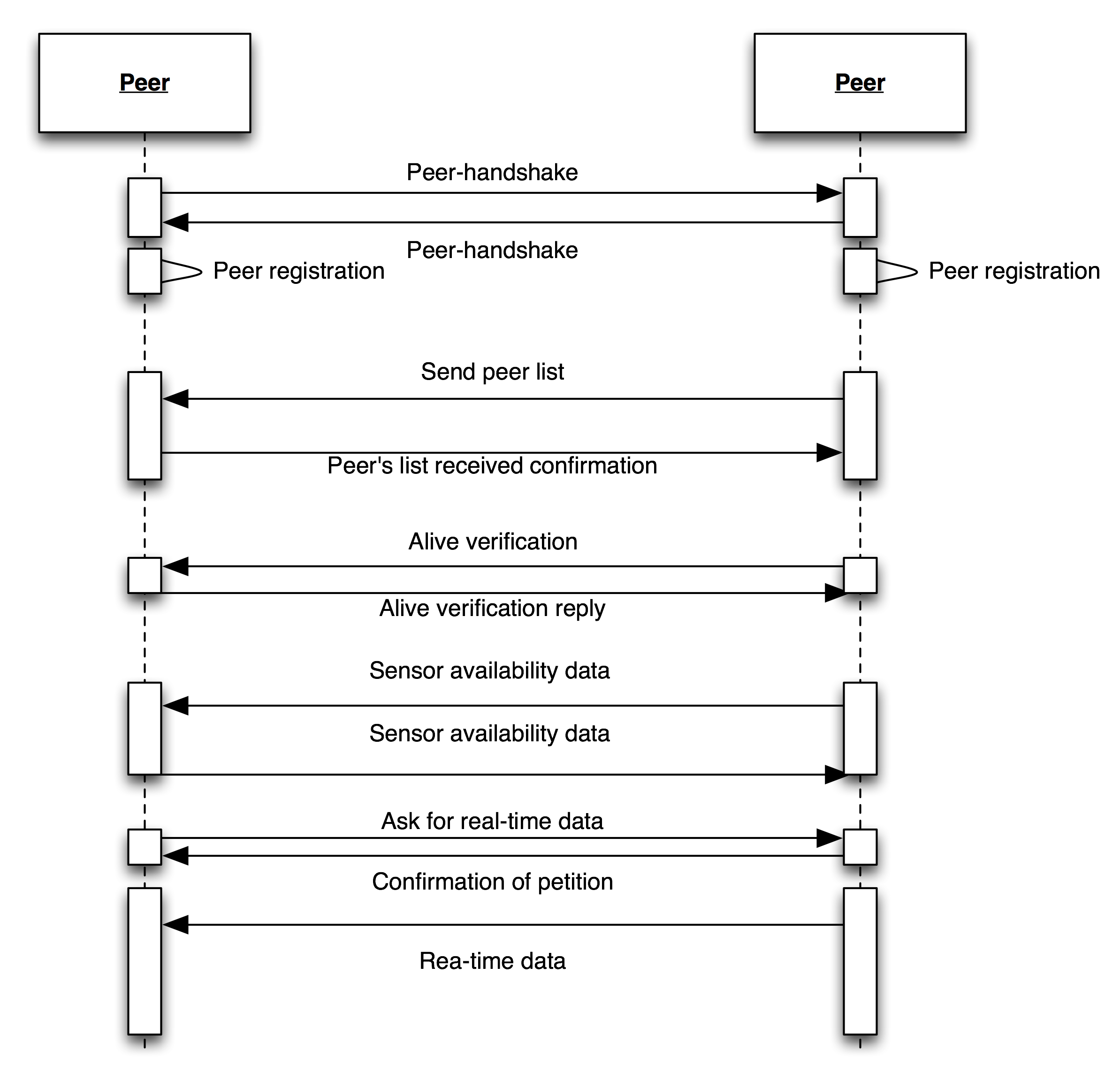}}
\caption{Real-time data sequence diagram.}
\end{figure}

The real-time data will be deliver until the node requesting it decides to stop the data stream\footnote{The data streams can be interrupted by other exceptions as connectivity or software issues.}. This data stream provides the real-time data generated in the remote \gls{AWS}. As in any other network solution the delay that the nodes can experience can affect the delivery of the data. Nevertheless, all the data messages are timestamped when the data was assembled within them. Because this timestamp is available, it will be feasible to implement an algorithm on the software side,  applying a correction factor to the timestamp based on the latency of the nodes, to fix this issue.

\subsection{Data on demand}

Apart from the the real-time data, a user can request data on demand. When a user requests data on demand, it creates individual requests with a specific timestamp. Based on these requests, the remote node will deliver an individual data message timestamped with the date and time provided, the requests and the weather data on that time. 

\begin{figure}[H]
\centerline{\includegraphics[width=1\textwidth]{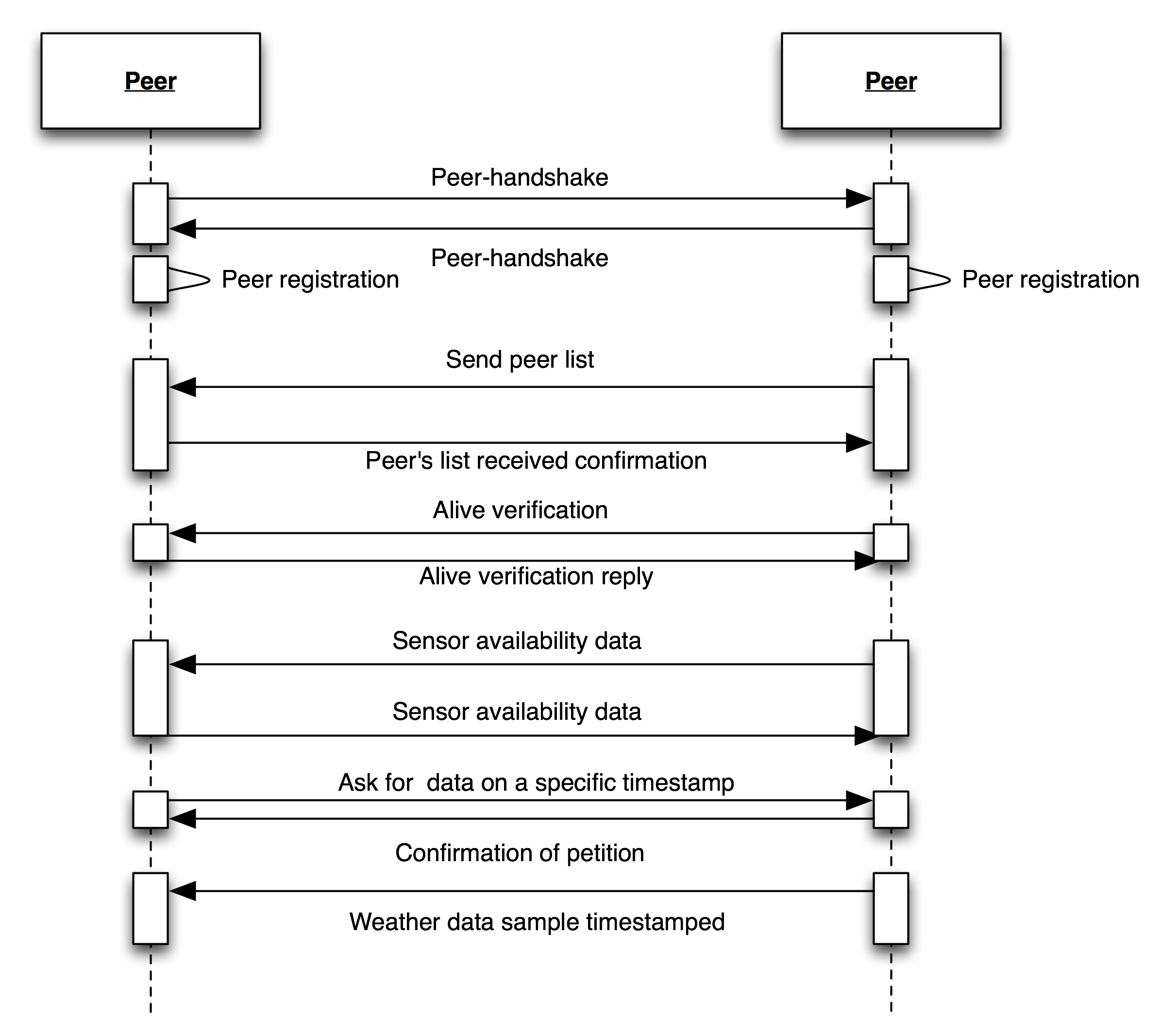}}
\caption{On demand data sequence diagram.}
\end{figure}
 
OpenWeather does not support the capability to request a range of dates or times on protocol level, meaning that it is not possible to retrieve isolated weather data samples form the node during a certain period of time. Instead, it is possible to implement on the software side the functionality to process a group of data messages with a certain timestamp. The justification of this limitation is based on the bandwidth availability in an \gls{AWS}. In contrast with individual weather data samples, a range of them can have a considerable size and this can cause significant obstacles for the \gls{AWS}: heavy \gls{CPU} load, bandwidth consumption, etc.

\section{Data messages}

A data message refers to the data transmitted using the OpenWeather protocol. A data message can contain multiple informational values, referring to weather data or data needed for protocol maintenance.

All data fields contained in an OpenWeather data message are considered to be encapsulated data represented through \gls{JSON} objects using \gls{UTF}-8\cite{UTF}\cite{rfc3629} as character encoding. According with the \gls{RFC} 4627\cite{rfc4627}, the definition of an \gls{JSON}'s object is:

  \emph{[...] An object is an unordered collection of zero or more name/value
   pairs, where a name is a string and a value is a string, number,
   boolean, null, object, or array.[...]}
 
Therefore, any data field contained in an OpenWeather data message, is an individual or group of \gls{JSON} objects or values. These objects are optimized according to the data that they contain. For instance, some data fields are \gls{JSON} objects containing other objects at the same time. The data optimization made in the protocol using these data structures, allows data encapsulation which makes enables a fast data the data processing from the network to software levels.

All OpenWeather data messages are formatted using \gls{JSON} syntax. Type of data contained in the data message is insignificant as it is structured in one \gls{JSON} object composed for different sub-objects. These objects are represented as data fields in terms of networking architecture. 

\begin{figure}[H]
\centerline{\includegraphics[width=0.6\textwidth]{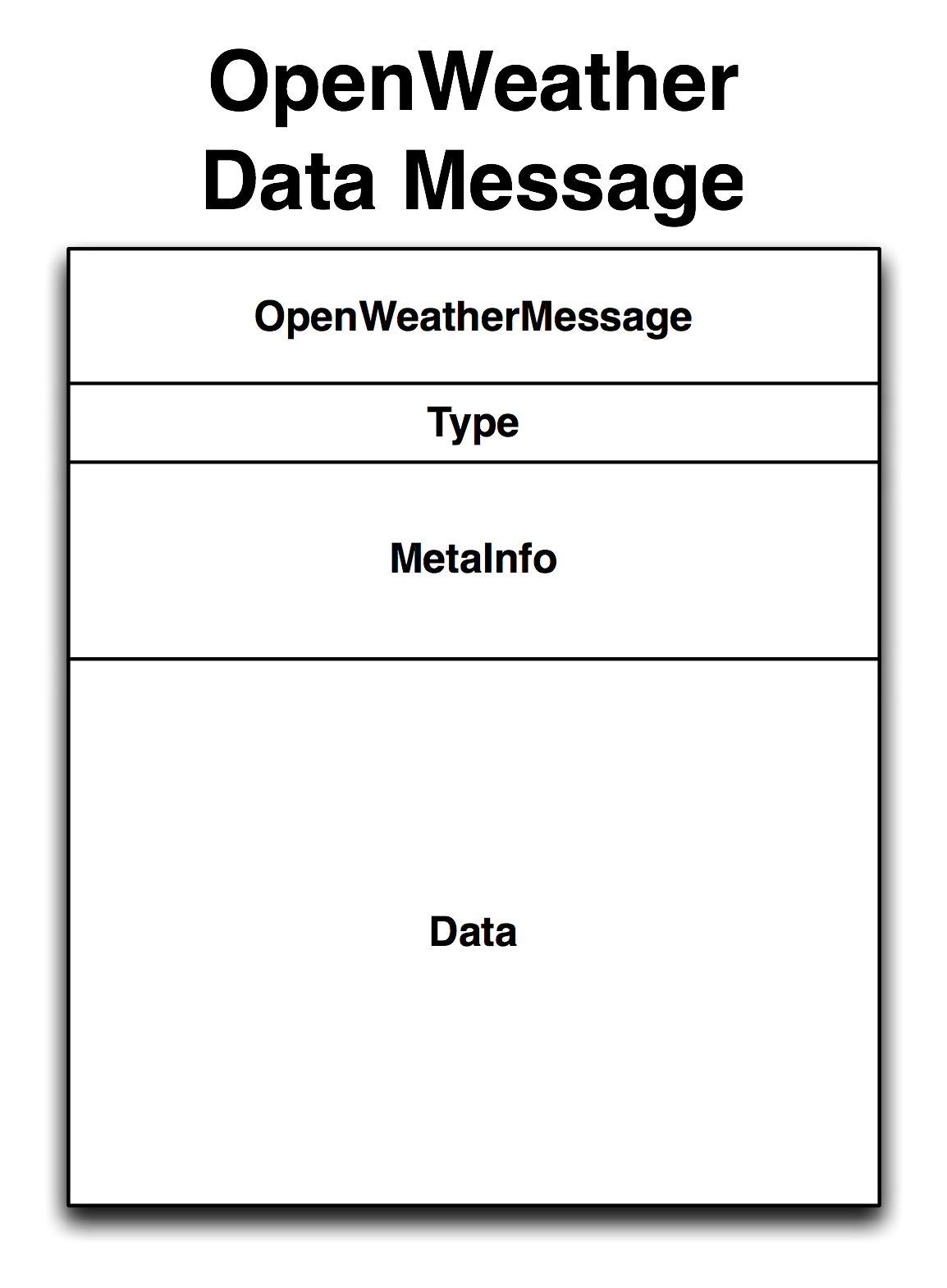}}
\caption{OpenWeather data message structure.}
\end{figure}

The parent object is denominated \textbf{OpenWeatherMessage}; this object \textbf{is present in all the data messages} inside of OpenWeather network. This parent contains two sub-objects; the MetaInfo object and the Data object or Info object. The MetaInfo object is a data field acting as \textbf{the header of the data message in OpenWeather protocol}\footnote{This object is added an individual data field named "Type", explained in the next section.}. Furthermore an OpenWeather data message contains the Data object or the Info object. The Data object is a data field containing all data related to the weather data that the data message transports. The Info object contains the information used internally by the OpenWeather protocol.

\subsection{Header}

OpenWeather uses a fixed\footnote{In terms of data fields provided.} header data field in all the data messages, in order to guarantee its functioning. The function of this header is to provide all necessary data parameters needed by the OpenWeather protocol in every data message. Though it requires some data repetition, its insignificant size of this header, compensates the disadvantages of its repetition during transmission.

Table \ref{headertable} shows the fields contained in the header:

\begin{table}[H]
\centering
\begin{tabular}{|l|l|l|p{8cm}|}
\hline
\begin{minipage}[t]{\linewidth}
	\begin{verbatim}
{
    "OpenWeatherMessage": {
        "Type" : "",
        "MetaInfo" : {
            "ID" : "", 
            "Peer-IP": "", 
            "Port": "",    
            "Location": "", 
            "Update-Interval": "", 
            "Peers-request":"", 
            "Keep-Alive":"", 
            "Bandwidth": "",   
            "Timestamp" : "", 
            "Version" : "",
        },
      \end{verbatim}
\end{minipage} \\
\hline
\end{tabular}
\label{headertable}
\caption{Header field (Header object) in a data message of OpenWeather.}
\end{table}

As the table \ref{headertable} exposes all data messages start with the term ''OpenWeatherMessage'', building \gls{JSON} parent object of the data message. Any data contained within the data message will belong to this parent object. Although this hierarchy does not impact the data message size, it provides significant assistance to the post processing of the data on the software side. This design is inspired by the same concepts use in \gls{XML} and \gls{XML} Schemas \cite{XML}, concerning the metadata fields. Nevertheless, OpenWeather does not providing any extra fields for metadata definition, meaning that the software utilizing OpenWeather, should recognize the expected format beforehand\footnote{\gls{XML} allows data type provision in the data itself. However, this practice increases the size of the data considerably.}. With this practice speed up and simplifies the parsing compare to \gls{XML}.

\subsection{Types of data messages}\label{types}

The second field contained in an OpenWeather's message is denominated \textbf{Type}. This field indicates which type of data is located within a data message through a numerical code and if it is related with weather data, peers exchange, protocol itself, etc.

Depending on the type of data message it will be in one of the following categories:

\begin{itemize}
\item Data messages for protocol maintenance only.
\item Data messages use to transport weather data only.
	\begin{itemize}
		\item Real-time data.
		\item Data on demand.
	\end{itemize}
\end{itemize}

\subsection{Protocol codes}

The "Type" field can contain a numerical value from \textbf{1..n}. This numerical value is known as the protocol code associated with the type of messages. The codes used are divided in categories and subcategories:

\begin{itemize}
\item Codes assigned to data messages used for protocol maintenance.
	\begin{itemize}
		\item {Protocol codes (From: 1..1xx)}
		\begin{itemize}
			\item {Requests}
			\item {Retrievals}
			\item {Status}
				\begin{itemize}
					\item {Success}
					\item {Error}
				\end{itemize}
		\end{itemize}
	\end{itemize}
	\end{itemize}
\begin{itemize}
\item Codes assigned to data messages for weather data exchange between peers:
	\begin{itemize}
		\item {Peer codes}
		\begin{itemize}
			\item {Requests}
				\begin{itemize}
					\item {Real-time data}: 200
					\item {Data on demand}: 201
				\end{itemize}
			\item {Retrievals}
					\begin{itemize}
					\item {Real-time data}: 300
					\item {Data on demand}: 301
				\end{itemize}
			\item {Status}
				\begin{itemize}
					\item {Success}: 500..599
					\item {Error}: 600..699
				\end{itemize}
		\end{itemize}
\end{itemize}
\end{itemize}

The numerical value is used by the software in order to recognize the data processing procedure.

All the protocol codes used in the prototype are available \textbf{in the appendix}.

\subsection{MetaInfo data field}\label{header}

The MetaInfo data field (MetaInfo \gls{JSON} object) defines fixed data fields transmitted in every data message. The purpose of these fields is to provide all information needed, in order to identify the peer's ID, its geographical location, \gls{IP} address, among other data. The use of this data throughout all data messages makes allows for easier implementation and extensibility of the \gls{P2P} architecture,  as it enables the software to be aware properties and status of a specific peer at all times.

The MetaInfo field contains the following data fields:

\begin{itemize}
\item Bandwidth
\item ID
\item Keep-Alive
\item Location
\item Peer-IP 
\item Peers-Request
\item Port
\item Timestamp 
\item Update-Interval
\item Version
\end{itemize}

The MetaInfo data field is structure as a \gls{JSON} object containing an array of elements\footnote{\textbf{Note:} in the following figures the expression "ARRAY DATA ELEMENTS" is used to refer the MetaInfo data fields.}. These elements are the fields mentioned above. Every element does reference to an specific parameter needed by OpenWeather protocol.

\begin{table}[H]
\centering
\begin{tabular}{|l|l|l|p{10cm}|}
\hline
\begin{minipage}[t]{\linewidth}
	\begin{verbatim}
{
"OpenWeatherMessage": {
        "Type" : 1,
        "MetaInfo" : {
	              ARRAY DATA ELEMENTS
        },
},
      \end{verbatim}
\end{minipage} \\
\hline
\end{tabular}
\caption{MetaInfo field in a data message of OpenWeather protocol.}
\end{table}

\begin{figure}[H]
\centerline{\includegraphics[width=0.5\textwidth]{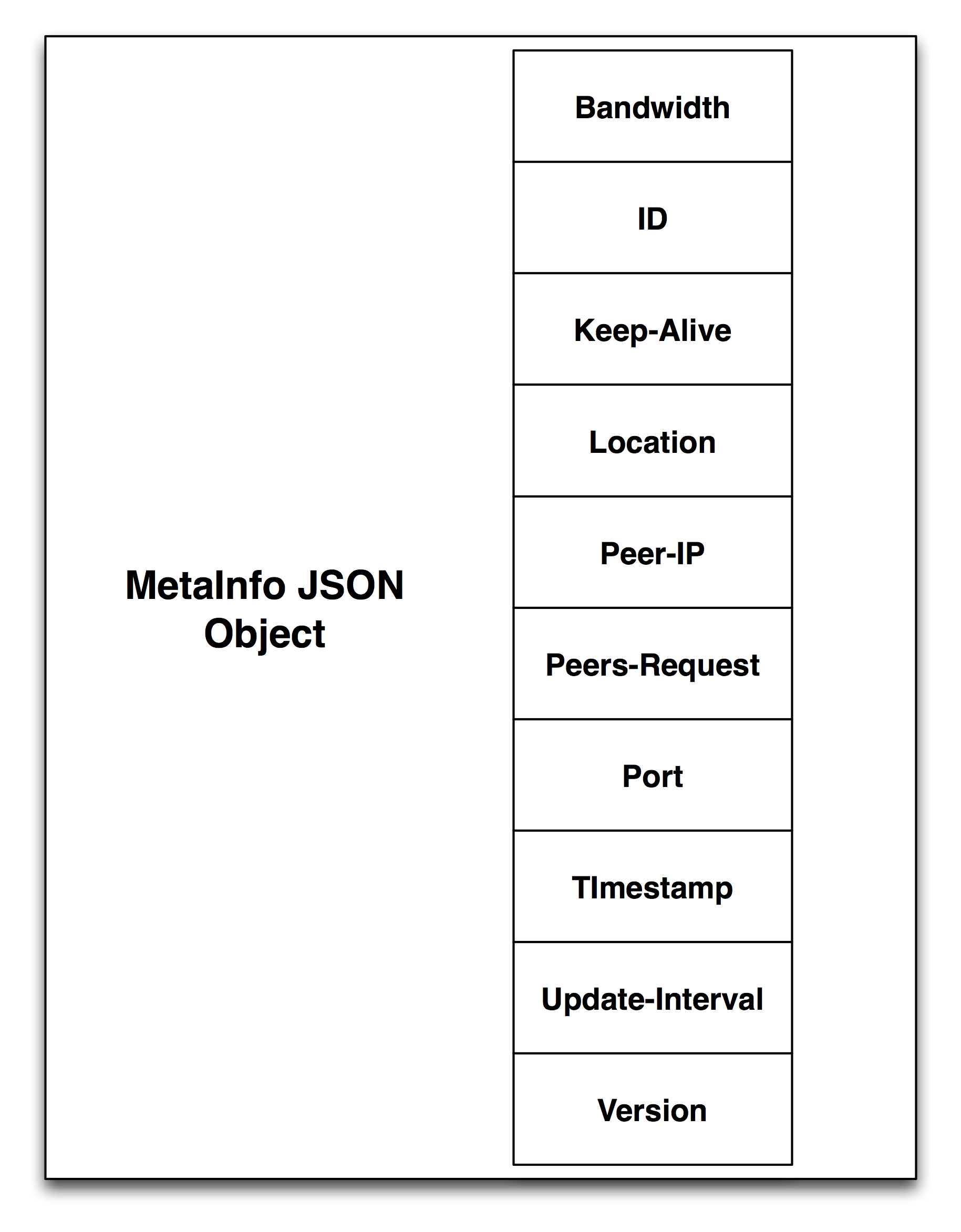}}
\caption{OpenWeather MetaInfo data field with data array elements.}
\end{figure}

\subsubsection{Bandwidth}

As any other network oriented software, the amount of bandwidth is a critical factor in its proper functionality. Most software solutions using \gls{P2P} architecture offer a dedicated section to control the bandwidth parameters. OpenWeather informs others nodes of the amount of bandwidth that a node has available while giving full control of the amount of bandwidth and connections and remote connections allow. As opposed to mainstream solutions, in which the node is only controls the amount of connections and bandwidth locally, the bandwidth control in OpenWeather can be managed both locally and remotely. To achieve this, the field "Bandwidth" is provided in every data message, informing the nodes what is the capacity of the node whereby are operating.

\begin{table}[H]
\centering
\begin{tabular}{|l|l|l|p{8cm}|}
\hline
\begin{minipage}[t]{\linewidth}
	\begin{verbatim}
{
    "OpenWeatherMessage": {
            ...
            "Bandwidth" : "4", // Correspondency 1 Megabit/s
            ...
},
      \end{verbatim}
\end{minipage} \\
\hline
\end{tabular}
\caption{Bandwidth field in a data message of OpenWeather}
\end{table}

The user must provide this parameter to configure its node.  Due to the a big amount of possibilities for bandwidth quality, this data field contains a numeric value that should be translated by the software to bits per second. Nevertheless, if the user considers that its bandwidth does not fit in the categories provided, it is possible to provide an integer number that will be translated by the software to bits per seconds. Thus, if the "Bandwidth" data field contains a numeric value higher than 6, the value will be translated for the software to bits per second. This feature allows the user to use a custom parameter.

\begin{table}[h]
\centering
\begin{tabular}{ | c | l |}
\hline
\textbf{Numeric value} & \textbf{Bandwidth equivalency} \\ \hline
0 & 56 \gls{KBITS}/s \\ \hline
1 & 128 \gls{KBITS}/s \\ \hline
2 & 256 \gls{KBITS}/s \\ \hline
3 & 512 \gls{KBITS}/s \\ \hline
4 & 1 \gls{MBITS}/s\\ \hline
5 & 10 \gls{MBITS}/s \\ \hline
6 & 100 \gls{MBITS}/s \\ \hline
\end{tabular}
\caption{Bandwidths equivalency in Bandwidth data field.}
\end{table}

\subsubsection{ID}

As explained in sections \ref{5.2.1}, every peer has an unique ID throughout OpenWeather's network. In fact, its properties make it theoretically unique in the world.The ID is generated based in the \gls{AWS} identification. The ID data field is thought to be representation of the \gls{AWS}, such representation is the result of the hash applied over some identification system for \gls{AWS}es\footnote{Several weather organizations provide this identification.}.

If the \gls{AWS} is not part of some identification system, its ID can be generated randomly by the software, however is highly recommended to provide an ID assigned for some organization as the \gls{CWOP} or \gls{NOAA}.\footnote{In the evaluation setup, the author uses a random ID.}.

\begin{table}[H]     
\centering
\begin{tabular}{| p{15.6cm} |}
\hline
\begin{minipage}[t]{\linewidth}
	\begin{verbatim}
{
    "OpenWeatherMessage": {
            ...
"ID" :"4f9a67e8496d69b8707858576ec12b8aa3fa5519c23a79ea071dc7dbc0c9b2e3",
            ...
},
      \end{verbatim}
\end{minipage} \\
\hline
\end{tabular}
\caption{ID's field in a data message of OpenWeather protocol.}
\end{table}

\subsubsection{Keep-Alive}

Due to possible node connection instability, it is necessary to implement a mechanism to identify the current connection status with a specific node is, on the application layer level. OpenWeather implements the field "Keep-Alive". This field provides the amount of time that the software must wait until the connection is close.

\begin{table}[H]
\centering
\begin{tabular}{|l|l|l|p{8cm}|}
\hline
\begin{minipage}[t]{\linewidth}
	\begin{verbatim}
{
    "OpenWeatherMessage": {
            ...
            "Keep-Alive" : "120000",
            ...
},
      \end{verbatim}
\end{minipage} \\
\hline
\end{tabular}
\caption{Keep-Alive field in a data messages of OpenWeather protocol.}
\end{table}

When a node stops sending data to other node/s, the connection will be closed when the sum of the timestamps of the last data messages received and the "Keep-Alive" value, is less than the current date and time.

The protocol assumes that if the node is not delivering data, is not useful to keep a connection with it. The same principle is applied in a number of network oriented software solutions. The value of this field is expressed in milliseconds, and by the default has a timeout of  \textbf{120000 milliseconds} (2 minutes). Though possible, the customization this parameter is not recommended, as it assumes responsibility between nodes when necessary.

\subsubsection{Location}

The "Location" field does reference to the geographical coordinates of the node, expressed in the \gls{UTM} system. This data field has two different functions:

\begin{itemize}
\item Identify the geographical location of the node.\footnote{Mandatory due to the nature of the data.}
\item Provide identificational information to other peers, does providing them with the updated information which store in the node's internal list. \footnote{This is explained deeper in section 7.2.}
\end{itemize}

\begin{table}[H]
\centering
\begin{tabular}{|l|l|l|p{8cm}|}
\hline
\begin{minipage}[t]{\linewidth}
	\begin{verbatim}
{
    "OpenWeatherMessage": {
            ...
            "Location" : "4597807 269999 30T",
            ...
},
      \end{verbatim}
\end{minipage} \\
\hline
\end{tabular}
\caption{Location field in a data messages of OpenWeather protocol.}
\end{table}

This field should be filled manually by the user. It is highly recommended to provide this parameter with as much accuracy as possible.

\subsubsection{Peer's IP address \& port}

The MetaInfo's field contains two fields dedicated to \gls{TCP}/\gls{IP}:
\begin{itemize}
\item Peer-IP
\item Port
\end{itemize}

The field "Peer-IP" contains the \textbf{public} \gls{IP} address assigned to the computer's network interface that is running the software supporting OpenWeather's protocol.This field can be an \gls{IP} address using 32-bit number (\gls{IP}v4) or 128 bit number (\gls{IP}v6). The introduction of this field is based on the requirement of the protocol to possess an updated address of the peer in order to able to connect to it. Though the field is labeled as "Peer-IP" not necessarily must be the numeric address. It is possible to implement the OpenWeather protocol to use hostname resolution based on \gls{DNS} requests\footnote{However as it is implicit in the use of \gls{DNS}, it will be required to have the hostname of the peers recorded in the name servers.}, with a few modifications on the software's side.

The field "Port" contains the port used in the \gls{TCP} to establish a connection with the peer. The default \gls{TCP} port number is \textbf{62535}\footnote{Port number choose according with the range of ports available for dynamic and/or private use published by IANA\cite{IANA}.}\footnote{We assume fixed ports and port forwarding techniques for this. The functioning of OpenWeather behind firewalls or/and \gls{NAT} is out of the scope of this thesis.}nevertheless any port can be used inside of \gls{TCP}'s range always that it does not conflict with other ports.

\begin{table}[H]
\centering
\begin{tabular}{|l|l|l|p{8cm}|}
\hline
\begin{minipage}[t]{\linewidth}
	\begin{verbatim}
{
    "OpenWeatherMessage": {
            ...
            "Peer-IP" : "140.186.70.148",
            "Port": "62535",
            ...
},
      \end{verbatim}
\end{minipage} \\
\hline
\end{tabular}
\caption{Peer-IP \& Port fields in a data message of OpenWeather protocol.}
\end{table}

Both fields, "Peer-IP" and "Port", are present in others \gls{P2P} architectures\footnote{Often denominated with different terms and syntax.}, the reason for this is that these fields facilitate a significant part of the software implementation and the network functionality. Adding these fields to all data messages, enables the software to keep the peer list updated and working between nodes and at the same time it facilitates the protocol session establishment.

\subsubsection{Peers-Requested}

The "Peers-Requested" field provides the number of peers that the node requests to other nodes in order to fill its internal list of peers. In a \gls{P2P} architecture it is critical to keep an updated list of peers to guarantee successful delivery of the data throughout the network. By default this field is set to \textbf{20}, with a possible range of \textbf{1..100}.

\begin{table}[H]
\centering
\begin{tabular}{|l|l|l|p{8cm}|}
\hline
\begin{minipage}[t]{\linewidth}
	\begin{verbatim}
{
    "OpenWeatherMessage": {
            ...
            "Peers-Requested" : "20",
            ...
},
      \end{verbatim}
\end{minipage} \\
\hline
\end{tabular}
\caption{Peers-Requested field in a data messages of OpenWeather protocol.}
\end{table}

\subsubsection{Timestamp}

As explained in chapter two, the success of weather prediction depends on different factors. One of the most important variables are the geographical location and the time and date, in which the weather data samples were collected. OpenWeather provides the field "Timestamp" to supply a solution for this condition. Every data message contains the timestamp in which the data was assembled. This provides a feasible mechanism to know when the weather data sample by the data message received was collected.

The data format used by OpenWeather protocol follows the \gls{RFC} 3339 (Date and Time on the Internet: Timestamps)\cite{rfc3339} and it follows the guidelines established by the \gls{ISO} 8601:2004\cite{ISO8601} as well. All data messages are timestamped using the \gls{UTC}.\footnote{The conversion to the original timezone of the data message can be managed through software.}

\begin{table}[H]
\centering
\begin{tabular}{|l|l|l|p{8cm}|}
\hline
\begin{minipage}[t]{\linewidth}
	\begin{verbatim}
{
    "OpenWeatherMessage": {
            ...
            "Timestamp" : "2011-05-29T12:10:23Z",
            ...
},
      \end{verbatim}
\end{minipage} \\
\hline
\end{tabular}
\caption{Timestamp field in a data message of OpenWeather.}
\end{table}

Note that OpenWeather protocol does not use the timestamp value for any purpose related with protocol operations. This Timestamp field is provided in order to fit the requirements of the weather data. Because the weather data requires precise stamping of the time in which it was acquire, this field is introduced. In addition, as in other real-time data systems, it is recommended to sync the time of the node using protocol such as \gls{NTP}, to guarantee the quality of the data. Such synchronization must be managed independently of OpenWeather.

\subsubsection{Update-Interval}

The "Update-Interval" field contains the time value, expressed in \textbf{milliseconds}, that other peers should wait before to requesting protocol information. This field can be used to manage data availability provided absence of network congestion.

\begin{table}[H]
\centering
\begin{tabular}{|l|l|l|p{8cm}|}
\hline
\begin{minipage}[t]{\linewidth}
	\begin{verbatim}
{
    "OpenWeatherMessage": {
            ...
            "Update-Interval" : "120000",
            ...
},
      \end{verbatim}
\end{minipage} \\
\hline
\end{tabular}
\caption{Update-Interval field in a data messages of OpenWeather protocol.}
\end{table}

By default this field is set to \textbf{120000 milliseconds} (2 minutes), however this parameter that can be customize by the user.

\subsubsection{Protocol versioning}

Following the same principles as \gls{HTTP} and other protocols, OpenWeather uses \textbf{<major>.<minor>} numbering scheme to indicate the versions of the protocol. The versioning is indicated in the "Version" field of the data header, adding the term ''OpenWeather''  and the character '/' before the numbering. 

\begin{table}[H]
\centering
\begin{tabular}{|l|l|l|p{8cm}|}
\hline
\begin{minipage}[t]{\linewidth}
	\begin{verbatim}
{
    "OpenWeatherMessage": {
            ...
            "Version" : "OpenWeather/1.0",
            ...
},
      \end{verbatim}
\end{minipage} \\
\hline
\end{tabular}
\caption{Version field in a data message of OpenWeather.}
\end{table}

\subsubsection{MetaInfo data field summary}

The table \ref{t6.13} shows the structure of the MetaInfo data field with all array elements already filled in with data:

\begin{table}[H]
\centering
\begin{tabular}{| p{15.6cm} |}
\hline
\begin{minipage}[t]{\linewidth}
	\begin{verbatim}
{
    "OpenWeatherMessage": {
        "Type" : 1,
        "MetaInfo" : {
"ID" :"4f9a67e8496d69b8707858576ec12b8aa3fa5519c23a79ea071dc7dbc0c9b2e3",
            "Peer-IP" : "140.186.70.148",
            "Port": "62535",
            "Location" : "4597807 269999 30T",
            "Update-Interval" : "120000",
            "Peers-Request" : "20",
            "Keep-Alive" : "120000",
            "Bandwidth" : "4",
            "Timestamp" : "2011-05-29T12:10:23Z",
            "Version" : "OpenWeather/1.0",
            },
            ...
}
\end{verbatim}
\end{minipage} \\
\hline
\end{tabular}
\caption{MetaInfo data field (MetaInfo object) in a data message of OpenWeather.}
\label{t6.13}
\end{table}

All the OpenWeather data messages will contain a header as the shown in the table \ref{t6.13}, fill in with the particular data of the node.

\subsection{Data field}

As part of the MetaInfo data field (MetaInfo object), OpenWeather data messages can contain a field named Data (Data object). 
This data field is a \gls{JSON} object composed from different sub-objects. The values or sub-objects having this object as a parent, are dedicated to transport weather data.

The Data field is necessary in order to complement the MetaInfo data field. The MetaInfo data field only provides information about the node itself. The data field contains the data that the node retrieves or request from others nodes. The type of data available in this data field can be:

\begin{itemize}
\item Real-time weather data
\item Data requested/delivery in demand (non real-time)
\end{itemize}

\begin{table}[H]
\centering
\begin{tabular}{|l|l|l|p{10cm}|}
\hline
\begin{minipage}[t]{\linewidth}
	\begin{verbatim}
{
"OpenWeatherMessage": {
        "Type" : 1,
        "MetaInfo" : {
	              ARRAY DATA ELEMENTS
        },
        "Data" : {
        	              ARRAY DATA  OBJECTS
        }
},
      \end{verbatim}
\end{minipage} \\
\hline
\end{tabular}
\caption{Data field in a data message of OpenWeather protocol.}
\end{table}

All phenomena data transmitted in OpenWeather uses the data units, specify in the \emph{Guide to Meteorological Instruments and Methods of Observation}\cite{GMIMO}, published by the \gls{WMO}\cite{WMO}. The author assumes that the protocol must follow this standard, because it is adopted by the major number of countries\footnote{Exceptions: United States, Liberia and Myanmar (Burma).}. Though some countries still keep local units for measurements, OpenWeather protocol does not take in consideration these use cases, nevertheless the implementation of the conversion between units, can easily be done on the software side.
 
All values or sub-objects containing information about weather data \textbf{will always have the Data object as a parent..}
The following sections explain how these different types of data are assembled in OpenWeather.

\subsubsection{Real-time data messages}

The section \ref{header} introduced the persistent data provided in every data message of OpenWeather. However, this data is provided in order to guarantee the protocol's functioning. A part of the header field, the data messages can contain weather data. This section explains how a real-time message is assembled. Note that the prototype used in the experimental setup \textbf{only supports} data extracted from the following phenomena:

\begin{itemize}
\item \textbf{PTU -Pressure, Temperature, Humidity}
	\begin{itemize}
		\item Air temperature
		\item Relative humidity
		\item Air pressure
\end{itemize}

\item \textbf{Wind}
	\begin{itemize}
		\item Direction (minimum, average, maximum)
		\item Speed (minimum, average, maximum)
	\end{itemize}
	
\item \textbf{Precipitation}
	\begin{itemize}
		\item Rain (accumulation, duration, intensity, peak)
		\item Hail (accumulation, duration, intensity, peak)
	\end{itemize}
\end{itemize}

These data have been chosen because it is available in most of the \gls{AWS} of semi-professional / end-user range. In addition, the data used in OpenWeather provides a functional prototype adapted to this thesis. The author highlights that none of these data fields (concerning weather data) are used claiming them to be a standard or a suggestion of it. As mentioned in section \ref{missingstd}, only a process of standardization can provide the correct data fields to use. Nevertheless, the use of these weather data fields is enough to develop a prototype.

Note that some data objects contain values in the data fields such as "minimum", "maximum" or "accumulation" among others, that are representing data collected in time intervals. Depending of the phenomenon these time intervals can be completely different. The recommend intervals of measurement are described in the \emph{"Guide to Meteorological Instruments and Methods of Observation"} \cite{GMIMO}, and theoretically they must be always the same independently of the brand of the weather instrument used.

\subsubsection{Pressure, temperature and humidity data}

The \gls{PTU}, are the most common data available in an \gls{AWS}, due to the close relation between the phenomena and the ease of its acquirability. Any modern \gls{AWS} will is equipped with necessary sensors to measure these phenomena.

The \gls{AWS}es collect this data in real-time, transforming the raw input data from the sensors into digital data. The workflow of this data is described in section \ref{3.1.4}. As other data in OpenWeather, it will be normalized in the layer implemented between the hardware layer and OpenWeather\footnote{Explained in section \ref{5.2}.}.

\begin{table}[hc]
\centering
\begin{tabular}{ | l | l | l | l |}
\hline    
0r2,Ta=18.7C,Ua=77.4P,Pa=1002.1H\\
\hline
\end{tabular}
\caption{\protect \gls{PTU} real-time data in the raw format used by the \protect \gls{AWS}.}
\end{table}

When the \gls{PTU} data is transformed to OpenWeather's format, it has the following format:

\begin{table}[H]
\centering
\begin{tabular}{|l|l|l|p{10cm}|}
\hline
\begin{minipage}[t]{\linewidth}
	\begin{verbatim}
{
"OpenWeatherMessage": {
        "Type" : 1,
        "MetaInfo" : {
	              ARRAY DATA ELEMENTS
        },
        "Data" : {
            "PTU" : {
                "Air-Temperature" : "", 
                "Relative-Humidity" : "", 
                "Air-Pressure":  ""
            },
},
      \end{verbatim}
\end{minipage} \\
\hline
\end{tabular}
\caption{PTU data field in a data message of OpenWeather protocol.}
\end{table}

The three data fields contained in the Data object are:

\begin{itemize}
\item \textbf{Air-Temperature}: expressed in degree Celsius (\textdegree C)
\item \textbf{Relative-Humidity}: expressed in percentage in base of relative humidity
\item \textbf{Air-Pressure}: expressed in Hectopascals (hPa)
\end{itemize}

These data fields are encapsulated as an array of data elements inside of the \gls{JSON} object \gls{PTU}. The table \ref{tptu} shows an example of the \gls{PTU} object filled with real-time data:

\begin{table}[H]
\centering
\begin{tabular}{|l|l|l|p{10cm}|}
\hline
\begin{minipage}[t]{\linewidth}
	\begin{verbatim}
        "Data" : {
            "PTU" : {
                "Air-Temperature" : "20.0", // Celsius: ºC
                "Relative-Humidity" : "59.5", // %RH 
                "Air-Pressure":  "1002.1" // Hectopascals: hPa
            },
      \end{verbatim}
\end{minipage} \\
\hline
\end{tabular}
\caption{PTU data field with real-time data in a data message of OpenWeather protocol.}
\label{tptu}
\end{table}

The frequency of reporting this data will depend on the configuration of the \gls{AWS}. Most of \gls{AWS} offer a time interval between \textbf{1 second} and \textbf{3 seconds}, to generate this data.

\subsubsection{Wind data}

The wind is other of the most popular phenomena to measure in \gls{AWS}es. The wind data contains two sub-objects: direction and speed. 
\begin{table}[H]
\centering
\begin{tabular}{|l|l|l|p{10cm}|}
\hline
\begin{minipage}[t]{\linewidth}
	\begin{verbatim}
{
"OpenWeatherMessage": {
        "Type" : 1,
        "MetaInfo" : {
	              ARRAY DATA ELEMENTS
        },
        "Data" : {
            ...
          "WIND" : {
                "Direction" : {
                    "min" : "",
                    "ave" : "",
                    "max" : ""
                },
                "Speed" : { 
                    "min" : "",
                    "ave" : "",
                    "max" : ""
                }
            },
      \end{verbatim}
\end{minipage} \\
\hline
\end{tabular}
\caption{Wind data field in a data message of OpenWeather protocol.}
\end{table}

At the same time these two objects are composed by three array data elements:

\begin{itemize}
\item \textbf{Direction}
	\begin{itemize}
	\item Minimum  (min): expressed in degrees
	\item Maximum (max): expressed in degrees
	\item Average   (avg): expressed in degrees
	\end{itemize}
\item \textbf{Speed}
	\begin{itemize}
	\item Minimum  (min): expressed in meters per second $m \over s$
	\item Maximum (max): expressed in meters per second $m \over s$
	\item Average   (avg): expressed in meters per second $m \over s$
	\end{itemize}
\end{itemize}

\begin{table}[H]
\centering
\begin{tabular}{|l|l|l|p{10cm}|}
\hline
\begin{minipage}[t]{\linewidth}
	\begin{verbatim}
{
"OpenWeatherMessage": {
        "Type" : 1,
        "MetaInfo" : {
	              ARRAY DATA ELEMENTS
        },
        "Data" : {
            ...
          "WIND" : {
                "Direction" : {
                    "min" : "217", // Degrees
                    "ave" : "217",// Degrees
                    "max" : "218"// Degrees
                },
                "Speed" : { 
                    "min" : "4.2",// m/s
                    "ave" : "4.2",// m/s
                    "max" : "4.5"// m/s
                }
            },
      \end{verbatim}
\end{minipage} \\
\hline
\end{tabular}
\caption{Wind data field with real-time in a data message of OpenWeather protocol.}
\end{table}

\subsubsection{Precipitation data}

Precipitation is the last phenomena that typically all the \gls{AWS}es measure. Inside of the concept of precipitations encompasses two different classes, rain and hail. Thus, the precipitation is structure two sub-objects containing an array of four data elements.

\begin{table}[H]
\centering
\begin{tabular}{|l|l|l|p{10cm}|}
\hline
\begin{minipage}[t]{\linewidth}
	\begin{verbatim}
{
"OpenWeatherMessage": {
        "Type" : 1,
        "MetaInfo" : {
	              ARRAY DATA ELEMENTS
        },
        "Data" : {
            ...
       "PRECIPITATION" : {
                "Rain" : {
                    "accumulation" : ""
                    "duration" : "", 
                    "intensity" : "", 
                    "peak" : "" 
                },
                "Hail" : {
                    "accumulation" : "", 
                    "duration" : "", 
                    "intensity" : "",
                    "peak" : ""
                }
            }
      \end{verbatim}
\end{minipage} \\
\hline
\end{tabular}
\caption{Precipitation data field in a data message of OpenWeather protocol.}
\end{table}

Both of them are measured with the same data fields:

\begin{itemize}
\item \textbf{Rain}
	\begin{itemize}
	\item Accumulation  (accumulation): expressed in millimeters
	\item Duration   (duration): expressed in seconds
	\item Intensity   (intensity): expressed in millimeters per hour
	\item Peak   (peak): expressed in millimeters per hour
	\end{itemize}
\item \textbf{Hail}
	\begin{itemize}
	\item Accumulation  (accumulation): expressed in hits per $cm^2$
	\item Duration   (duration): expressed in seconds
	\item Intensity   (intensity): expressed in hits per $cm^2$
	\item Peak   (peak): expressed in hits per $cm^2$
	\end{itemize}
\end{itemize}

Compared with the \gls{PTU} or wind, precipitation may be absent in the current weather. It means that the measurement of these phenomena will happen only when it is present. Despite this, OpenWeather always delivers the precipitation data field in the real-time data messages\footnote{A zero value is assigned to the data fields when the phenomena are not present.}.

\begin{table}[H]
\centering
\begin{tabular}{|l|l|l|p{10cm}|}
\hline
\begin{minipage}[t]{\linewidth}
	\begin{verbatim}
{
"OpenWeatherMessage": {
        "Type" : 1,
        "MetaInfo" : {
	              ARRAY DATA ELEMENTS
        },
        "Data" : {
            ...
       "PRECIPITATION" : {
                "Rain" : {
                    "accumulation" : "12" // mm
                    "duration" : "34", // seconds
                    "intensity" : "12", // mm/h
                    "peak" : "9" // mm/h
                },
                "Hail" : {
                    "accumulation" : "2", //hits/cm^2
                    "duration" : "78", //seconds
                    "intensity" : "1", // hits/cm^2h
                    "peak" : "1" // hits/cm^2h
                }
            }
      \end{verbatim}
\end{minipage} \\
\hline
\end{tabular}
\caption{Precipitation data field with real-time in a data message of OpenWeather protocol.}
\end{table}

\subsubsection{Data field overview}

The table \ref{dataoverview} indicates the Data field structure with all objects and their array elements filled with data:

\begin{table}[H]
\centering
\begin{tabular}{|l|l|l|p{10cm}|}
\hline
\begin{minipage}[t]{\linewidth}
	\begin{verbatim}
{
"OpenWeatherMessage": {
        "Type" : 1,
        "MetaInfo" : {
            ARRAY DATA ELEMENTS
        },
        "Data" : { 
            "PTU" : {
                "Air-Temperature" : "20.0", // Celsius: C 
                "Relative-Humidity" : "59.5", // %RH 
                "Air-Pressure": "1002.1" // Hectopascals: hPa
            },
            "WIND" : {
                "Direction" : {
                    "min" : "217", // Degrees
                    "ave" : "217",// Degrees
                    "max" : "218"// Degrees
                },
                "Speed" : { 
                    "min" : "4.2",// m/s
                    "ave" : "4.2",// m/s
                    "max" : "4.5"// m/s
                }
            },
            "PRECIPITATION" : {
                "Rain" : {
                    "accumulation" : "12" // mm
                    "duration" : "34", // seconds
                    "intensity" : "12", // mm/h
                    "peak" : "9" // mm/h
                },
                "Hail" : {
                    "accumulation" : "2", //hits/cm^2
                    "duration" : "78", //seconds
                    "intensity" : "1", // hits/cm^2h
                    "peak" : "1" // hits/cm^2h
                }
            }
        }
    }
}   
      \end{verbatim}
\end{minipage} \\
\hline
\end{tabular}
\caption{Real-time data message of OpenWeather protocol.}
\label{dataoverview}
\end{table}

\subsubsection{Data on demand}

As highlighted in section \ref{5.2.2}, OpenWeather possesses the capability to transport data on demand (not being the data generated in real-time)\footnote{This data can be stored in the \gls{AWS} itself.}. In this use case, this data is only delivered by the nodes when the user requests it. To achieve this operation, OpenWeather uses the object's hierarchy, to know which kind of data the user is requesting. The protocol encodes such data in OpenWeather data header, after that it is interpreted by the software to localize the data requested from the \gls{AWS}.

Note that the levels of hierarchy can be as deep as it is required. Nevertheless, the prototype only offers the possibility to retrieve the same data as in real-time.\footnote{Mark with a different timestamp inside of an\gls{AWS} or datalogger.}

\begin{figure}[H]
\centerline{\includegraphics[width=1\textwidth]{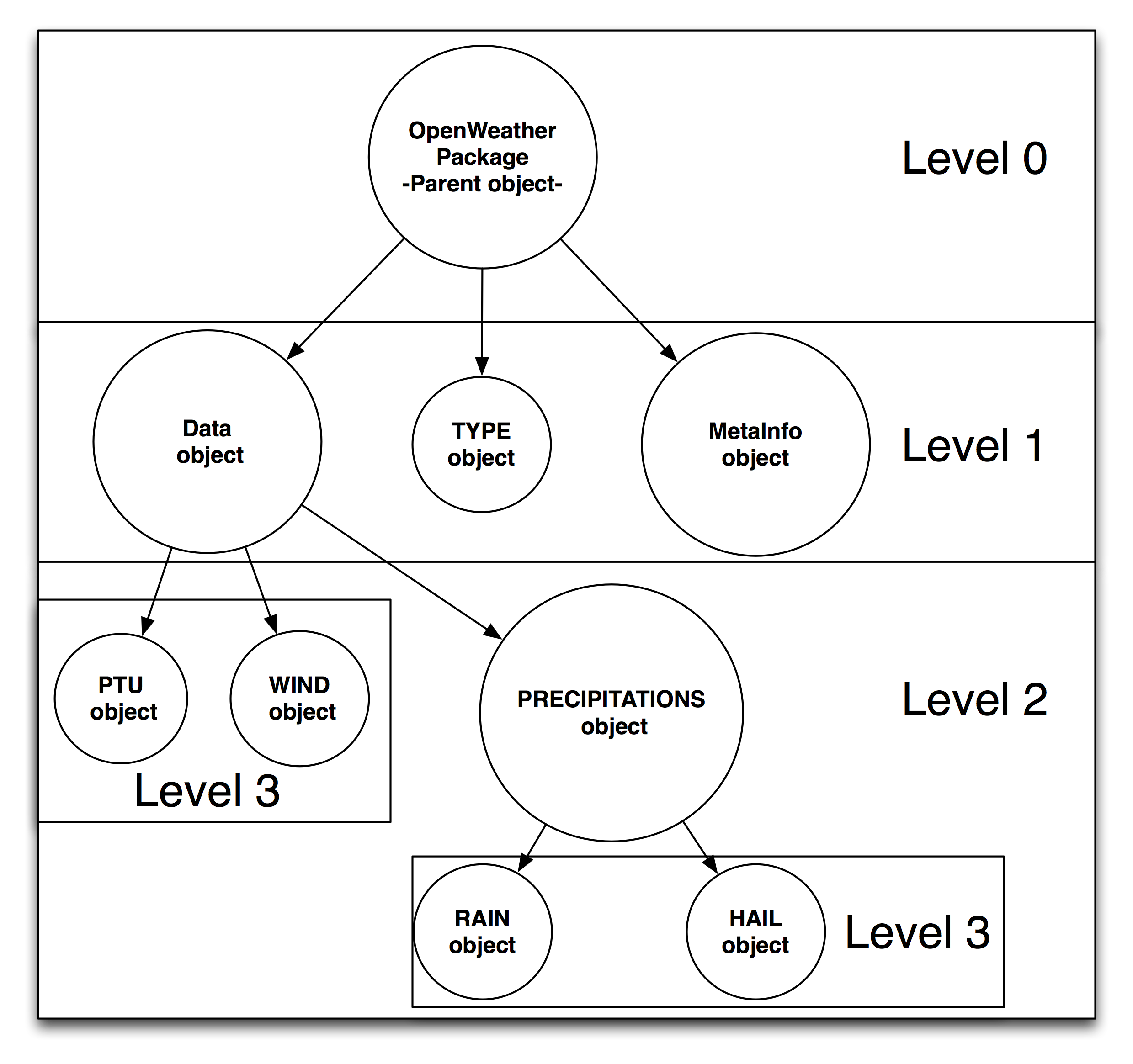}}
\caption{OpenWeather's MetaInfo data field with the data array elements.}
\end{figure}

Through the different levels established in the object's hierarchy, it is easy to find the information that the user expects.

As explained in section \ref{types}, OpenWeather uses numerical codes to identify the types of data messages. In this case the data on demand must be requested by a user (node), thus the protocol's code will be \textbf{201}\footnote{Review the protocol codes reference.}.

The data message will contain a \gls{JSON} object containing an array of data elements. The data field is named "Retrieve", it contains the data requested, indicated by the letter 'D' as a variable to be reference for data objects requested (\gls{PTU}, wind or precipitation). In addition a timestamp\footnote{This variable follows exactly the same standards used in the Timestamp field used in the MetaInfo data field.} is added to the request in order to specify in which sample is interested the user.\footnote{It is possible to change this field value in order to adapt it to request samples from a range of time.}

\begin{table}[H]
\centering
\begin{tabular}{|l|l|l|p{10cm}|}
\hline
\begin{minipage}[t]{\linewidth}
	\begin{verbatim}
{
"OpenWeatherMessage": {
        "Type" : 201,
        "MetaInfo" : {
            ARRAY DATA ELEMENTS
        },
        "Data" : {
            "Retrive" : {
                ["D":"PTU","D":"WIND","D":"PRECIPITATION"],
                "Timestamp": "2011-05-29T12:10:23Z"
            }
        }
    }
}   
      \end{verbatim}
\end{minipage} \\
\hline
\end{tabular}
\caption{Real-time data message of OpenWeather protocol.}
\end{table}

This request will return the \gls{PTU}, wind and precipitations recorded in the timestamp provided. The next data message received by the node in response of this will have exactly the same format as a real-time data message, \textbf{except the code and the timestamp in the header}; they will provide referencing to the response for the data on demand in the date and time specified.

\subsection{Internal protocol data}

As any other \gls{P2P} architecture, OpenWeather needs a certain amount of internal data to keep working. Commonly, this data is focused in peer's information as hostnames and ports used by the nodes. OpenWeather uses a mechanism to exchange list of peers between nodes, to guarantee the well-functioning of OpenWeather network. The information provided in these data messages can have different purposes. The author reserves this type of data for future implementations, nevertheless the protocol has been implemented to be able to transfer list of peers and keep updated the nodes with them.

The data messages used for this purpose are categorized as protocol dedicated, as explained in section \ref{types} these data messages can be requests, retrievals  or status information.

Opposed to weather data messages, the internal data messages do not have a Data object, but instead are composed by an Info object. This info object contains the data fields referencing the information required by the protocol.

The type of data message —code \textbf{101}—, notifies to the node that it must return a list of peers. Because this message also contains the MetaInfo object, the receiver is inform of all the information necessary to deliver the best peer list to the node in the same requests.

In the case of a list of peers, the Info object will contain a list of variables composed by an array of data elements with the \gls{IP} address of the nodes, the port and the bandwidth available in it:

\begin{table}[H]
\centering
\begin{tabular}{| p{15.6cm} |}
\hline
\begin{minipage}[t]{\linewidth}
	\begin{verbatim}
{
"OpenWeatherMessage": {
        "Type" : 101,
        "MetaInfo" : {
            ARRAY DATA ELEMENTS
        },
        "Info": {
"Peer-ID" : ["Peer-IP":"226.134.231.73","Port": "62535","Bandwidth":"2"],
"Peer-ID" : ["Peer-IP":"116.234.231.13","Port": "62535","Bandwidth":"1"],
"Peer-ID" : ["Peer-IP":"186.214.211.53","Port": "62535","Bandwidth":"5"],
"Peer-ID" : ["Peer-IP":"182.124.221.23","Port": "62535","Bandwidth":"6"],
"Peer-ID" : ["Peer-IP":"190.144.231.13","Port": "62535","Bandwidth":"1"]
        }
}
\end{verbatim}
\end{minipage} \\
\hline
\end{tabular}
\caption{Peer's list exchange in OpenWeather protocol.}
\label{tpeer}
\end{table}

The table \ref{tpeer} shows the response of the data message, providing a list of peers. Note that the "Peer-ID" will contain the unique ID of the peers. After the requester gets this data message, the software should update the internal list of peers with the new data and to deliver a status data message to the node that provides the list of peers to confirm the correct retrieval of the data.

\subsubsection{Services availability}

OpenWeather offers a mechanism to know which services are available in an \gls{AWS}. A node requesting data from these services, must send a data message with code \textbf{102}, to obtain a response with the services remotely available in the node.

\begin{table}[H]
\centering
\begin{tabular}{|l|l|l|p{10cm}|}
\hline
\begin{minipage}[t]{\linewidth}
	\begin{verbatim}
{
"OpenWeatherMessage": {
        "Type" : 102,
        "MetaInfo" : {
            ARRAY DATA ELEMENTS
        }
    }
}
\end{verbatim}
\end{minipage} \\
\hline
\end{tabular}
\caption{Services list availability request.}
\end{table}
After this data message is received by the remote node, it will reply with another data message, providing the list of the services:

\begin{table}[H]
\centering
\begin{tabular}{| p{15.6cm} |}
\hline
\begin{minipage}[t]{\linewidth}
	\begin{verbatim}
{
"OpenWeatherMessage": {
        "Type" : 101,
        "MetaInfo" : {
            ARRAY DATA ELEMENTS
        },
        "Info": {
            "Services" : { "PTU":"RO","WIND": "RO","PRECIPITATION":"RO"}
        }
}
\end{verbatim}
\end{minipage} \\
\hline
\end{tabular}
\caption{Peer's list exchange in OpenWeather protocol.}
\end{table}

One array of data is delivered in the reply:

\begin{itemize}
\item Services array: indicating the type of service and its availability.\footnote{R is equal to "real time data" and O to "data on demand". Both can be present or isolated.}
\end{itemize}

With this information the software knows which services can be checked on the remote node and which kind of data —real-time or on demand— can be retrieved from them.

\section{Protocol considerations}

The following sections describe some aspects of OpenWeather related with other protocols or future features of it.

\subsubsection{OpenWeather and other protocols}

We can find dozens of protocols available, using \gls{P2P} architectures and/or optimizations in the data delivery. Nevertheless, the author could not find any protocol suitable enough to fit in the characteristic required by the \gls{AWS}es. Protocols as Bittorrent\cite{BITORRENT}, have a proven track delivering large amount of data and scaling their networks properly. FastTrack\cite{FASTRACK} has been successful achieving similar results as Bitorrent. However, almost all the \gls{P2P} protocols are oriented to transfer files or real-time data with a big size (such as video or voice streams). In addition, these protocols are designed focusing in nodes with common computational capabilities (such as desktops or small servers), not considering embedded system inside of their purpose (being difficult to handle the necessary resources to implement these protocols on an embedded system).

Other alternatives as \gls{HTTP}, were considered by the author as solutions for this thesis. Nevertheless, \gls{HTTP} still has a big dependency of the centralized model. At the same time, \gls{HTTP} works under synchronously mode, something that will limit the real-time capabilities needed for the \gls{AWS}es.

Finally, because the use of \gls{FTP} (a generic protocol) is under use for weather instruments, the author considered much more interesting to research a custom solution for the \gls{AWS}es.

Nevertheless, several concepts have been taken from the mentioned protocols. OpenWeather uses the same philosophy as \gls{HTTP}, providing in the header of the data message all the information needed. The same approach as \gls{HTTP} has been chosen to identify the type of data messages. Through protocol codes the data message is identified in a category / purpose, being simple to extend the amount of protocol operations, just creating new identifiers through the codes. Moreover, the protocol uses \gls{JSON} as data format, being text-based as \gls{HTTP}. Concepts such node ID, peers-requested or update-interval have been taken from protocols as Bitorrent\cite{BITORRENT} or FastTrack\cite{FASTRACK} . These properties allow OpenWeather to implement methodologies tested in other \gls{P2P} networks with successful results.

\subsubsection{Aggregation of data between nodes}

As in other \gls{P2P} networks, the scalability of the OpenWeather network can be an issue. Although OpenWeather does not implement an aggregation technique between the nodes, it is ready to be adapted to it. The nodes conforming to OpenWeather protocol could require the capability to request and retrieve data using indirect paths to the end node. These paths could be found using the connections already established with other nodes.

The aggregation of the data will be executed using the same data format as common on OpenWeather protocol, thus, the data messages will use \gls{JSON} format plus the required fields in the data message to provide such functionality. The same operations of the protocol will be available through aggregation. In addition, the protocol will require the implementation of new operations for internal use.

We need to consider the nature of the weather data networks when we chose the aggregation technique. As it is described in section \ref{3.1.2}, the amount of bandwidth is commonly limited in an \gls{AWS}. Several techniques have been developed to aggregate information from different sources having in consideration connectivity and bandwidth availability issues. These techniques are classified based in how they aggregate and route the data \cite{Ogston09peer-to-peeraggregation}. 

In case that the aggregation is required in OpenWeather, it should be a combination of gossiping and tree-based methods, in order to provide a feasible way to aggregate data between nodes. The reason for this combination is that both methodologies have one specified purpose. Gossiping techniques are focused into offer robust communications, meanwhile, tree-based techniques are focused in to have better performance transferring data. Because a weather network needs to guarantee the flow of the data and at the same time the availability of the data as soon as possible, a research combining both techniques must be performed in order to find suitable solutions for such environment.

Notwithstanding, the OpenWeather specification available in this thesis provides the capability to request and retrieve the list of peers of a remote node. The combination of this list of peers and the Keep-Alive value of them, can be used to build a tree-based structure with the nodes that have a established session.
Through this tree, OpenWeather can be able to find new paths to other nodes. This will require the implementation of a internal operation of the protocol, providing the capability to make queries to other nodes, in order to build new paths. In addition, the tree-based structure will not be enough to guarantee the robustness necessary for the weather data transmission. Hence, it will be required to find the compatibility of this technique with gossiping methodologies, implementing an algorithm inside the protocol that periodically and randomly tries to update  the table of nodes available, and the paths of them.

Finally, we need to highlight that the aggregation of data is a complicated area, not being possible to treat it in this thesis.

\subsubsection{Compatibility with centralized models}

In chapters two and three we introduced the different techniques and topologies used by weather organizations to acquire weather data. The centralized model was explained, showing how the nodes have a strong dependency of one common point. This setup is the current solution chosen by weather organizations, and almost all big weather data networks are builded based on such infrastructure.
3
Thus, it is needed to consider the compatibility of OpenWeather with this topology. Although OpenWeather is designed to have the \gls{AWS}es as independent nodes, infrastructures using the centralized model, can provide a node doing a bridge between the collection point and OpenWeather network. It will be required to develop the methods to retrieve the data from the subnet of the collection point. As it was mentioned in section \ref{5.1.2} and the example of \gls{HTTP} and OpenWeather, it is possible to encapsulate data to other protocols with the proper adaption.

Because every weather organization has their own setups and methodologies, an independent study will be required in order to design a bridge from the collection point models to the \gls{P2P} architecture of OpenWeather.

\section{Summary}

In this chapter the core architecture of OpenWeather was presented. The definitions establish by the protocol have been explained. 
The roles of nodes and their identification is presented to justify how OpenWeather can be adapted for future use in a different system for \gls{AWS} identification.

We have explained the architecture of OpenWeather. Justifying the use of the \gls{AWS}es as indivudal nodes conforming a \gls{P2P} network. The protocol functionality is analyzed, explaining how the different operations perform. The main characteristics of OpenWeather have been exposed.

The structure used in OpenWeather data messages has been analyzed, explaining how \gls{JSON} is used as syntax to encapsulate the data. In addition, the application of object hierarchy on data has been explained. All data fields, which compose data messages were defined technically. 

The protocol codes and their categories have been described, justifying their numeration and purposes.

The differences between real-time data messages, data messages on demand and internal data messages, have been justified, putting attention in how the different data messages have a common structure and use. Finally an example of all the types of data messages implemented in the protocol are explained, providing enough information to implement a functional prototype of it.

\pagebreak

%% file: chapter7.tex
\chapter{Experimental evaluation setup}

In this  chapter, the experimental setup used to implemented the proof of concept of the OpenWeather protocol is explained. A generic \gls{AWS} has been setup to test the protocol with real-time data. The software architecture implementing the functionality of the protocol is introduced as well.
The purpose of this chapter is to introduce the general guidelines followed by the implementation of a prototype of OpenWeather protocol and to analyze the tests cases performed using it.

\section{Scenario}
The \gls{AWS} utilized in the experimental setup is the model WXT520, manufactured by Vaisala Oyj. Along with other \gls{AWS}es sharing these characteristics, it is able to measure the following phenomena:

\begin{itemize}
\item Liquid Precipitation
	\begin{itemize}
	\item Hail
	\item Rain
	\end{itemize}
\item Relative Humidity
\item Wind
	\begin{itemize}
	\item Direction
	\item Speed
	\end{itemize}
\item Air Temperature
\item Barometric Pressure
\end{itemize}

The geographical location of the \gls{AWS} is \textbf{N 60º 11' 15.6'' E 024º 50' 14.8''}\footnote{\gls{UTM}: Zone: 35
Easting: 380076 Northing: 6674276. Municipality of Otaniemi, Espoo, Finland.}. The \gls{AWS} has been connected to a computer, in which the software developed to implement OpenWeather protocol is installed. The \gls{AWS} has been configured following the manufacturer suggestions, emulating a normal installation environment. The digital interface configured in the \gls{AWS} is a  \gls{RS232} port, offering a maximum amount of bit rate of 116 \gls{KBITS}/s.
\begin{figure}[H]
\centerline{\includegraphics[width=1\textwidth]{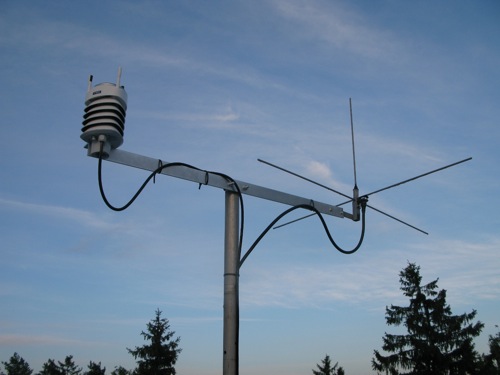}}
\caption{\protect \gls{AWS} installed to simulate a real scenario.}
\end{figure}

The \gls{AWS} is plugged in continually 24 hours and installed on a mast of 2 meters length. The \gls{RS232} port provides the data acquired in the \gls{AWS} to computer a that operates an implementation of OpenWeather protocol.

Thus, the \gls{AWS} used to implement the protocol has not been modified to adapt it to OpenWeather, all the adaptions realized have been made through a software implementation. This fact allows the verification of the adaptability of the protocol to the current technology without no major modifications to the \gls{AWS}.

\subsubsection{Evaluation setup}

The evaluation setup consists four nodes. All of them run a copy of the prototype, thus acting as nodes.
Nevertheless, only one node is connected to a functional \gls{AWS}, the other three simulate the weather data input\footnote{Generated randomly based on the same patterns as a normal \gls{AWS}.}.

The table \ref{t8.1} shows nodes specifications:

\begin{table}[h]
\centering

    \begin{tabular}{ | l | l | l | l | l |}
    \hline
    \textbf{CPU} & \textbf{Memory} & \textbf{Network connection} & \textbf{Operating  system}  & \textbf{Hostname} \\ \hline
	2.4GHz & 4GB & 100Mbps & GNU/Linux & Node 1 \\ \hline
	2.2GHz & 1GB & 100Mbps & GNU/Linux & Node 2  \\ \hline
	900GHz & 1GB &128Kbps & GNU/Linux & Node 3 \\ \hline
	1GHz & 1GB & 56Kbps  & GNU/Linux & Node 4 \\ \hline

\end{tabular}
  \caption{Hardware and \protect \gls{OS} specifications of the evaluation setup.}
  \label{t8.1}
\end{table}

\begin{figure}[H]
\centerline{\includegraphics[width=1\textwidth]{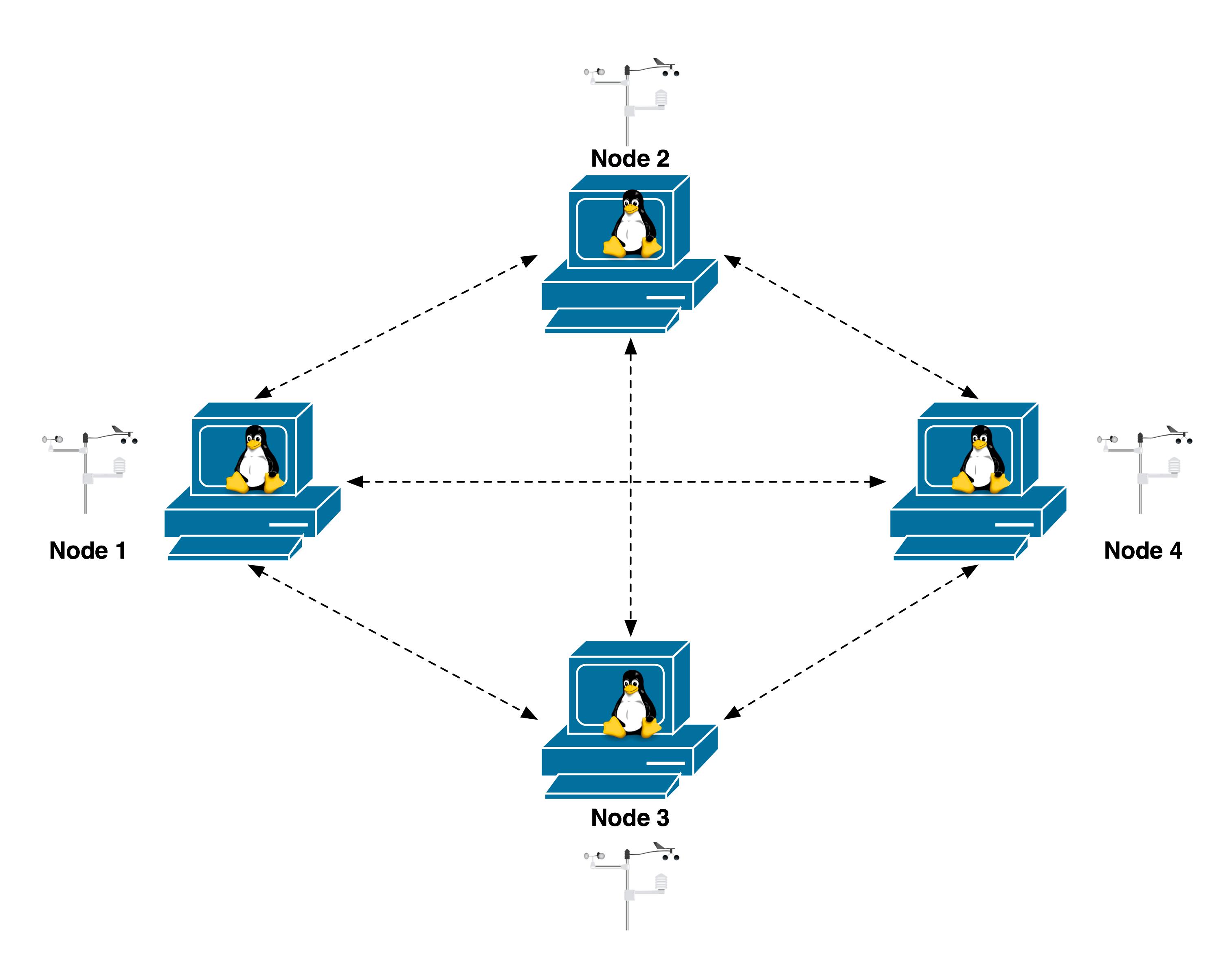}}
\caption{Network topology used in the evaluation setup.}
\end{figure}

All the nodes posses network visibility among them, with maximum network latency less than \textbf{75 milliseconds}. The bandwidth in node number two and four has been limited (\gls{RTT}) to 128\gls{KBITS}/s and 56\gls{KBITS}/s respectively. These restrictions emulate the network limitations mentioned in chapter three.

The purpose of this setup is to create an environment that simulates common conditions experience during weather data acquisition. All the nodes use OpenWeather protocol to exchange data between them. This environment provides the necessary resources to test and verify the characteristics of the protocol, such data message size, times of response, etc.

\section{Prototype implementation}

In order to verify the feasibility and the functionality of OpenWeather, the author developed a proof of concept of the protocol, to test and verify its feasibility as alternative protocol for weather data transmission. This implementation provides the necessary data to independently evaluate the protocol.

\subsection{Technologies used}

OpenWeather is designed to have an emphasis on the data structures used in the software implementation. In addition, the object hierarchy used to structure the data makes the implementation of the protocol easier by using an object oriented programming language.

Thus, C++ has been chosen as the primary language used in the prototype. The C++ standard library is used to write the intermediary layer (in combination with some Python scripts). Because the target of this protocol can have end users which are not familiar with command-line applications, a functional \gls{GUI} has been implemented. The Qt framework\cite{QT} has been chosen to implement the \gls{GUI}, together with QJson\cite{QSJON} for the data representation.

\subsection{Software Architecture}

The prototype requires to be implemented supporting the functionality described in the \gls{P2P} architecture taxonomies\cite{rfc5694}. Thus, the  nodes should posses the capability to request services, and at the same time, offer services to others nodes. This requirement conditions the node to behave as a client and server at the same time.

To realize this architecture, the concept of \textbf{local peer} is introduced. The local peer refers to the node itself; representing the \gls{AWS} entity in the network; nevertheless, as described in section \ref{5.2.1}, a node without an \gls{AWS} can be part of the network as well.

Because the node behaves as  client and server at the same time, the software implementation is designed to maximize the utilization of the common resources between both modalities. Thus, the implementation of the classes have been done using abstract interfaces, not mattering if the data to process has been received through the client or server module.

\subsubsection{Common implementation}

The prototype implementing OpenWeather protocol has been optimized for the data structures and object hierarchy explained in chapter six. The handling of \gls{JSON} data through \gls{TCP} sockets is the basis of the implementation.
The prototype focuses its core functionality in to take advance of the most optimal way of sockets management and data manipulation.

\begin{figure}[H]
\centerline{\includegraphics[width=0.8\textwidth]{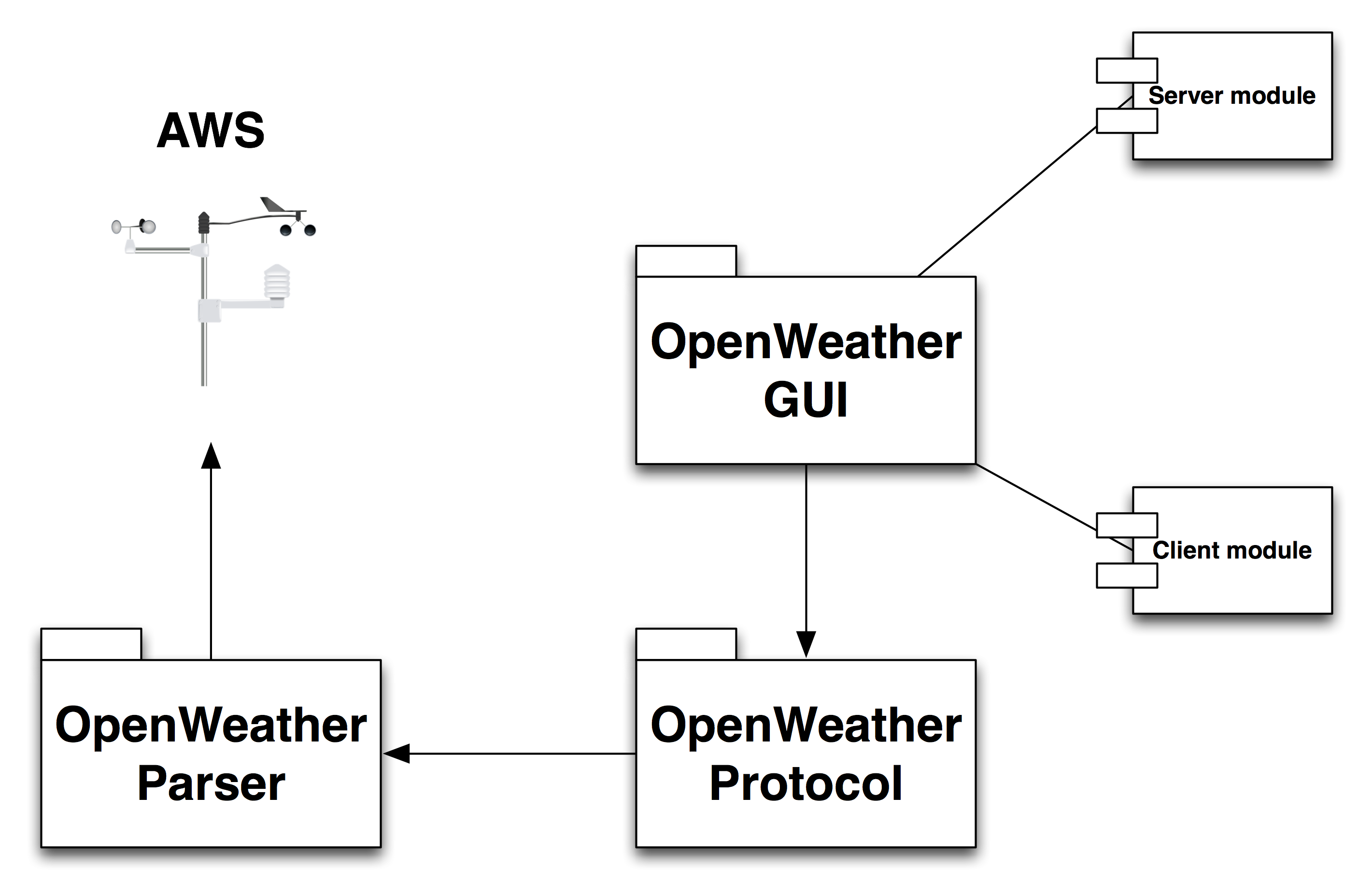}}
\caption{Software prototype conceived.}
\end{figure}

The prototype is structured in three parts:

\begin{itemize}
\item The GUI providing access to certain functionalities of OpenWeather protocol.
\item The network level implementation of OpenWeather protocol.
\item The intermediary middle layer adapting the WXT520 to OpenWeather protocol.
\end{itemize}

Despite this modularity in the components, everything is assembled in one application.
The prototype implements the client and the server modules  internally. Both modules have access to the core implementation of OpenWeather protocol, and at the same time the application is linked with the OpenWeather parser (\emph{libopenweatherparser}). 

The implementation of the protocol has been made based on the objects hierarchy explained in chapter six. Thus, the representation of the OpenWeather data only involves the transformation of \gls{JSON} objects into primitive data types.

\begin{figure}[H]
\centerline{\includegraphics[width=1\textwidth]{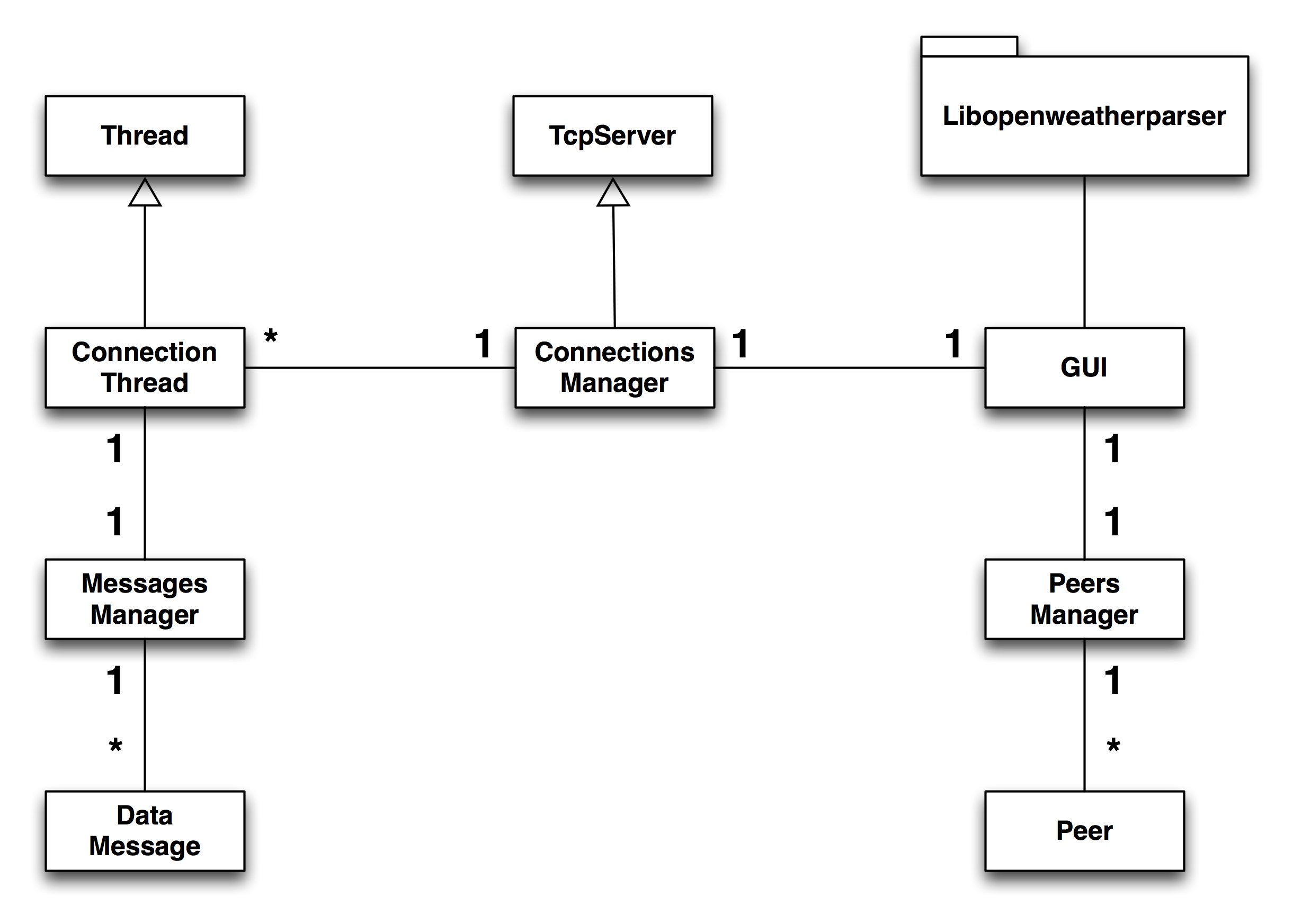}}
\caption{\protect \gls{UML} diagram of the prototype.}
\label{tuml}
\end{figure}

The figure \ref{tuml} shows a general overview of the classes implemented in the prototype in order to make functional the OpenWeather protocol. All the classes developed are able to manage the data in both modes (client and server), being possible to retrieve and delivery data using the same internal software mechanisms, with complete transparency for the end user.

\subsubsection{Client module}

The software implements certain parts fully pertaining to client operations. Client operations are identified those that involve the data request to other nodes. When the software is using OpenWeather to retrieve data from other nodes, we denominate that it is working under client mode.

The client module of the software allows the following operations:

\begin{itemize}
\item Request session establishment - peer handshake
\item Request real-time and/or data on demand
\item Request the service availability in remote node/s
\end{itemize}

\subsubsection{Server module}

As requirement of the \gls{RFC} 5694\cite{rfc5694}, an application implementing a \gls{P2P} architecture must be able to offer services. To achieve this, the prototype implements one part that provides the server functionality. A socket listening to the \gls{TCP} port used in OpenWeather is created when the prototype software is executed. Thus, the software allows other peers to connect to it, providing exactly the same features that client mode is able to request. Because the OpenWeather protocol is designed to not distinguish between the nodes and the services that they offer, the implementation of the server module is nearly identical to the client mode.

The server module o allows the following operations:

\begin{itemize}
\item Session establishment
\item Delivery of real-time and/or on data on demand
\item Delivery of the service available on the local node
\end{itemize}

\subsubsection{GUI}

The graphical interface aims to provide the possibility to use the protocol\footnote{A set of screenshots took from the GUI is available in the appendix.}. The GUI allows fully utilization of the \gls{AWS} data interface to check the data received, to connect to OpenWeather, and to perform the operations described in the chapters six (connect to other peers, delivery real-time data samples or retrieve the services available in the remote peers).

\begin{figure}[H]
\centerline{\includegraphics[width=0.5\textwidth]{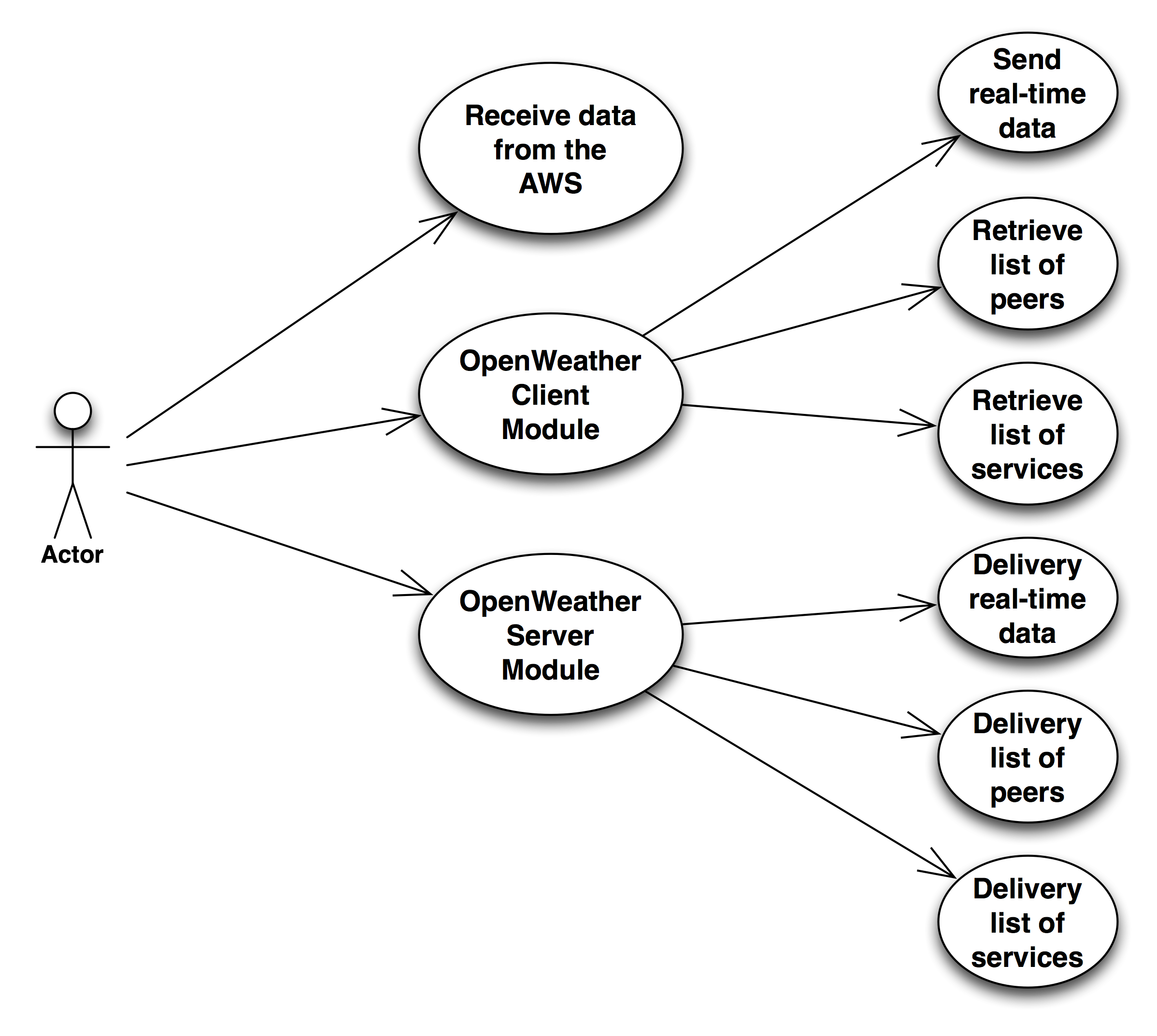}}
\caption{Prototype use case diagram.}
\end{figure}

The \gls{GUI} has single instances of the \textbf{ConnectionsManager} and the \textbf{MessagesManager} classes. Both classes provide the functionality required to handle peers and connections. In addition, the library \textbf{\emph{libopenweatherparser}}, provides the middle layer explained in section \ref{5.2}. This library is linked to the \gls{AWS}, providing the RAW data collected from its digital interface, and converting it from the vendor's format to OpenWeather's format.

\subsubsection{Connections manager}

The ConnectionsManager class is in charge of handling the sockets, managing all the connections of the node. 
In addition, this class controls the socket used to allow remote nodes to connect to the local peer (server module).

All sockets are handled using threads, thus, all the connections are managed independently in a secondary plane, not blocking the \gls{GUI} or not interfering with other connections. This implementation allows the prototype to manage multiple connections with multiple peers without performance issues.

\subsubsection{Peers manager}

The PeersManager class is in charge of the peers. The purpose of this class is to provide a control system of the peers that the node can connect to and their information; at the same time this class manages the local peer and the services that it offers to the remote peers.

This class gets updated information when a the data messages received contain data related with the peers (protocol internal traffic), for instance if some peer updates its metainformation or just confirms the receival of message.

\subsubsection{Messages manager}

The MessagesManager class handles the OpenWeather data messages. This class is able to generate data messages based on the specifications of OpenWeather protocol. Every connection containing a data message is able to access it. This class provides the core functionality of the protocol, being able to understand the protocol codes and based on them, executing the operations needed in order to achieve the expected result.

All data messages are assembled and disassembled in this class, because as OpenWeather requires  \gls{JSON} as its primary data format, this class provides mechanisms to generate and validate the data format of the messages.

\subsubsection{Libopenweatherparser}

This library has been developed in order to create a bridge between the \gls{AWS} and the prototype. The data format used by the vendor in the \gls{AWS} has been implemented in the library, creating the functionality to convert the vendor's format to the OpenWeather data format. This library is thought to normalize the data from one to multiple vendors, offering primitive data types ready to be assembled in \gls{JSON} objects as output.

\begin{figure}[H]
\centerline{\includegraphics[width=1\textwidth]{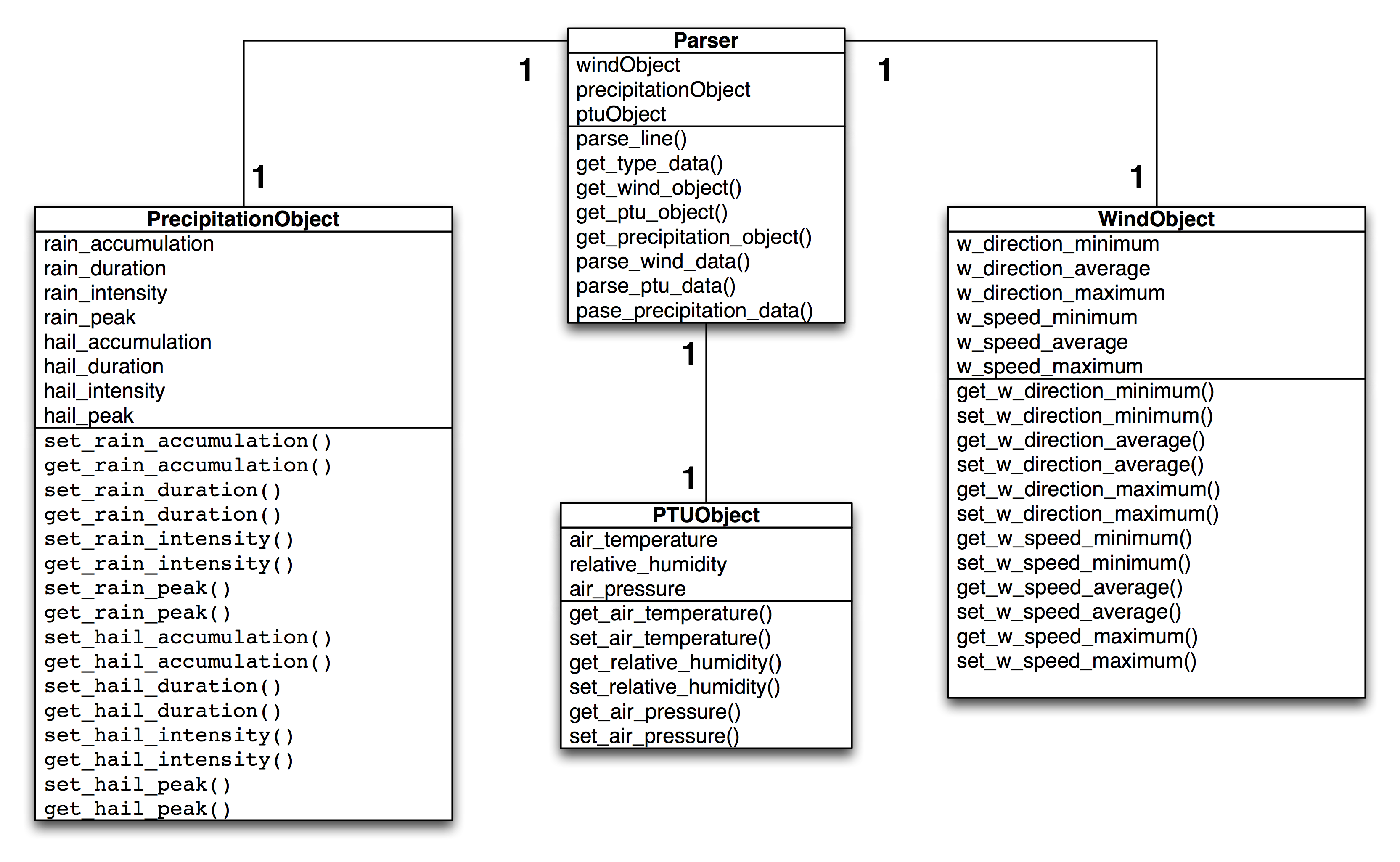}}
\caption{\protect \gls{UML} diagram of the library.}
\end{figure}

The library acts as an intermediary layer. Should the vendors choose to implement the OpenWeather format, the requirement of this library will be dropped. However, since the source code the \gls{AWS} operating system is not available, it is not currently possible to implement the OpenWeather protocol integrated with the vendors software without their cooperation.

\section{Testing}
The prototype provides the capability to perform the operations described on chapter six. The main goal of the testing is to analyze if the implementation of the protocol achieves its purpose and the results that its generates.

The scenario used for the testing is described in the previous section. The following sections explain the utilization of different nodes used to transmit weather data using the OpenWeather protocol.

The methodologies followed to evaluate the behavior of the protocol are based in the analysis of the network traffic between nodes and the verification of the protocol operations. The tool used to capture the data messages is Wireshark\cite{WIRESHARK}. This tool provides enough information to verify the operations of the protocol in the network layer.

The following protocol operations have been implemented in the prototype:

\begin{itemize}
\item Session establishment - peer handshake
\item Service discovery
\item Real-time data retrieval
\end{itemize}

\subsubsection{Implementation considerations}

Although the chapter six specifies more operations, as data on demand or peer list exchange, they have not be implemented due to their similarity in the architecture and data messages size, with the test cases executed.

The Keep-Alive functionality has not being implemented in the prototype, because this feature is just an extra check performed for OpenWeather to double assure the connectivity and the response of the node in the application layer, and it does not influence the functionality of the prototype.

All nodes have been synchronized according with date-time through \gls{NTP} \cite{rfc1305} with the \textbf{ntp1.funet.fi} server before to execute any operation. This synchronization has been performed in order to guarantee the accuracy of the measurements. Nevertheless, as it is explained in the Timestamp section, it is highly recommended to sync the clock of the nodes to guarantee the quality of the weather data.

The RAW \gls{ASCII} representation of the data messages appears in different order compare to the specifications. This is due to the software re-orders the data elements by alphabetical order (always inside of the objects hierarchy).

All the data messages are keeping similar space constrains between the data elements inside the \gls{JSON} object. This is causing a known additional increase of the data message size, this size can be reduced even more, suppressing theses spaces. In addition, the migration to a binary representation of the data messages using \gls{BSON}\cite{BSON} should be straightforward\footnote{Though will cause conflicts with the endianess.}.

The execution of the test has been done 50 times, extracting the average from it. The times of the sequence are including the execution of the software operations in both sides.

\subsection{Test 1: Handshake between nodes}

The purpose of this test is to validate the operation described in section \ref{7.3.1} \textbf{—Session establishment \& Peer handshake—}. In this test the peers involved are exchanging information about themselves, in order to establish the session.

The \emph{Node 1} will send a handshake data message to the \emph{Node 2}. This data message contains all the MetaInfo data field filled with the data of the \emph{Node 1}, the protocol code used is \textbf{100}.

\subsubsection{Sequence}

The scenario assumes that the \emph{Node 1} knows the \gls{IP} address, \gls{TCP} port and ID of the \emph{Node 2}, because it was obtained from some list of peers received from other nodes.

The following sequence happens in the network layer:

\begin{enumerate}

\item \emph{Node 1} sends a data message  containing all its metainformation to the remote \emph{Node 2}, connected through the port specified and requests session establishment.

\item \emph{Node 2} receives a data message delivery by \emph{Node 1}, containing all its metainformation and requesting the session establishment.

\item \emph{Node 2} sends a data message to \emph{Node 1}, providing all its meta-information and confirming the session establishment with the protocol code \textbf{101}.

\item The session is established between both nodes.

\end{enumerate}

The following sequence happens in the software layer:

\begin{enumerate}

\item The button session-establishment generates the connection sequence to the node chosen (\emph{Node 2}).

\item A thread is created, establishing a \gls{TCP} connection to the chosen node. The messages manager assembles data message with all metainformation of the local node and with the protocol code \textbf{100}.

\item The data message is delivery through the socket managed by the thread.

\item The messages manager in the \emph{Node 2} receives a data message and creates a thread to handle it.

\item The messages manager called by the thread, parses the data message and identifies its protocol code.

\item A response is generated based on the protocol code of the data message, and is deliver to \emph{Node 2}.

\item Because the operation is the session establishment the peers manager gets executed in both sides, updating the peer information (if needed) of the peers.
\end{enumerate}

\subsubsection{Analysis}

\begin{itemize}
\item The data session captured with Wireshark involves \textbf{7} \gls{TCP} segments. 
\item The data message (OpenWeather) generated by \textbf{\emph{Node 1}}, has a size of \textbf{375} bytes.
\item The data message (OpenWeather) generated by \textbf{\emph{Node 2}}, has a size of \textbf{375} bytes.
\item The total size of the OpenWeather data message is \textbf{750} bytes.
\item The total size of the sequence (\gls{TCP}/\gls{IP} and OpenWeather) is \textbf{1227} bytes. 
\end{itemize}
 
The RAW \gls{ASCII} representation of the data message data capture is shown in table \ref{raw1}.

\begin{table}[H]
\centering
\begin{tabular}{|l|l|l|p{8cm}|}
\hline    
\textbf{\emph{Node 1}}\\
\hline
\begin{minipage}[t]{\linewidth}
\begin{verbatim}
{ "OpenWeatherMessage" : { "MetaInfo" : { "Bandwidth" : 6, "ID" :
"33c11957579d1093e931bd540536b40e90339dbded8e2a2ce4e
64c480c8132bc", "Keep-Alive" : 120000, "Location" : "6672224
385565 35V", "Peer-IP" : "172.21.25.16", "Peers-Requested" : 
20, "Port" : 62535, "Timestamp" : "2011-07-20T16:51:29", "Update
-Interval" : 120000, "Version" : "OpenWeather/1.0" }, "Type" : 100 } 
}

\end{verbatim}
\end{minipage} \\
\hline    
\textbf{\emph{Node 2}}\\
\hline
\begin{minipage}[t]{\linewidth}
\begin{verbatim}
{ "OpenWeatherMessage" : { "MetaInfo" : { "Bandwidth" : 6, "ID" : 
"11f1cb9fb5bc57cf7905dc26c3ef045ae7b54d5ff1c7e233ff2d31be
4977bd18", "Keep-Alive" : 120000, "Location" : "6672224 385565
35V", "Peer-IP" : "172.21.25.20", "Peers-Requested" : 20, "Port" 
: 62535, "Timestamp" : "2011-07-20T16:51:29", "Update-Interval" 
: 120000, "Version" : "OpenWeather/1.0" }, "Type" : 101 } 
}
\end{verbatim}
\end{minipage} \\
\hline
\end{tabular}
\caption{Data messages transmitted between \emph{Node 1} and \emph{Node 2}.}
\label{raw1}
\end{table}

The \gls{TCP} flow between both nodes using OpenWeather is the following:

\begin{table}[H]
\begin{minipage}[t]{\linewidth}
\begin{verbatim}
| 172.21.25.16          172.21.25.20    |               
|         SYN       |                   |Seq = 0 Ack = 1303623571
|(39239)  ------------------>  (62535)   |
|         SYN, ACK  |                   |Seq = 0 Ack = 1
|(39239)  <------------------  (62535)   |
|         ACK       |                   |Seq = 1 Ack = 1
|(39239)  ------------------>  (62535)   |
|         PSH, ACK - Len: 375           |Seq = 1 Ack = 1
|(39239)  ------------------>  (62535)   |
|         ACK       |                   |Seq = 1 Ack = 376
|(39239)  <------------------  (62535)   |
|         PSH, ACK - Len: 375           |Seq = 1 Ack = 376
|(39239)  <------------------  (62535)   |
|         ACK       |                   |Seq = 376 Ack = 376
|(39239)  ------------------>  (62535)   |
\end{verbatim}
\end{minipage}
\caption{\protect \gls{TCP} flow sequence between \emph{Node 1} and \emph{Node 2}.}
\end{table}
The time of execution of this \gls{TCP} sequence is \textbf{65 milliseconds} on average.

Both nodes have delivered the data successfully, achieving the session establishment as result of the sequence. 

\subsubsection{Discussion}

The measurements show that OpenWeather requires a small amount of data for the session establishment. In addition, a low response time is needed to complete the operation. It achieves the goal to provide a mechanism to establish session even with really low bandwidth availability. This small size of data can be easily handled by the memory and processor of an \gls{AWS}. As the protocol specification requires, the session establishment provides all the necessary information to both nodes, to proceed requesting other data, after the peer registration happens in the software side.

\subsection{Test 2: Service discovery}

The purpose of this test is to validate the operation described in section \ref{7.3.2} \textbf{—Service discovery—}. In this test the peers involved are exchanging information about service availability, in order to know which services could be requested.

The \emph{Node 3} will send a service discovery data message to \emph{Node 4}. This data message contains all the MetaInfo data field filled with the data of the \emph{Node 3}, in addition the protocol code used is \textbf{102}.

\subsubsection{Sequence}

The scenario assumes that \emph{Node 3} and \emph{Node 4} have established the session, following exactly the same steps than mentioned in section \ref{7.3.1}.

The following sequence happens in the network layer:

\begin{enumerate}

\item \emph{Node 3} sends a data message  containing all its metainformation to the remote host of the \emph {Node 4}, using the session already established between them.

\item \emph{Node 4} receives a data message delivered by \emph{Node 3}, containing all its metainformation and requesting the services available on it.

\item \emph{Node 4} sends a data message to the \emph{Node 3}, providing all its metainformation and delivering a data message with all the services available on it through the protocol code \textbf{103}.

\item \emph{Node 3} receives the list of services available in the \emph{Node 4}.

\end{enumerate}

The following sequence happens in the application layer:

\begin{enumerate}

\item The button services discovery generates the  connection sequence to the node chosen (\emph{Node 4}).

\item The thread previously created by the session, uses the \gls{TCP} connection established  to the chosen node. The messages manager assembles a data message with all the metainformation of the local node and sends through connection with the protocol code \textbf{102}.

\item The data message is delivery through the socket managed by the thread.

\item The connections manager in \emph{Node 3} receives a data message and creates a thread to handle it.

\item The messages manager is called by the thread, parses the data messages and identifies its protocol code.

\item A response is generated based on the protocol code of the data message, and is deliver to the \emph{Node 4}.

\item Due to the operation being service discovery, the peers manager gets executed in \emph{Node 4}, checking the services available on it and providing their information into the data message.

\end{enumerate}

\subsubsection{Analysis}

\begin{itemize}
\item The data session captured with Wireshark involves \textbf{7} \gls{TCP} segments. 
\item The data message (OpenWeather) generated by the \textbf{\emph{Node 3}}, has a size of \textbf{375} bytes.
\item The data message (OpenWeather) generated by the \textbf{\emph{Node 4}}, has a size of \textbf{458} bytes.
\item The total size of the OpenWeather data message is \textbf{833} bytes.
\item The total size of the sequence (\gls{TCP}/\gls{IP} and OpenWeather) is \textbf{1310} bytes. 
\end{itemize}
 
The RAW \gls{ASCII} representation of the data message captured is shown table \ref{raw2}

\begin{table}[H]
\centering
\begin{tabular}{|l|l|l|p{8cm}|}
\hline    
\textbf{\emph{Node 3}}\\
\hline
\begin{minipage}[t]{\linewidth}
\begin{verbatim}
{ "OpenWeatherMessage" : { "MetaInfo" : { "Bandwidth" : 1, "ID" :
"654b7b521acc7549bf6854b1113d44e6433bf94a1b4caf4327e33
e9bc89b4025", "Keep-Alive" : 120000, "Location" : "6672224 385
565 35V", "Peer-IP" : "172.21.25.35", "Peers-Requested" : 20, 
"Port" : 62535, "Timestamp" : "2011-07-24T12:04:09", "Update-
Interval" : 120000, "Version" : "OpenWeather/1.0" }, "Type" : 102 }
}

\end{verbatim}
\end{minipage} \\
\hline    
\textbf{\emph{Node 4}}\\
\hline
\begin{minipage}[t]{\linewidth}
\begin{verbatim}
{ "OpenWeatherMessage" : { "Info" : { "Services" : { "PRECIPITATION"
: "RO", "PTU" : "RO", "WIND" : "RO" } }, "MetaInfo" : { "Bandwidth" : 
0, "ID" : "3b1f665e0d622aab7b2e71b29d966dd2a22c5d427f337585
09d4205720de9d2e", "Keep-Alive" : 120000, "Location" : "6672224 
385565 35V", "Peer-IP" : "172.21.25.40", "Peers-Requested" : 20, 
"Port" : 62535, "Timestamp" : "2011-07-24T12:04:09", "Update-
Interval" : 120000, "Version" : "OpenWeather/1.0" }, "Type" : 103 } 
}
\end{verbatim}
\end{minipage} \\
\hline
\end{tabular}
\caption{Data messages transmitted between \emph{Node 3} and \emph{Node 4}.}
\label{raw2}
\end{table}

The \gls{TCP} flow between both nodes using OpenWeather is the following:

\begin{table}[H]
\begin{minipage}[t]{\linewidth}
\begin{verbatim}
| 172.21.25.35             172.21.25.40 |                
|         SYN       |                   |Seq = 0 Ack = 2259331907
|(50550)  ------------------>  (62535)   |
|         SYN, ACK  |                   |Seq = 0 Ack = 1
|(50550)  <------------------  (62535)   |
|         ACK       |                   |Seq = 1 Ack = 1
|(50550)  ------------------>  (62535)   |
|         PSH, ACK - Len: 375           |Seq = 1 Ack = 1
|(50550)  ------------------>  (62535)   |
|         ACK       |                   |Seq = 1 Ack = 376
|(50550)  <------------------  (62535)   |
|         PSH, ACK - Len: 458           |Seq = 1 Ack = 376
|(50550)  <------------------  (62535)   |
|         ACK       |                   |Seq = 376 Ack = 459
|(50550)  ------------------>  (62535)   |
\end{verbatim}
\end{minipage}
\caption{\protect \gls{TCP} flow sequence between \emph{Node 3} and \emph{Node 4}.}
\end{table}
The time of execution of this \gls{TCP} sequence is \textbf{84 milliseconds} on average.

Both nodes have delivered the data successfully, achieving the service discovery as the result of the sequence.

\subsubsection{Discussion}

The service discover operation has bigger data message size than the session establishment. Nevertheless, this operation considered fairly small in size, and it fitting to the environment with low bandwidth available. As the session establishment, the service discovery is a common operation inside of the protocol. Its fast delivery is critical, thus, in order to provide the services available as soon as possible to the requester.

\subsection{Test 3: Real-time data retrieval}

The purpose of this test is to validate the operation described in section 7.3.3 \textbf{—Real-time data retrieval—}. In this test the peers involved are exchanging real-time weather data.

The \emph{Node  4} will send a real-time data request to \emph{Node 1}. This data message contains all the MetaInfo data field filled with the data of \emph{Node  4},  in addition the protocol code used is\textbf{200}.

\subsubsection{Sequence}

The scenario assumes that \emph{Node 4} and \emph{Node 1}, have established the session, following exactly the same steps mentioned in section \ref{7.3.1}.

The following sequence happens in the network layer:

\begin{enumerate}

\item \emph{Node 4} sends a data message containing all its meta-information to the remote host of the \emph{Node 1}, using the session already established between them.

\item \emph{Node 1} receives a data message delivered by \emph{Node 4}, containing all its metainformation and requesting real-time data.

\item \emph{Node 1} sends a data message  to \emph{Node 4}, providing all its metainformation and delivering a data message with the latest weather data available on its \gls{AWS}, assigning the protocol code \textbf{201}.

\item \emph{Node 4} receives the latest real-time data sample available in \emph{Node 1}.

\end{enumerate}

The following sequence happens in the application layer:

\begin{enumerate}

\item The button assign to request real-time data, generates the connection sequence to the node chosen (\emph{Node 1}).

\item The thread previously created by the session, uses the \gls{TCP} connection established  to the chosen node. The messages manager assembles a data message with all the metainformation of the local node and assigning the protocol code \textbf{200}.

\item The data message is delivered through the socket managed by the thread.

\item The connections manager in \emph{Node 1} receives a data message and creates a thread to handle it.

\item The messages manager called by the thread, parses the data messages and identifies its protocol code.

\item A response is generated based on the protocol code of the data message. Since this response involves real-time weather data, a call is made to the \textbf{libopenweatherparser}, to obtain the latest real-time data available in the \gls{AWS}. After that, the data is deliver to \emph{Node 4}.

\end{enumerate}

\subsubsection{Analysis}

\begin{itemize}
\item The data session captured with Wireshark involves \textbf{7} \gls{TCP} segments.
\item The data messages (OpenWeather) generated by \textbf{\emph{Node 4}}, has a size of \textbf{375} bytes.
\item The data messages (OpenWeather) generated by \textbf{\emph{Node 1}}, has a size of \textbf{814} bytes.
\item The total size of the OpenWeather data messages is \textbf{1189} bytes.
\item The total size of the sequence (\gls{TCP}/\gls{IP} and OpenWeather) is \textbf{1666} bytes. 
\end{itemize}
 
The RAW \gls{ASCII} representation of the data messages is shown in table \ref{raw3}.

\begin{table}[H]
\centering
\begin{tabular}{|l|l|l|p{8cm}|}
\hline    
\textbf{\emph{Node 4}}\\
\hline
\begin{minipage}[t]{\linewidth}
\begin{verbatim}

{ "OpenWeatherMessage" : { "MetaInfo" : { "Bandwidth" : 0, 
"ID" : "3b1f665e0d622aab7b2e71b29d966dd2a22c5d427
f33758509d4205720de9d2e", "Keep-Alive" : 120000, "
Location" : "6672224 385565 35V", "Peer-IP" : "172.21.
25.40", "Peers-Requested" : 20, "Port" : 62535, "Timest
amp" : "2011-07-25T14:15:35","Update-Interval" : 
120000, "Version" : "OpenWeather/1.0" },"Type" : 200 } }

\end{verbatim}
\end{minipage} \\
\hline    
\textbf{\emph{Node 1}}\\
\hline
\begin{minipage}[t]{\linewidth}
\begin{verbatim}
{ "OpenWeatherMessage" : { "Data" : { "PRECIPITATION" : {
 "Hail" : { "accumulation" : "0", "duration" : "0", "intensity" : "0"
, "peak" : "0" }, "Rain" : { "accumulation" : "0", "duration" : "0",
"intensity" : "0", "peak" : "0" } }, "PTU" : { "Air-Pressure" : "10
14.1", "Air-Temperature" : "19.1", "Relative-Humidity" : "69.4" 
}, "WIND" : { "Direction" : { "ave" : "160", "max" : "160", "min"
: "160" }, "Speed" : { "ave" : "1.7", "max" : "1.8", "min" : "1.7"
} } }, "MetaInfo" : { "Bandwidth" : 6, "ID" :"33c11957579d10
93e931bd540536b40e90339dbded8e2a2ce4e64c480c8132
bc", "Keep-Alive" : 120000, "Location" : "6672224 385565 35V
", "Peer-IP" : "172.21.25.16", "Peers-Requested" : 20, "Port"
: 62535, "Timestamp" : "2011-07-25T14:15:35", "Update-Inter
val" : 120000, "Version" : "OpenWeather/1.0" }, "Type" : 300 }
}
\end{verbatim}
\end{minipage} \\
\hline
\end{tabular}
\caption{Data messages sent between \emph{Node 3} and \emph{Node 4}.}
\label{raw3}
\end{table}

The \gls{TCP} flow between both nodes using OpenWeather is the following:

\begin{table}[H]
\begin{minipage}[t]{\linewidth}
\begin{verbatim}
| 172.21.25.20             172.21.25.40 |
|         SYN       |                   |Seq = 0 Ack = 1015394402
|(49983)  ------------------>  (62535)   |
|         SYN, ACK  |                   |Seq = 0 Ack = 1
|(49983)  <------------------  (62535)   |
|         ACK       |                   |Seq = 1 Ack = 1
|(49983)  ------------------>  (62535)   |
|         PSH, ACK - Len: 374           |Seq = 1 Ack = 1
|(49983)  ------------------>  (62535)   |
|         ACK       |                   |Seq = 1 Ack = 375
|(49983)  <------------------  (62535)   |
|         PSH, ACK - Len: 814           |Seq = 1 Ack = 375
|(49983)  <------------------  (62535)   |
|         ACK       |                   |Seq = 375 Ack = 8153
|(49983)  ------------------>  (62535)   |
\end{verbatim}
\end{minipage}
\caption{\protect \gls{TCP} flow sequence between \emph{Node 1} and \emph{Node 2}.}
\end{table}
The time of execution of this \gls{TCP} sequence is \textbf{96 milliseconds} on average.

Both nodes have delivered the data successfully, achieving the transmission of a real-time weather sample as result of the sequence.

\subsubsection{Discussion}

Though this real-time data sample does not contain rain or hail data (both are delivered with a 0 value), we can observe how the \gls{PTU} and the wind data (together with the MetaInfo data field) are up to \textbf{1.5} \gls{kB}. 
Even with this data size, it will fit in the memory available in an \gls{AWS} described in section \ref{3.1.2}. Assuming that an \gls{AWS} has between 32-64 \gls{kB} of volatile memory, and taking half of its memory for internal use of \gls{AWS} operating system, there is still enough memory to handle real-time data samples using the OpenWeather protocol.

\section{Summary}

In this chapter the scenario and software architecture used to evaluate OpenWeather has been introduced.
We tested three different use cases of the protocol with the prototype developed. 

In all the use cases executed, the protocol is taking advantage of its properties and achieving a successful result. 

Though the prototype implements the partial functionality of the OpenWeather protocol, it shows how the \gls{P2P} can be implemented in applications oriented to weather transmission. In addition, the small sizes of the data messages and the robustness of the data transmission offered by \gls{TCP}, provide enough confidence to confirm that the protocol can be implemented in the environments with low bandwidth availability.

Our goal was to verify a feasible implementation of the OpenWeather protocol and verify its functionality with a real scenario. Both purposes have been achieved.

Finally, the use of a real scenario and the integration of the prototype with it, proves how a modern \gls{AWS} can be adapted to OpenWeather protocol with a few modifications through a software adaption. This fact supports that the current technology can be adapted to new methodologies to transmit the weather data, without a modification in the electronics or industrial design of the \gls{AWS}.

\pagebreak

%% file: chapter8.tex
\chapter{Conclusions}

In this thesis we exposed the basis of weather observation, how different organizations around the world are collecting and studying enormous amounts data of different phenomena. From the very beginning the industry has been building really complex instruments to measure these phenomena. Many people, from individuals to scientists, are spending their time and resources to part take in the worldwide observation of weather. It is a fact that we need to understand the weather in order to better understand our planet and implicitly, to increase our quality of life.

We have analyzed how the instruments used for such purposes and their limits restrict our knowledge expansion. We described how the industry has been improving these instruments in many different  ways. Areas such as the industrial design of the instruments or their internal electronics, have been experiencing tremendous improvements during the last decades, thus allowing the industry to offer weather measure instruments of strong robustness and high accuracy.

Based on the study of these instruments and the scientific discussion of those using them, such the SMEAR project\cite{SMEAR}, we have come to a conclusion that methods used in them can be improved significantly concerning real-time weather data transmission.

Through the analysis of the different architectures used to collect the weather data, we found several points related to technologies used on network level that need to be changed in order to achieve a successful delivery of real-time data.

We explained how the industry have been introducing new digital interfaces in order to adapt the \gls{AWS} to the new standards. Nevertheless, although the digital interfaces have been upgraded, the protocols used to transmit the data through them have certain particularities such the use of vendor data specific formats.

In addition, the analysis performed in different instruments and the network technologies that they use, has indicated that the data format and the protocol standards used are of low compatibility with capabilities such real-time data acquisition or data exchange.

The mainstream methodologies currently used to transmit the weather data, such the \gls{FTP} or the use of \gls{CSV} as data formats, are limiting the possibility to deliver data with frequency and accuracy high enough to consider it real-time data. Nevertheless, these methodologies are currently considered the state of the art and thought to be sufficient for performing in current architectures used to acquire data.

Though some organizations as \gls{NOAA} or \gls{ICAO}, have been creating some data formats for certain purposes (such air navigation or \gls{CWOP}),nowadays , the global standard still not adapted for the weather industry. The \gls{WMO}, conscious of this situation, started a process of standardization for weather data representation in 2002. At the moment, this process still under development without any official standard published.

The absence of a standard data format and a protocol to transmit it, is avoiding the possibility to take advantage of all the capabilities that an \gls{AWS} can offer, more specifically the real-time data acquisition. Although the weather organizations have access to weather data samples updated with small frequencies of time, programs as \gls{GOS} or \gls{GDPFS}, are seeking to establish the basis of future systems for weather observation, providing features as real-time capabilities and compatibility between data formats.

All the issues mentioned previously have been considered during the development of OpenWeather. As solution for the problem statement, OpenWeather aims to provide all the features necessary to take advantage of the weather instruments concerning their capabilities to accomplish weather data transmission in real-time.

Based on the architecture used to collect weather data, we use its topology to adapt it to the \gls{P2P} architecture. Thus, we transform any \gls{AWS} in a node offering services to other nodes. To achieve such behavior, we developed the OpenWeather protocol from scratch, conceiving it will all the necessary properties to make it \gls{P2P} and at the same time, adapting its core functionality to the weather data requirements. Being conscious of the absence of standards in such area, OpenWeather has been designed adopting as much standards as possible into its architecture, such the use of standard measurement units or date-time format.

As a result, OpenWeather provides a new way to transmit weather data and to interact with the \gls{AWS}es. 

The implementation of the protocol in a software prototype and its posteriorly use, verify its feasibility in order to translate the protocol specifications to a functional software implementation to be tested in a more complex scenario.

In the experimental setup we verify that OpenWeather —in its implementation as prototype— works in a scenario using the same technologies that are currently common among weather observation experts. The prototype implemented gives us the possibility to communicate with other nodes, executing the protocol operations designed to achieve the weather data transmission. In addition, the \gls{P2P} functionality of the protocol has been tested, verifying that the \gls{AWS}es can be treated as independent nodes, requesting and offering services at the same time, and still achieving a successful weather data transmission without a centralized collection point.

We identify as requirement the adaption of the intermediary layer developed to other vendor's data formats, in order to make compatible OpenWeather with different weather instruments from different brands.

Although we described how nodes using the OpenWeather protocol could be able to gather data between them, such functionality has not being implemented in the prototype. Hence, future research should be performed in order to evaluate the capabilities of the protocol to scale in large networks. In addition, the implementation of weather data networks using scalable methodologies, should be study together with their connectivity technologies. Thus, the possibility to use other protocols on the \gls{AWS}es to transport data instead of \gls{TCP}, should be considered, looking for protocols more optimized for low bandwidth availability.

Through the execution of the test cases, we analyzed the results of the protocol in the scenario given. These results show how the protocol can fit in the technical specifications of an \gls{AWS}, making possible to use it in future adaptations.

The main goal of this thesis has been to study state of the affairs in weather observation systems, their technologies and methodologies, trying to find ways of their improvement. OpenWeather fits that goal. Through the prototype we can show how the weather data transmission can be improved in several aspects from network topology to data structure use.

This topic suggests deeper research, as it could provide a solid basis for future implementation of a global real-time weather observation with a high capability in data exchange operations. In addition, in this thesis we have not treated security matters related with the weather data transmission. Despite the nature of the weather data, a complete solution has to consider security threats. Thus, an independent study is required to evaluate how the weather data transmission can be protected. Although, it would be possible to use cryptographic protocols such as \gls{TLS} together with OpenWeather, such combination will have an impact on the bandwidth used to transmit weather data. In addition, \gls{ACL} mechanisms could be considered to assure the identity of the nodes and their locations, in order to guarantee their legitimacy. Moreover, weather data networks can be an objective of \gls{DOS} or \gls{DDOS} attacks. Although this should be treated independently of OpenWeather protocol, future adaptions of it should have these threats in consideration to provide methodologies to lead with them.

The involvement of organizations such \gls{WMO} and the vendors,  is critical to make this happen, possibly in cooperation with standardization organizations for communication protocols such the \gls{IETF}.In addition, any adaption of the industry to protocols designed and adapted for a most efficient use of resources available, will provide an improvement in their products, providing new ways to use their instruments to understand the weather phenomena.

Finally the author believes that the understanding of the weather phenomena will be accompanied by open and scalable network technologies. Thus, the OpenWeather protocol could be a first step to make it happen.

\pagebreak

%% file: appendix.tex
\phantomsection
\addcontentsline{toc}{chapter}{Appendix I \& Appendix II}
\sectionwp{Appendix I}
\begin{table}[H]
\centering
    \begin{tabular}{ | l | l | l | l |}
    \hline
    \textbf{Protocol code} & \textbf{Description} & \textbf{Category} \\ \hline
    
    100 & HANDSHAKE & Protocol codes - Requests\\ \hline

    101 & HANDSHAKE-S & Protocol codes - Status\\ \hline

    102 & SERVICES-AVAILABLE & Protocol codes - Requests\\ \hline

    103 & SERVICES-AVAILABLE-R & Protocol codes - Retrievals\\ \hline

    104 & SERVICES-AVAILABLE-S & Protocol codes - Status\\ \hline

    104 & LIST-PEERS & Protocol codes - Requests\\ \hline

    105 & LIST-PEERS-R & Protocol codes - Retrievals\\ \hline

    106 & LIST-PEERS-S & Protocol codes - Status\\ \hline

    200 & REAL-TIME-DATA & Peer codes - Requests\\ \hline

    201 & ON-DEMAND-DATA & Peer codes - Requests\\ \hline

    300 & REAL-TIME-DATA-R & Peer codes - Retrievals\\ \hline

    301 & ON-DEMAND-DATA-R & Peer codes - Retrievals\\ \hline

    500 & REAL-TIME-DATA-S & Peer codes - Status\\ \hline

    501 & ON-DEMAND-DATA-S & Peer codes - Status\\ \hline

    \end{tabular}    
\end{table}
\thispagestyle{empty}
\pagebreak

\sectionwp{Appendix II}
\thispagestyle{empty}

\begin{figure}[H]
\centerline{\includegraphics[width=1\textwidth]{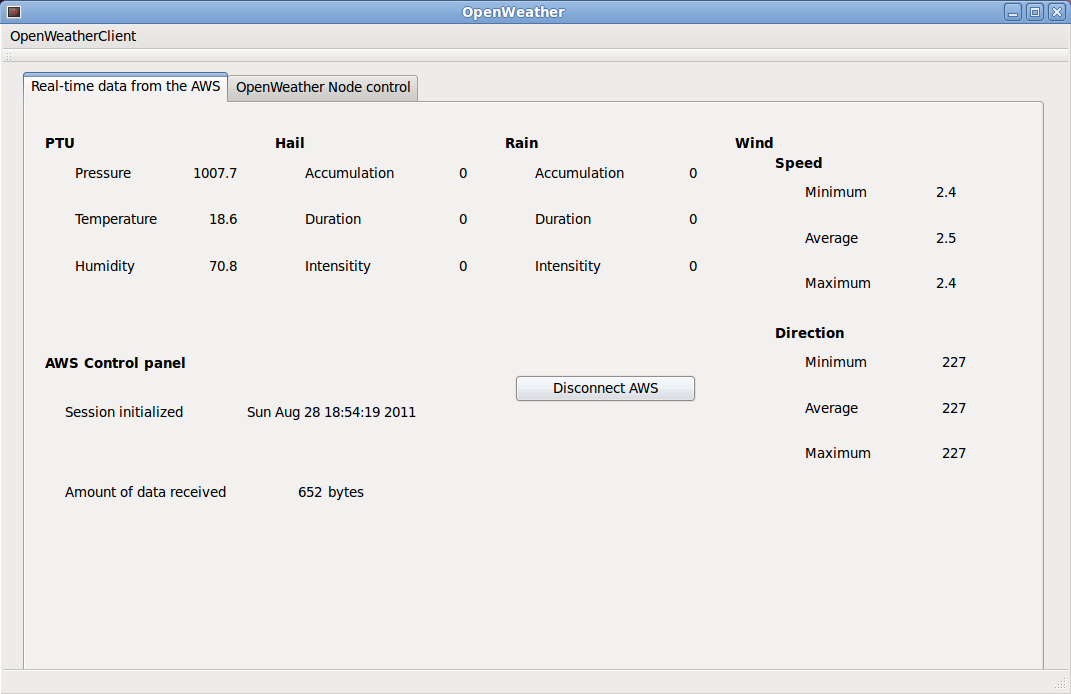}}
\caption{\protect \gls{GUI} of the OpenWeather prototype -AWS control-.}
\end{figure}
 
\begin{figure}[H]
\centerline{\includegraphics[width=1\textwidth]{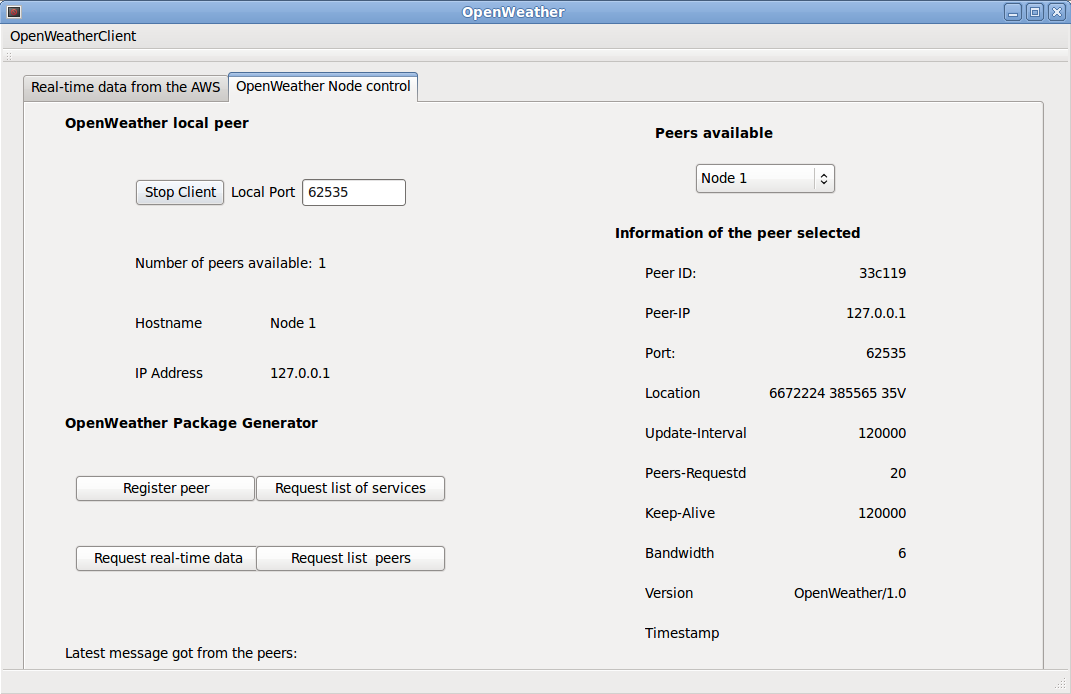}}
\caption{\protect \gls{GUI} of the OpenWeather prototype -Node control-.}
\end{figure}

\begin{figure}[H]
\centerline{\includegraphics[width=1\textwidth]{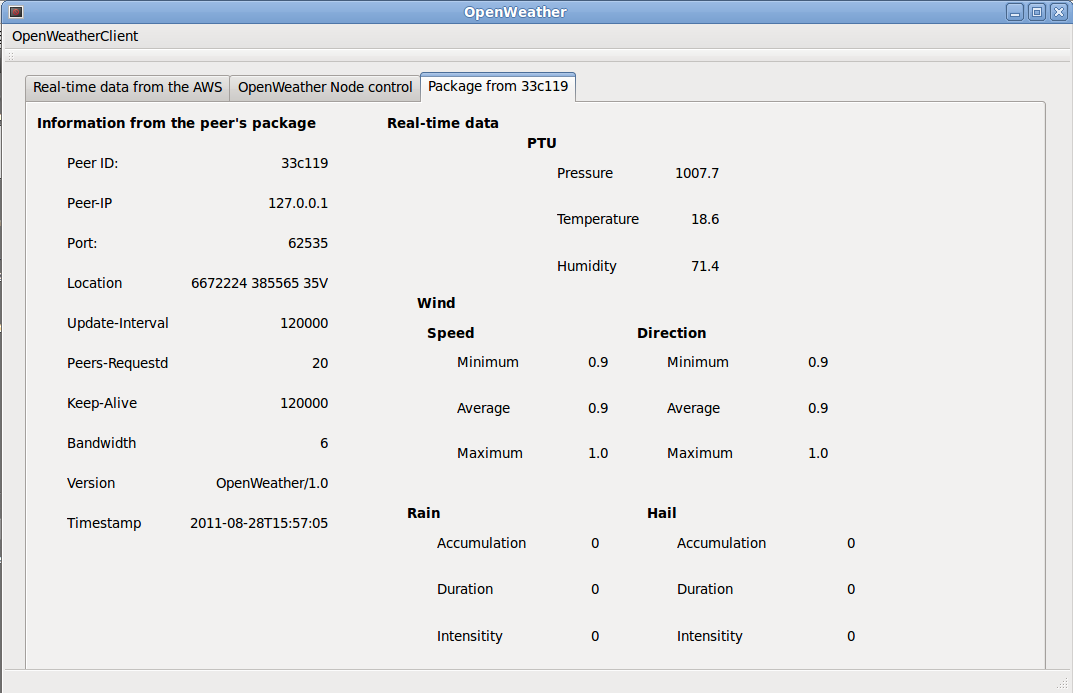}}
\caption{\protect \gls{GUI} of the OpenWeather prototype -Data message visualizer-.}
\end{figure}
\thispagestyle{empty}
\pagebreak